\RequirePackage{fix-cm}
\documentclass[twocolumn]{svjourArxiv}          
\smartqed  
\usepackage{graphicx}
\usepackage{amsmath}
\usepackage{hyperref}	
\hypersetup{colorlinks=true,linkcolor=blue,citecolor=blue,filecolor=blue,urlcolor=blue}
\usepackage{multirow}
\usepackage{subfigure}

\DeclareMathSymbol{\Omega}{\mathalpha}{operators}{"0A}
\newcommand\degr{\hbox{$^\circ$}}
\newcommand{\doi}[1]{\textsc{doi}: \href{http://dx.doi.org/#1}{\nolinkurl{#1}}}

\newcommand{\fref}[1]{Fig.~\ref{#1}}

\journalname{Nonlinear Dynamics}

\begin{document}

\title{On the predictability of Galileo disposal orbits
}
%



\author{David J. Gondelach \and Roberto Armellin \and Alexander Wittig}

\authorrunning{D. J. Gondelach, R. Armellin, A. Wittig} 

\institute{D.J. Gondelach \at
              Surrey Space Centre, University of Surrey, GU2 7XH, Guildford, United Kingdom \\
              \email{d.gondelach@surrey.ac.uk}, ORCID: 0000-0002-8511-9523
           \and
           R. Armellin \at
              Surrey Space Centre, University of Surrey, GU2 7XH, Guildford, United Kingdom \\
              \email{r.armellin@surrey.ac.uk}
           \and
           A. Wittig \at
              Astronautics Research Group, University of Southampton, Southampton, SO17 1BJ, United Kingdom \\
              \email{a.wittig@soton.ac.uk}
}

\date{~}

\maketitle

\begin{abstract}
The end-of-life disposal of Galileo satellites is needed to avoid collisions with operational spacecraft and to prevent the generation of space debris. Either disposal in stable graveyard orbits or disposal into the atmosphere exploiting eccentricity growth caused by lunisolar resonances are possible. However, there is a concern about the predictability of MEO orbits because of possible chaotic behaviour caused by the overlap of resonances. In this work, we investigate if Galileo disposal orbits are predictable and if safe disposal is possible under initial uncertainties. For this, we employ finite-time Lyapunov exponents (FTLE) and sensitivity analysis and compare these two methods regarding their practicality for analysing the predictability of disposal orbits. The results show that FTLE are not suitable for determining if an orbit behaves chaotically or not on the time scale of interest. Sensitivity analysis, on the other hand, can be used to quantify the effect of uncertainties on the orbital evolution and to determine if safe disposal is possible. In addition, we show that reliable re-entry disposal is feasible with a limited $\Delta V$ budget. However, safe disposal into a graveyard orbit is not always feasible considering uncertainties in the disposal manoeuvre and dynamical model when the available $\Delta V$ for disposal is low.

\end{abstract}

\section{Introduction}

Most satellites of the four main global satellite navigation systems (GPS, GLONASS, Galileo and BeiDou) are located in the Medium Earth orbit (MEO) region between 19,100 and 23,222 km altitude and with inclinations between 54.8$\degr$ and 64.8$\degr$ \cite{radtke2015impact}. After their operational lifetime, it is important to put these satellites in stable graveyard orbits or to dispose them into the atmosphere to avoid collisions with operational spacecraft or between themselves, which would generate space debris. 

The MEO region is also a location where lunisolar resonances occur, which mainly depend on the orbit inclination \cite{cook1962luni,hughes1980,ely1997stability}. These resonances can cause the eccentricity to grow secularly and as a result an initially near-circular MEO orbit can re-enter in the atmosphere within 200 years \cite{chao2004long,rossi2008,Deleflie2011}. On the one hand, this instability in eccentricity is undesirable for safe disposal in graveyard orbits, because disposed satellites must keep a safe distance from the operational spacecraft. On the other hand, the instability can be exploited to lower the perigee and dispose satellites by re-entering in the Earth's atmosphere, removing them completely from space \cite{radtke2015impact,sanchez2015study,Alessi2016}. So, depending on the type of disposal, we are looking for either minimum change in eccentricity for graveyard orbits or maximum change in eccentricity for re-entry disposal orbits.

In addition, due to the overlap of different lunisolar resonances, chaos can occurs in the MEO region \cite{ely1997stability,rosengren2015chaos,daquin2016,gkolias2016}. This means that the evolution of MEO orbits can be very sensitive to the initial conditions and a tiny perturbation of the initial state may cause the evolution of the orbit to be completely different. This makes it impossible to reliably predict the orbit and has lead to concern about the predictability of orbits in the MEO region \cite{daquin2016,rosengren2017galileo}.

Unwanted orbital changes can be corrected by performing orbital manoeuvres, however, disposed satellites do not have this option. Therefore, all possible orbital evolutions due to uncertainties must be considered beforehand. In case of disposal orbits, we are especially interested in the evolution of the eccentricity that determines the perigee and apogee altitude of the disposal orbit, and thus dictates the flight domain of a graveyard orbit and whether a spacecraft will re-entry (since the semi-major axis remains almost constant in the absence of drag, only varying due to tesseral resonances).

Besides, when designing a disposal orbit, we have to consider the practical aspects of carrying out a disposal.
For example, we need to take into account the amount of fuel available to manoeuvre the spacecraft into the disposal orbit. During the design, considering an extensive portion of the phase space as possible disposal options is not useful, because large parts of the phase space cannot be reached with a limited fuel budget. In addition, from an operational point of view, it is not feasible to delay a disposal manoeuvre for a long time, e.g. to wait until a favourable configuration with respect to the Moon is established (as suggested in e.g. \cite{rosengren2017galileo}), because the spacecraft may fail to operate.

Recently,  Mistry and Armellin \cite{Mistry2015} and Armellin and San-Juan \cite{armellin2018optimal} approached the design of graveyard and re-entry disposal orbits for Galileo satellites by optimizing the orbits regarding the amount of $\Delta V$ required for the disposal manoeuvre. In this way, they found practically feasible disposal options in terms of required propellant. However, these designs did not take into consideration the predictability of the disposal orbits, that is, the orbital evolution of the disposal options due to uncertainties was not considered.

The predictability of Galileo disposal orbits was addressed by Rosengren et al \cite{rosengren2017galileo}. In their work they analysed the chaos and stability of orbits by studying the dynamics and computing the Fast Lyapunov indicator (FLI) and Lyapunov time for different orbits in the initial phase space. They concluded that many orbits that seems stable within 200 years are actually unstable in 500 years and the Lyapunov time is generally much shorter than the propagation time. 
However, as noted by Rosengren et al \cite{rosengren2017galileo}, the correlation between the Lyapunov time and the effective predictability time horizon is not always clear or can be hard to establish (see e.g. \cite{milani1992example} and \cite{siegert2016}) and was left open for future work. In addition,  Rosengren et al \cite{rosengren2017galileo} suggests that the orbital evolution of chaotic orbits must be studied in statistical terms (see e.g. \cite{laskar2009existence}), because a single trajectory is not representative and one must look at the evolution of ensembles of trajectories.
This raises the question how we should study the predictability of orbits? And how we can determine if an orbit is predictable on the time scale of interest?

In this paper we study the predictability of Galileo disposal orbits using different approaches to investigate which method works well to analyse the predictability of orbits and to determine if Galileo disposal options are reliable or not. Here, with reliable we mean that the uncertainty in eccentricity remains small enough such that re-entry of the satellite is guaranteed (re-entry disposal) or the satellite keeps a safe distance from operational orbits (graveyard disposal). 
We investigate the use of sensitivity analysis and chaos indicators to analyse the predictability of disposal orbits. The sensitivity analysis is performed numerically and using Taylor differential algebra (DA).
We compare the results of the different approaches to address the practicality of using chaos indicators and to discuss the predictability and reliability of the studied disposal orbits.

The aims of this work are:
\begin{enumerate}
\item investigate the reliability of low $\Delta V$ disposal options;
\item assess the practicality of chaos indicators and sensitivity analysis for analysing the predictability of disposal orbits.
\end{enumerate}

We focus on four Galileo disposal orbits that were found by optimizing the disposal manoeuvre. Three of these orbits dispose the spacecraft via re-entry \cite{armellin2018optimal} and the other is a graveyard orbit \cite{Mistry2015}. We investigate if these orbits are chaotic on the time scale of interest and if we should be concerned about their predictability.

The paper is set up as follows. First, we clearly define the terms stable and chaotic and introduce the theory of chaos indicators. After that, the methods used to compute chaos indicators and perform sensitivity analysis are discussed. We briefly describe the dynamical models used in this work and introduce the test cases. In Section~\ref{sec:results} the results are presented and these are subsequently discussed in Section~\ref{sec:discussion}. Finally, conclusions are drawn about the practicality of chaos indicators and sensitivity analysis and about the reliability of the investigated disposal orbits.

\section{Theory}

The two main aspects of Galileo disposal orbits that we are interested in are their stability and chaoticity. Both features are related to the orbital evolution, however, by stability we refer to the amount of change of the orbital elements while by chaoticity we refer to the sensitivity of the orbit to the initial conditions.

\subsection{Reference time scale}
To more clearly define what we mean by stable and chaotic, we shall first define a reference time scale. On this time scale we will investigate the stability and chaoticity of orbits. Anything that happens beyond the reference time period is not considered in this study. For re-entry disposal the reference time scale is simply the time to re-entry. For a graveyard orbit the reference time scale depends on the viewpoint of the investigator. An astronomer may be interested in the orbital evolution for thousands or even millions of years, whereas an engineer may only be interested in a few hundreds of years. In this paper, we set the reference time period for a graveyard orbit to 200 years. For a re-entry orbit we consider a reference time of 100 years, which is the time to re-entry for our test cases.

\subsection{Stability}
We say that an orbit is \emph{stable} if it does not change significantly (e.g. not more than a certain threshold) in the reference period of time. More specifically, in case of near-Earth orbits, we are mainly interested in the change in semi-major axis $a$, eccentricity $e$ and inclination $i$, since the right ascension of the ascending node $\Omega$, argument of perigee $\omega$ and mean anomaly $M$ change secularly anyway due the Earth's oblateness and Keplerian motion. An orbit is therefore considered stable if the variation in $a$, $e$ and $i$ is limited. The threshold on the variation is here defined by the requirements for the graveyard orbit; the variation in $a$, $e$ and $i$ must be small enough for the orbit not to interfere with other operational orbits. For the re-entry scenario, on the other hand, we seek orbits that are unstable such that their eccentricity grows and the perigee is lowered until they reach re-entry altitude.

\subsection{Chaoticity}
We say that an orbit is \emph{chaotic} if its divergence from sufficiently close neighbouring orbits during the reference time is exponentially fast.
If an orbit is chaotic, then any generic tiny deviation in the initial state grows exponentially fast and results in a potentially completely different orbital evolution over time. Because in practice the initial state is never known exactly, chaos makes it impossible to accurately predict the future evolution of an orbit for all time. On the other hand, if an orbit is \emph{regular}, then a tiny deviation in the initial state will not result in a completely different orbital evolution and a single orbit prediction is representative for the evolution of neighbouring orbits.

To recap, in this work stability relates to the amount of change in orbital elements of a single orbit over time, whereas chaoticity relates to the predictability of the orbit.

We remark that chaoticity is a local property: an orbit can exhibit chaotic behaviour in some small neighborhood around it, while globally the deviation of nearby orbits from the reference orbit remains bounded. Intuitively, nearby orbits behave chaotically within a tube around the reference orbit. This observation already hints at an important requirement for using chaoticity as a measure of predictability that will become more clear later: chaoticity must be established over a sufficiently large region of phase space to be a relevant measure in practice.

Since large-scale chaos is often observed to result in large transport of the orbital elements, researchers frequently investigate the chaotic behaviour of orbits when actually searching for stable orbits. In some cases, this may be a useful approach, but it is not based on sound theory as we can have chaotic orbits that are stable (i.e. chaotic orbits that show no significant long term change in orbital elements, see e.g. \cite{milani1992example}) and regular orbits that are unstable.

In case of disposal orbits, we are mainly interested in the stability and chaotic behaviour of the \emph{eccentricity} of the orbit. We want the eccentricity to either remain small, and thus stable, as in case of a graveyard orbit, or to grow, and thus be unstable, to let the spacecraft re-enter in the atmosphere. On the other hand, in both cases, we want the orbit to be predictable, so we can be sure that initial uncertainties do not cause the spacecraft to fail to re-enter (for re-entry disposal) or cause the orbit to become unstable (for graveyard disposal). Therefore, we always want the orbit to behave regularly on the time scale of reference.

\subsection{Chaos indicators}
A common approach to analyse whether an orbit is regular or chaotic is to compute chaos indicators for different orbits in the domain of interest \cite{skokos2016chaos}. These chaos indicators generally relate to a measure of the divergence between infinitely close neighbouring orbits. If the divergence is exponential then the trajectory is said to be chaotic, else the orbit is regular. The larger the divergence, the stronger is the chaos, so we can also distinguish between strongly chaotic and weakly chaotic orbits. 

To study the divergence between neighbouring orbits, we start from the dynamics of the spacecraft. These can be written as a time-dependent system of first-order differential equations:
\begin{equation}
	\dot{\boldsymbol{x}}(t,\boldsymbol{x}_0) = f(\boldsymbol{x}(t,\boldsymbol{x}_0),t)
    \label{eq:ODEsystem}
\end{equation}
with
\begin{equation}
	\boldsymbol{x}(t_0,\boldsymbol{x}_0) = \boldsymbol{x}_0
\end{equation}
The solutions of the system \eqref{eq:ODEsystem} evolve or flow along their trajectories in the phase space. The map that takes a point $\boldsymbol{x}_0$ in the initial domain at $t_0$ to its location at time $t$ is the flow map:
\begin{equation}
	\phi^t(\boldsymbol{x}_0) = \boldsymbol{x}(t,\boldsymbol{x}_0)
\end{equation}

Now consider the distance $|\delta \boldsymbol{x}(t)|$ between two neighbouring orbits starting at $\boldsymbol{x}_0$ and $\boldsymbol{x}_0+\delta \boldsymbol{x}_0$. This distance will change over time and its evolution can be approximated by a linearisation of the ODE around the reference orbit as:
\begin{equation}
	|\delta \boldsymbol{x}(t)| \approx e^{\mu t} |\delta \boldsymbol{x}_0| = e^{t/T_L} |\delta \boldsymbol{x}_0|
    \label{eq:exponentialGrowthOfDeviation}
\end{equation}
where $\mu$ is one of the Lyapunov exponents (LE) and $T_L$ the corresponding Lyapunov time. The LE depends on the orientation of the initial $\delta \boldsymbol{x}_0$. If $\delta \boldsymbol{x}_0$ is in the direction of maximum growth then the corresponding $\mu$ is the maximum Lyapunov exponent (MLE). The MLE can be computed as \cite{skokos2016chaos}:
\begin{equation}
	\mu = \lim\limits_{t \to \infty} \lim\limits_{\delta \boldsymbol{x}_0 \to 0} \frac{1}{t} \ln \frac{|\delta \boldsymbol{x}(t)|}{|\delta \boldsymbol{x}_0|} 
    \label{eq:MLE}
\end{equation}
If the MLE is positive then neighbouring trajectories separate exponentially fast and the trajectory is said to be chaotic.

It is worth noting that the existence of these limits is not guaranteed in all systems \cite{Ott2008}. One common criterion ensuring existence is the existence of an invariant measure under the flow on a compact set: in this case the set of points where the limit does not exist has zero measure \cite{Oseledec1968}. For symplectic systems, such as Hamiltonian dynamics, the first criterion is trivially satisfied as the phase space volume is preserved due to Liouville's theorem. However, even then in many cases it is not easy to show that the relevant dynamics happen on a compact set.

In general, when they exist the Lyapunov exponents depend on the orbit along which they are computed. We can therefore think of them as a measure of the chaoticity of a particular orbit. In certain types of systems where the invariant measure is also ergodic, however, the Lyapunov exponents are independent of the initial condition.

\subsubsection{Finite Time Lyapunov Exponent}
Since it is in general not possible to propagate a trajectory for infinite time (as required to compute the MLE using Eq.~\ref{eq:MLE}), we can approximate the MLE in finite time. If the chosen finite time is sufficiently large, and all the limits in Eq.~\eqref{eq:MLE} exist, this procedure yields a good approximation of the MLE.

In this work, we use the finite-time Lyapunov exponent (FTLE)\footnote{A good introduction to the FTLE and its use can be found online at \url{http://shaddenlab.berkeley.edu/uploads/LCS-tutorial/contents.html}.}. The FTLE is a scalar value which characterizes the amount of stretching about the trajectory of point $\boldsymbol{x}_0$ after time $t$ \cite{SHADDEN2005}.
The FTLE is computed as follows.

First, we compute the (right) Cauchy-Green deformation tensor $\Delta$ by multiplying the gradient of the flow $\frac{d\phi^{t}}{d\boldsymbol{x}_0}$, i.e., the Jacobian of the flow map with respect to the initial condition, by its transpose:
\begin{equation}
	\Delta = \frac{d\phi^{t}}{d\boldsymbol{x}_0}^{T}  \frac{d\phi^{t}}{d\boldsymbol{x}_0}
    \label{eq:CauchyGreenTensor}
\end{equation}
where $T$ indicates that we take that transpose. This tensor provides a measure of the square of local change in distances due to expansion or contraction.

The maximum stretching is given by the square root of the maximum eigenvalue $\lambda_{\mathrm{max}}$ of $\Delta$ and therefore the FTLE is defined as:
\begin{equation}
	\mathrm{FTLE}(t, \boldsymbol{x}_0) = \frac{1}{t} \ln{\sqrt{\lambda_{\mathrm{max}}(\Delta)}}
    \label{eq:FTLE}
\end{equation}

\subsubsection{Fast Lyapunov Indicator}
Another commonly used numerical chaos indicator in celestial mechanics is the Fast Lyapunov Indicator (FLI)\footnote{A thorough introduction to the FLI and its use can be found in \cite{lega2016}.} that measures the maximum stretching of a tangent vector $\boldsymbol{v}_0$ \cite{FROESCHLE1997}. The FLI is defined as:
\begin{equation}
	\mathrm{FLI}(t,\boldsymbol{x}_0,\boldsymbol{v}_0) = \sup\limits_{\tau \le t} \ln \frac{|\boldsymbol{v}(\tau)|}{|\boldsymbol{v}_0|} \label{eq:FLI}
\end{equation}

The value of the FLI depends on the orientation of the initial deviation vector $\boldsymbol{v}_0$. Therefore, different initial deviation vectors should to be integrated over time to find the largest stretching \cite{lega2016}. The FLI has the advantage that it is numerically easy to integrate as it does not rely on an explicit representation of the gradient of the flow. It also measures the largest expansion of the tangent vector anywhere along the trajectory, not just at the final point. The FTLE has the advantage that it does not use an initial deviation vector, but by definition uses the direction of maximum growth derived from the gradient of the flow.

\subsubsection{Lyapunov Time}
The inverse of the MLE is the Lyapunov time $T_L$, which is the average time in which two nearby trajectories diverge by a factor $e$:
\begin{equation}
	T_L(t, \boldsymbol{x}_0) =\frac{1}{\mathrm{MLE}} \approx  \frac{1}{\mathrm{FTLE}(t, \boldsymbol{x}_0)} = \frac{t}{\ln{\sqrt{\lambda_{\mathrm{max}}(\Delta)}}}
\end{equation}
This timescale is often interpreted as a limit of the predictability of the trajectory. This interpretation, however, is rather arbitrary and not always correct in practice as will be shown later. In particular, when approximating the MLE by the FTLE, additional complications arise due to the choice of $t$: during the finite time before the FTLE converges to the MLE, also the approximate Lyapunov time can vary greatly.

\subsection{Sensitivity}
The chaos indicators FTLE and FLI provide an estimate of the sensitivity of an orbit with respect to infinitesimally small initial uncertainties. These indicators are based on linearised dynamics and do not take into account non-linear behaviour.
In practice, however, the uncertainties are not infinitesimal. If these finite uncertainties in the initial conditions are known, or can be estimated, we can study the orbital evolution of a set of orbits starting in the initial uncertainty domain to analyse the impact of the uncertainties on the orbit including non-linear effects. In this way, we can analyse if the disposal requirements are satisfied in case of initial state errors or compute the probability of disposal success based on initial uncertainties. In addition, such a sensitivity analysis can reveal possible chaotic behaviour of the orbit. Our approach to carry out this sensitivity analysis is discussed in Section~\ref{subsec:sensitivityAnalysisMethod}.

\subsection{Coordinate Dependence}\label{sec:coordinatetheory}
When applied to practical engineering problems, the question of the coordinates and units used becomes relevant. In the case of astrodynamics, orbits are typically represented in either Cartesian coordinates or a set of orbital elements describing the geometry of the orbit. 

The MLE, by definition, is independent of the choice of (time-independent) coordinates. This is because in Eq.~\eqref{eq:exponentialGrowthOfDeviation} any change in the magnitude of $\delta\boldsymbol x_0$ and $\delta\boldsymbol x(t)$ due to the coordinate transformation is bounded independent of time. Since the MLE is the average exponential growth factor over time, any influence of the coordinate transformation disappears as $t\rightarrow\infty$. 

The FTLE, and to a somewhat lesser degree the FLI, on the other hand, depend on the choice of coordinates. In particular a non-linear change of coordinates, such as from Cartesian to orbital elements, can change the value of these indicators significantly.

For the FTLE in particular, we find that the Jacobian can either be calculated in one set of coordinates $\boldsymbol{X}$ directly, or it can be transformed from another set of coordinates $\boldsymbol{x}$ a posteriori. The a-posteriori coordinate transformation is achieved by multiplying the Jacobian of the flow by the Jacobian of the forward coordinate transformation at the initial point, and the Jacobian of the inverse transformation at the final point as follows:
\begin{equation}
	\frac{d\phi^{t}}{d\boldsymbol{X}_0} = \left[\frac{d\boldsymbol{X}_f}{d\boldsymbol{X}_0}\right] = \left[\frac{d\boldsymbol{X}_f}{d\boldsymbol{x}_f}\right] \left[\frac{d\boldsymbol{x}_f}{d\boldsymbol{x}_0}\right] \left[\frac{d\boldsymbol{x}_0}{d\boldsymbol{X}_0}\right]
    \label{eq:FLTEcoords}
\end{equation}
Simple scaling of units can be handled in the same way.

\section{Methods}
The methods used to compute chaos indicators and to perform sensitivity analysis as well as a brief introduction to Taylor differential algebra is given in the following. We refer the reader to the provided references for further details on each technique.

\subsection{Differential Algebra}
DA is an automatic differentiation method which enables the automatic computation of derivatives of functions in a computer environment \cite{berz1999modern}. Based on truncated power series algebra, the algebraic operations for floating-point numbers are replaced by operations for Taylor polynomials, such that any function $f$ of $v$ variables can straightforwardly be expanded into its Taylor polynomial up to arbitrary order. The differential algebra used in this work is the DA Computational Engine (DACE) \cite{rasotto2016differential} that includes all core DA functionality and a C++ interface\footnote{The open-source software package is available at \url{https://github.com/dacelib/dace}.}.

One of the key applications of DA is the automatic high order expansion of the solution of an ODE initial value problem with respect to the initial conditions \cite{berz1999modern,armellin2010asteroid}. Using DA ODE integration, the Taylor expansion of the flow in $x_0$ can be propagated forward in time, up to any final time. The main benefit of the use of DA to expand the flow is that we can expand up to arbitrary order and there is no need to write and integrate variational equations.

The high-order Taylor expansions are only accurate in a domain close to the expansion point. To ensure that the computed Taylor polynomial is sufficiently accurate, we can estimate the truncation error of the expansion by estimating the size of the order $n+1$ coefficients of the Taylor polynomial as explained in \cite{wittig2015propagation}.
Using the estimated truncation error, the distance from the expansion point where the Taylor series has a specific error can be estimated. For states inside the estimated domain the high-order map has a truncation error that is approximately smaller than the specified error.

\subsection{DA computation of chaos indicators}
\label{subsec:chaosIndicatorComp}
For the computation of FTLE, we require the Jacobian of the flow map $\frac{d\phi^{t}}{d\boldsymbol{x}_0}$. With the use of DA, this Jacobian can be computed straightforwardly using the first-order Taylor expansion of the flow with respect to the initial state without the need to write the variational equations. The values of the partial derivatives are simply given by the first-order coefficients of the Taylor polynomial. Once the Jacobian of the flow map is computed using DA, the FTLE is obtained via Eqs. \eqref{eq:CauchyGreenTensor} and \eqref{eq:FTLE}.

As mentioned in Section \ref{sec:coordinatetheory}, The FTLE can be computed using different coordinates and due to its finite time nature its value changes. In this work, we investigate the use of two different element sets, namely:
\begin{enumerate}
\item Classical orbital elements (CEO): $(a,e,i,\Omega,\omega,M)$ with $a$ the semi-major axis, $e$ the eccentricity, $i$ the inclination, $\Omega$ the right ascension of the ascending node, $\omega$ the argument of pericentre and $M$ the mean anomaly.
\item Modified equinoctial elements (MEE): $(p,f,g,h,k,L)$ with $p = a(1-e^2)$, $f = e\cos{(\omega+\Omega)}$, $g = e\sin{(\omega+\Omega)}$, $h = \tan{(i/2)}\cos{\Omega}$, $k = \tan{(i/2)}\sin{\Omega}$ and $L = \Omega+\omega+\nu$ where $\nu$ is the true anomaly \cite{Walker1985}.
\end{enumerate}

Unless otherwise mentioned, we use the following units in this work: radian for $i$, $\Omega$, $\omega$ and $M$, the semi-major axis $a$ is scaled by its initial value ($a/a_0$) such that it is unitless and equals 1 and the eccentricity is dimensionless by default. 

The coordinate transformation is carried out according to \eqref{eq:FLTEcoords}. The Jacobian of the coordinate transformation is obtained by computing the first-order Taylor expansion of the state in transformed coordinates $\boldsymbol{X}$ with respect to the state in old coordinates $\boldsymbol{x}$ using DA.

An important remark regarding the computation of chaos indicators is that we only consider the first five orbital elements and neglect the behaviour of the fast angular variable (i.e. we do not consider the divergence in mean anomaly and true longitude). The averaged dynamics do not depend on the fast angle and the long-term behaviour of the fast angle is not of interest. In addition, the evolution of the fast angle is sensitive to changes in the orbit size and therefore the amount of stretching between neighbouring orbits can be dominated by stretching in the fast angle while the difference in in-orbit position is not important. 

We will also compute the FTLE considering only stretching in a single or just two orbital elements to determine the divergence in these elements. For this, only the rows of the Jacobian corresponding to changes in a single or two elements of interest are considered and consequently we obtain one or two non-zero eigenvalues, respectively, of which the largest is used to compute an FTLE. For example, if we consider only the divergence in $e$, then we take the row of the Jacobian containing the partial derivatives of $e$, multiply this row with its transpose and compute the eigenvalue to obtain an FTLE that considers only $e$.

Finally, besides using the Jacobian to compute the FTLE, it can also be used to calculate the FLI by mapping an initial deviation vector over time (without the need to integrate the variational equations) as follows:
\begin{equation}
	\boldsymbol{v}(t) = \frac{d\phi^{t}}{d\boldsymbol{x}_0} \boldsymbol{v}_0
    \label{eq:mapDeviationVector}
\end{equation}

\subsection{Sensitivity analysis}
\label{subsec:sensitivityAnalysisMethod}
To analyse the evolution of the orbit under initial uncertainties, we perform a sensitivity analysis. This analysis can be performed either numerically or using Taylor differential algebra.

The numerical sensitivity analysis is carried out by numerically computing the orbital evolution for many different initial conditions in the initial uncertainty domain. The extremes of the propagated orbits are assumed to indicate the bounds of the uncertainty domain over time. This is equivalent to a Monte Carlo simulation except for the fact that in this work the points in the uncertainty domain are selected on a grid and not picked randomly. A drawback of this technique is that to carry out a reliable analysis many initial conditions need to be propagated, which can be time consuming.

Alternatively, we can estimate the bounds of the uncertainty domain over time using a single high-order Taylor expansion $\mathcal{T}_{\phi}$ of the flow $\phi$. For this, the flow about the nominal initial condition is expanded with respect to the initial uncertainties using DA. The high-order expansion is computed such that the initial uncertainty domain corresponds with the domain [-1,1] in the Taylor polynomial space (by scaling the expansion variables). An outer bound of the range of the Taylor expansion at different times is then estimated by applying interval arithmetic \cite{moore2009introduction}. In this way, the estimated range is guaranteed to include all possible values of the polynomial, which, however, not necessarily includes all values of the function that is approximated. Hence, we obtain the bounds of the domain of possible trajectories due to initial uncertainties. 

The estimation of the bounds using interval arithmetic is efficient, but the drawback is that the computed bounds overestimate the true range of the Taylor polynomial. In addition, the Taylor expansion needs to be accurate in the whole uncertainty domain to ensure that the estimated bounds are accurate. This can be checked heuristically by estimating the truncation error of the Taylor polynomial in the uncertainty domain. 

In this work, we assume that the uncertainties in the initial state are due to manoeuvre errors, that is, errors in the magnitude and direction of the applied $\Delta \boldsymbol{V}$. Therefore, we expand the flow with respect to the disposal manoeuvre magnitude $\Delta V$ and direction angles $\alpha$ and $\delta$ (see also Section~\ref{sec:testcases}), so we get: $\mathcal{T}_{\phi} = \mathcal{T}_{\phi}(\Delta V,\alpha,\delta)$. In this way, we can directly compute the orbital evolution of the disposal orbit due to manoeuvre errors. Moreover, the Taylor expansion has only three expansion variables, which reduces the computational effort compared to expanding with respect to six state variables.

Besides using the high-order expansion of the flow to compute bounds, we can also use it to accurately estimate the evolution of individual orbits in the uncertainty domain. This is computationally more efficient than numerically propagating the orbit if the flow expansion has already been computed, especially when many orbits need to be propagated \cite{gondelach2017semi}. This technique could therefore be applied to perform sensitivity analysis instead of using the numerical approach.

\section{Dynamical model}
\label{sec:dynamics}

The dynamical model used in this work is the one implemented in HEOSAT, a semi-analytical propagator developed to study the long-term evolution of satellites in highly elliptical orbits \cite{Lara2018HEOSAT}. The perturbations included in the model are zonal and lunisolar perturbations, solar radiation pressure (SRP) and drag. Short-periodic terms have been removed by expressing the gravitational terms in Hamiltonian form and averaging over the mean anomaly of the satellite using Deprit's perturbation algorithm \cite{deprit1969} based on Lie transformations. The averaging techniques applied for developing HEOSAT's dynamical model are described in \cite{Lara2018HEOSAT}. 

In this work, we will only consider the effects due to the Earth, Sun and Moon gravitation and SRP, whereas drag is neglected because it only acts on the spacecraft for a short period of time before re-entry. The main characteristics of the perturbation model and averaging procedures are as follows:
\begin{itemize}
  \item The zonal-term Hamiltonians are simplified by removing parallactic terms (via Elimination of the parallax \cite{deprit1981elimination,lara2013proper,lara2014delaunay}) and short-periodic terms are eliminated by Delaunay normalization. This is carried out up to second order of the second zonal harmonic, $J_2$, and to first order for $J_3-J_{10}$.
  \item The disturbing potentials of the Sun and Moon (point-mass approximation) are expanded using Legendre series to obtain the Hamiltonians for averaging out the short-periodic terms \cite{Lara2012third}. Second- and sixth-order Legendre polynomials are taken for the Sun and Moon potentials, respectively.
  \item The averaged equations of motion due to SRP are obtained by analytically averaging Gauss equations over the mean anomaly using Kozai's analytical expressions for perturbations due to SRP \cite{kozai1963srp}. For this, a spherical satellite and constant solar flux along the orbit (i.e. no shadow) are assumed.
\end{itemize}

The propagation is carried out in the True of Date reference system and Chapront’s analytical ephemeris is used for the Sun and Moon positions \cite{chapront1988,chapront2003}.

Three different models based on the described dynamics are used:
\begin{enumerate}
\item Simple gravitational model: includes only first-order $J_2$ and second-order Legendre polynomials for both Sun and Moon potentials;
\item Full gravitational model: includes second order $J_2$ and first order $J_3-J_{10}$, and second- and sixth-order Legendre polynomials for the Sun and Moon potentials, respectively;
\item Complete model: Full gravitational model plus SRP;
\end{enumerate}
According to Daquin et al \cite{daquin2016}, a single-averaged dynamical model considering only the second-order Legendre polynomial terms in the expansions of both the geopotential and lunisolar perturbations, i.e. the simple gravitational model, is sufficient to ``capture nearly all of the qualitative and quantitative features of more complicated and realistic models \cite{daquin2016}'', such as the full gravitational model. Therefore, we will use the simple model to compute chaos indicators when studying the chaotic behaviour of many orbits in the phase space to reduce the required computational effort. On the other hand, when focusing on a specific disposal orbit, we will use the full gravitational model that was also used to find the optimal disposal orbits \cite{armellin2018optimal}. Finally, to analyse the impact of model uncertainties, we use the complete model for comparison.
 
The propagator was implemented in Taylor Differential Algebra \cite{gondelach2017semi} to enable automatic high-order expansion of the flow. The high-order Taylor expansion allows efficient computation of the orbital evolution of a whole set of orbits starting close to the propagated orbit. In addition, the first-order partial derivatives of the propagated state with respect to the initial state are used to compute chaos indicators, see Section~\ref{subsec:chaosIndicatorComp}. A DA-version of a 7/8 Dormand-Prince Runge-Kutta scheme (8th order solution for propagation, 7th order solution for step size control) is used for the numerical integration with an error tolerance of $10^{-10}$.

The average computation times for propagating an orbit for 100 years using the simple and full gravitational model numerically and in DA is shown in Table~\ref{tab:compTimes}. The simulations were run on a computer with an Intel Core i5-6500 processor running at 3.20 GHz using 16 GB of RAM. We either compute a 1st-order DA expansion w.r.t. the initial orbital elements $(a,e,i,\Omega,\omega)$ to compute chaos indicators or compute a 5th-order Taylor expansion w.r.t. the disposal manoeuvre parameters $(\Delta V,\alpha,\delta)$. Due to the singularity at zero eccentricity in the dynamical model, propagations take longer when the eccentricity becomes very small ($e<10^{-5}$) as the stepsize decreases strongly to ensure accurate results. When propagating in the DA framework, the simple model is approximately 18 times faster than the full model. Finally, the complete model that also includes SRP requires about $2\%$ more computation time than the full model.

\begin{table*}[htbp]
  \centering
  \caption{Average computation time for propagating an orbit for 100 years using the simple or full model numerically or in DA.}
    \begin{tabular}{lccc}
    \hline
          & \multicolumn{3}{c}{Average computation time [s]} \\
    \cline{2-4}
    Model      & \multicolumn{1}{c}{Numerical} & \multicolumn{1}{c}{1st-order DA,} & \multicolumn{1}{c}{5th-order DA,} \\
          &       & \multicolumn{1}{c}{5 DA variables} & \multicolumn{1}{c}{3 DA variables} \\
    \hline
    Simple gravitational model &   1.82    & 25.5  & 275 \\
    Full gravitational model &   9.75    &   460    & 5140 \\
    \hline
    \end{tabular}%
  \label{tab:compTimes}%
\end{table*}%

\section{Test cases}
\label{sec:testcases}

The test cases are three re-entry disposal orbits and one graveyard orbit for the Galileo satellites with NORAD IDs 37846, 40890 and 41175, and 38858, respectively. The initial conditions of the disposal orbits were obtained through optimization by minimizing the required $\Delta V$ for the disposal manoeuvre while ensuring a re-entry within 100 years \cite{armellin2018optimal} or a minimum distance of 100 km from the Galileo constellation for 100 years \cite{Mistry2015} for the re-entry and graveyard disposal, respectively. 
The disposal manoeuvres are characterized by changes in semi-major axis, eccentricity and argument of perigee while leaving the inclination and ascending node almost unchanged, since plane changes are costly in terms of $\Delta V$. The orbital states before and after the disposal manoeuvre are shown in Table~\ref{tab:manoeuvreStateUncertainty}. This table also shows the applied manoeuvre $\Delta V$ and its direction indicated by $\alpha$ and $\delta$, which are the in-plane and out-of-plane angles of the thrust vector with respect to the velocity vector, and the manoeuvre date.
Note that, the disposal into a graveyard orbit actually requires two manoeuvres; one that increases the semi-major axis and one that inserts the spacecraft into the near-circular graveyard orbit. In Table~\ref{tab:manoeuvreStateUncertainty} only the second manoeuvre is shown. For details about the optimization of the disposal orbits see \cite{Mistry2015} regarding the graveyard orbit (38858) and \cite{armellin2018optimal} for the re-entry disposal orbits (37846, 40890 and 40890).

To investigate the predictability of the orbits using sensitivity analysis, we consider uncertainties in the disposal manoeuvre execution; we assume $1\%$ uncertainty in the magnitude of the applied $\Delta V$ and 1$\degr$ uncertainty in the applied thrust directions, $\alpha$ and $\delta$, which are typical manoeuvre uncertainties \cite{feldhacker2016incorporating}. Table~\ref{tab:manoeuvreStateUncertainty} shows the maximum absolute error in the initial orbital elements due to manoeuvre errors. The largest errors can be found in $e$ and $\omega$ and, noticeably, the errors in $a$ and $e$ increase with applied $\Delta V$. 

\begin{table*}[htbp]
  \renewcommand{\arraystretch}{1.2}
  \centering
  \caption{Manoeuvre magnitude and direction, orbital state before and after manoeuvre, and maximum error in state after manoeuvre assuming manoeuvre execution uncertainties of $1\%$ in $\Delta V$ and 1$\degr$ in $\alpha$ and $\delta$. JD means Julian date.}
    \begin{tabular}{clrrrrrr}
    \hline
    Object &  & \multicolumn{1}{c}{$a$ [km]} & \multicolumn{1}{c}{$e$ [-]} & \multicolumn{1}{c}{$i$ [deg]} & \multicolumn{1}{c}{$\Omega$ [deg]} & \multicolumn{1}{c}{$\omega$ [deg]} & \multicolumn{1}{c}{$M$ [deg]} \\
    \hline
          & \multicolumn{7}{c}{Manoeuvre: $\Delta V= 3.44$ m/s, $\alpha = 5.23\degr$, $\delta = 17.67\degr$, $\mathrm{JD}=2457273.22175$} \\ \cline{2-8}
    38858 & Before $\Delta V$ & 29650.06 & 0.001869 & 54.974 & 210.010 & 221.901 & 171.430 \\
    graveyard      & After $\Delta V$ & 29702.84 & 0.000395 & 54.988 & 210.020 & 293.465 & 99.847 \\
          & Max error & 0.92  & 2.54E-5 & 1.17E-3 & 9.40E-4 & 4.466 & 4.465 \\
    \hline
          & \multicolumn{7}{c}{Manoeuvre: $\Delta V= 86.7$ m/s, $\alpha = 1.87\degr$, $\delta = -3.0\degr$, $\mathrm{JD}=2457517.89887$} \\ \cline{2-8}
    40890 & Before $\Delta V$ & 29601.77 & 0.000426 & 57.256 & 323.093 & 40.862 & 292.704 \\
    re-entry      & After $\Delta V$ & 31086.33 & 0.047913 & 57.183 & 323.137 & 333.904 & -0.369 \\
          & Max error & 18.49 & 0.000566 & 0.030 & 0.018 & 0.709 & 0.372 \\
    \hline
          & \multicolumn{7}{c}{Manoeuvre: $\Delta V= 128.5$ m/s, $\alpha = 3.83\degr$, $\delta =-18.75\degr$, $\mathrm{JD}=2457535.41865$} \\ \cline{2-8}
    41175 & Before $\Delta V$ & 29598.90 & 0.000163 & 54.943 & 202.432 & 270.326 & 288.351 \\
    re-entry      & After $\Delta V$ & 31737.42 & 0.067472 & 55.529 & 202.672 & 196.333 & 1.908 \\
          & Max error & 39.77 & 0.001179 & 0.046 & 0.019 & 0.699 & 0.607 \\
    \hline
          & \multicolumn{7}{c}{Manoeuvre: $\Delta V= 173.3$ m/s, $\alpha = -23.87\degr$, $\delta = -1.13\degr$, $\mathrm{JD}=2457540.57102$} \\ \cline{2-8}
    37846 & Before $\Delta V$ & 29601.79 & 0.000526 & 55.560 & 82.396 & 1.597 & 314.777 \\
    re-entry      & After $\Delta V$ & 32479.72 & 0.089906 & 56.074 & 81.807 & 307.488 & 7.635 \\
          & Max error & 57.77 & 0.001573 & 0.050 & 0.057 & 0.807 & 0.692 \\
    \hline
    \end{tabular}%
  \label{tab:manoeuvreStateUncertainty}%
\end{table*}%

\section{Results}
\label{sec:results}

\subsection{FTLE analysis}
To investigate if the disposal orbits are chaotic, we first compute the FTLE of different orbits in the initial phase space to see if the disposal orbits are located in a chaotic region of the phase space. For this, different sections of the phase space are investigated by varying the initial $e$, $i$, $\Omega$ or $\omega$ to generate FTLE plots, also called stability maps. In addition, the FTLE is computed at different times and using different coordinates to analyse if and how the FTLE depends on time and coordinates. The results are compared with the behaviour of the eccentricity, which is the key orbital parameter for successful re-entry or graveyard disposal.

\subsubsection{Re-entry}
\begin{figure*}[tbp]
     \centering
     \subfigure[][FTLE using COE after 100 years]{\includegraphics[width=0.45\textwidth,trim={1cm 7.8cm 12.5cm 1cm},clip]{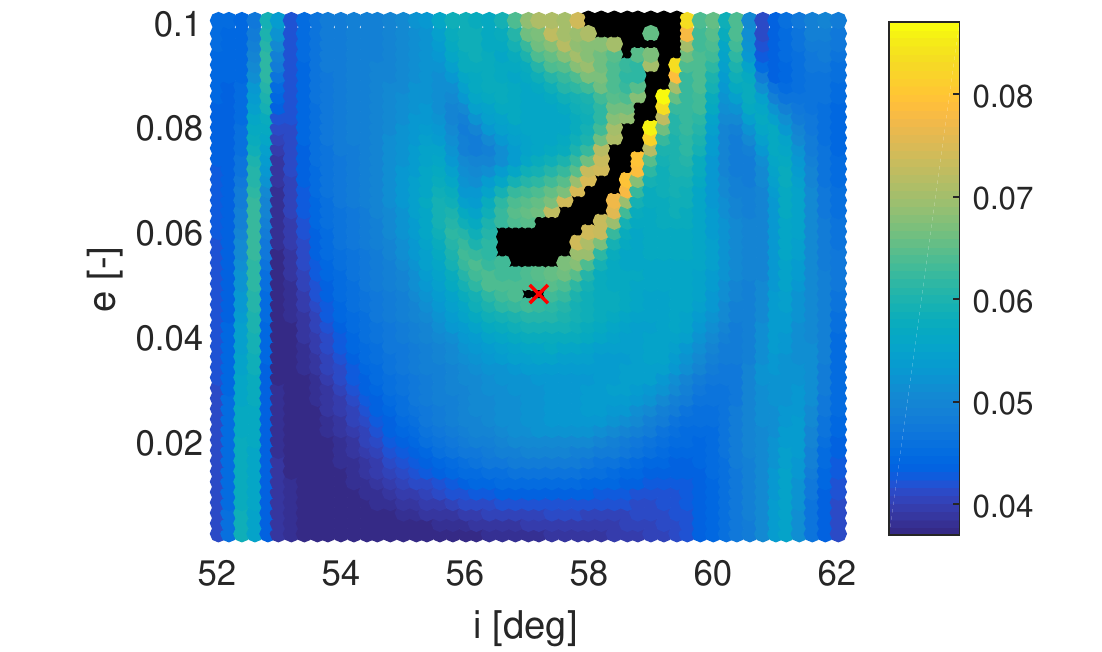} \label{fig:40890_EccIncl_FTLE_COE_100y}}
     \hspace{8pt}%
     \subfigure[][FTLE using COE after 200 years]{\includegraphics[width=0.45\textwidth,trim={12.5cm 7.8cm 1cm 1cm},clip]{input40890_case1_200y_1stZonal1stSunMoon_DAexpansionAEIOo_1stO_every10years_compact_sc1e-3_Ecc0-0_1_Incl52-62_FTLEtime_maxEcc_100-200y3.pdf} \label{fig:40890_EccIncl_FTLE_COE_200y}}
     
     \subfigure[][FTLE using MEE after 100 years]{\includegraphics[width=0.45\textwidth,trim={1cm 7.7cm 12.5cm 1cm},clip]{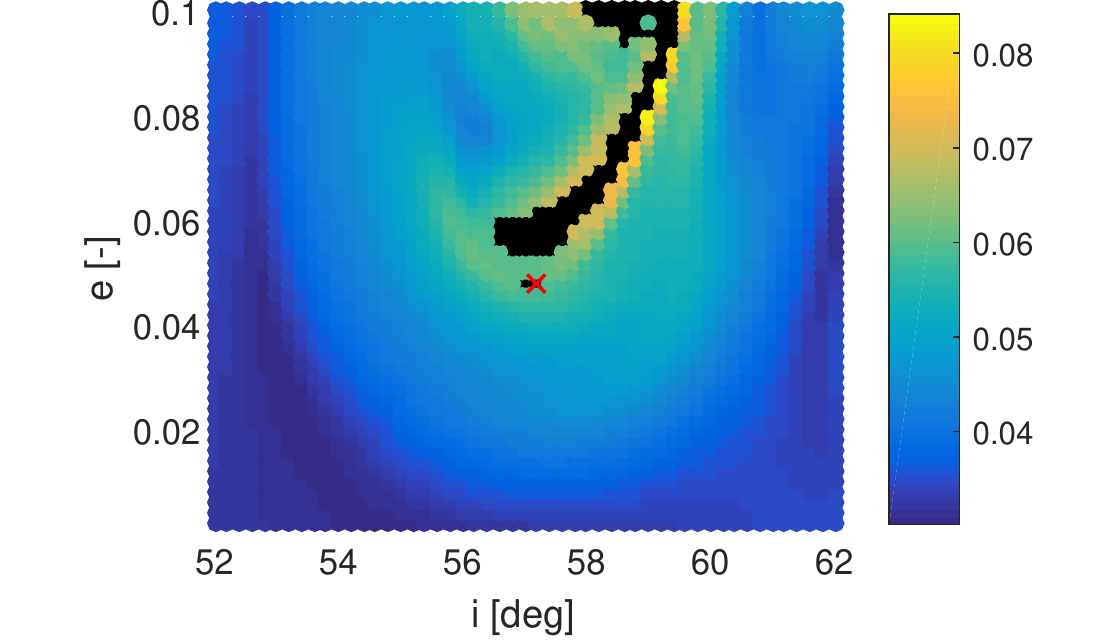} \label{fig:40890_EccIncl_FTLE_MEE_100y}}
     \hspace{8pt}%
     \subfigure[][FTLE using MEE after 200 years]{\includegraphics[width=0.45\textwidth,trim={12.5cm 7.7cm 1cm 1cm},clip]{input40890_case1_200y_1stZonal1stSunMoon_DAexpansionAEIOo_1stO_every10years_compact_sc1e-3_Ecc0-0_1_Incl52-62_MEE_FTLEtime_maxEcc_100-200y.pdf} \label{fig:40890_EccIncl_FTLE_MEE_200y}}
     
     \subfigure[][Maximum eccentricity after 100 years]{\includegraphics[width=0.45\textwidth,trim={1cm 0.5cm 12.5cm 7.8cm},clip]{input40890_case1_200y_1stZonal1stSunMoon_DAexpansionAEIOo_1stO_every10years_compact_sc1e-3_Ecc0-0_1_Incl52-62_MEE_FTLEtime_maxEcc_100-200y.pdf} \label{fig:40890_EccIncl_maxEcc_100y}}
     \hspace{8pt}%
     \subfigure[][Maximum eccentricity after 200 years]{\includegraphics[width=0.45\textwidth,trim={12.5cm 0.5cm 1cm 7.8cm},clip]{input40890_case1_200y_1stZonal1stSunMoon_DAexpansionAEIOo_1stO_every10years_compact_sc1e-3_Ecc0-0_1_Incl52-62_MEE_FTLEtime_maxEcc_100-200y.pdf} \label{fig:40890_EccIncl_maxEcc_200y}}
     
     \caption{FTLE computed using COE and MEE and considering only $e$ and $i$, and maximum eccentricity for different initial $i$ and $e$ for case 40890 after 100 (left) and 200 years (right)}
     \label{fig:40890_eccIncl_COE}
\end{figure*}

\fref{fig:40890_eccIncl_COE} shows the FTLE computed using COE or MEE and the maximum eccentricity after 100 and 200 years for different initial eccentricity and inclination for re-entry case 40890. The red cross indicates the initial condition of the disposal orbit and the black dots indicate orbits that have entered the Earth, such that the dynamics are not valid any more.
Note that all the plots have different colour scales, so yellow, green and blue indicate different FLTE values in different plots.

First of all, as expected the re-entry orbit is located in a region of large eccentricity growth that results in re-entry, see Figs. \ref{fig:40890_EccIncl_maxEcc_100y} and \ref{fig:40890_EccIncl_maxEcc_200y}. In addition, independent of the choice of coordinates, the FTLE plots show chaotic regions that correspond to large eccentricity growth. However, the FTLE plot for COE after 100 years also shows some regions with increased FTLE (green regions at $i=52.5\degr$ and $i=61\degr$) where the eccentricity change is small (compare the FTLE in \fref{fig:40890_EccIncl_FTLE_COE_100y} with the maximum eccentricity after 100 and 200 years in Figs. \ref{fig:40890_EccIncl_maxEcc_100y} and \ref{fig:40890_EccIncl_maxEcc_200y}). These green regions are not visible when the FLTE is computed using MEE elements. 

\begin{figure*}[tbp]
     \centering
     \subfigure[][FTLE after 100 years]{\includegraphics[width=0.85\textwidth,trim={1.5cm 0cm 0cm 0cm},clip]{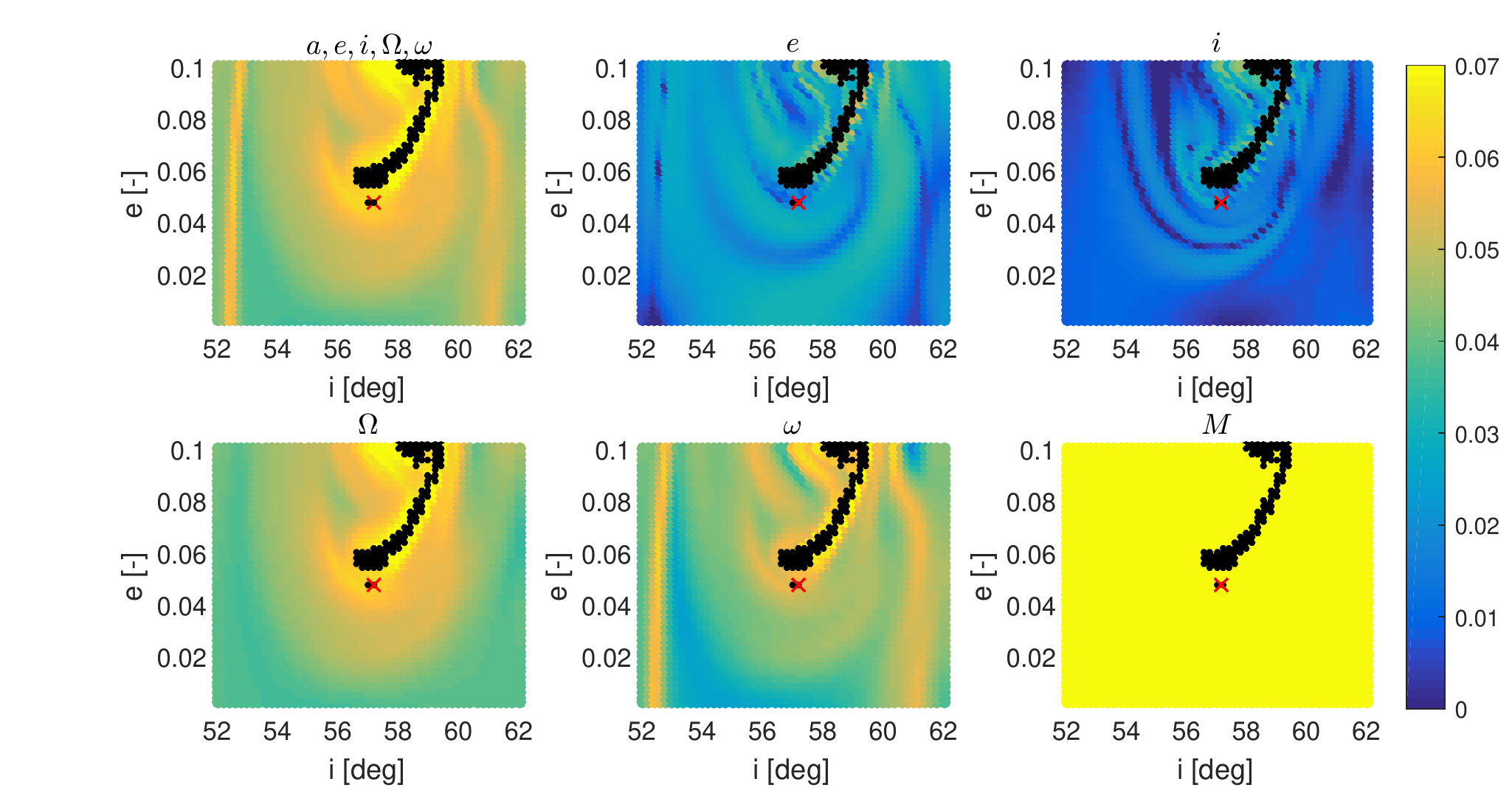} \label{fig:40890_EccIncl_FTLE_COE_aeiOw_100y}}
     \hspace{8pt}
     
     \subfigure[][FTLE after 200 years]{\includegraphics[width=0.85\textwidth,trim={1.5cm 0cm 0cm 0cm},clip]{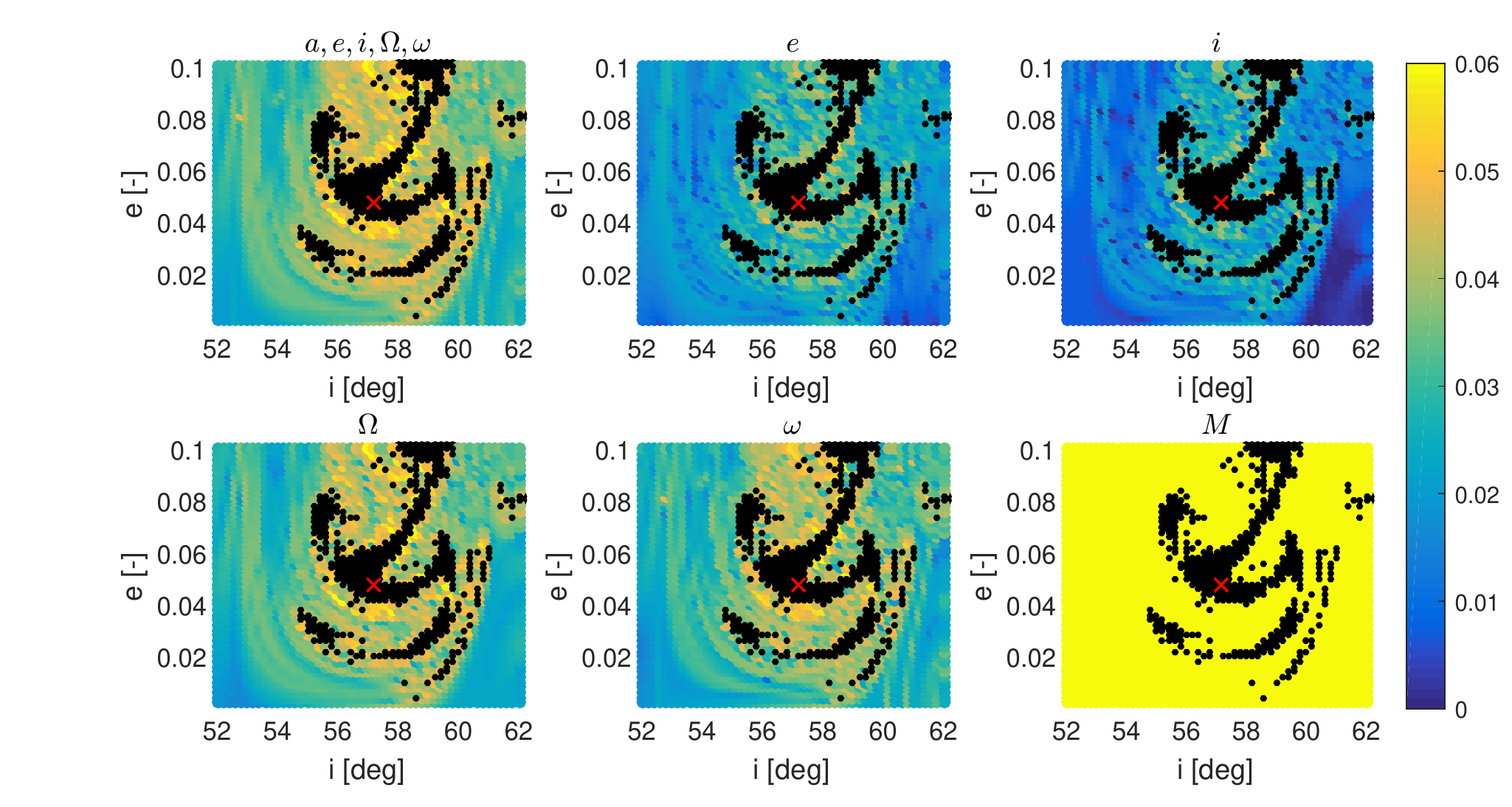} \label{fig:40890_EccIncl_FTLE_COE_aeiOw_200y}}
     
     \caption{FTLE computed using COE considering $(a,e,i,\Omega,\omega)$ or only $e$, $i$, $\Omega$, $\omega$ or $M$ for different initial $i$ and $e$ for case 40890 after 100 (left) and 200 years (right). The range of the FTLE color scale has been limited to improve the visibility of dynamical structures in the phase space. (Note that the plots considering $a,e,i,\Omega,\omega$ are the same as Figs. \ref{fig:40890_EccIncl_FTLE_COE_100y} and \ref{fig:40890_EccIncl_FTLE_COE_200y} but using a different colour scale.)}
     \label{fig:40890_eccIncl_FTLE_COE_aeiOw}
\end{figure*}

To determine the origin of the high FTLE values, we plotted the FTLE that only consider the divergence in a single orbital element, namely in $e$, $i$, $\Omega$, $\omega$ or $M$\footnote{Note that because $a$ is constant there is no divergence in $a$, so the evolution of $a$ does not contribute to the value of the FTLE.}, in \fref{fig:40890_eccIncl_FTLE_COE_aeiOw}. Here the colour scales in all plots are the same to facilitate comparison (note that the colour scales are truncated to improve the visibility of dynamical structures in the plots). \fref{fig:40890_EccIncl_FTLE_COE_aeiOw_100y} indicates that the green regions in \fref{fig:40890_EccIncl_FTLE_COE_100y} are caused by the behaviour of $\omega$. 
The MEE coordinates do not explicitly contain the argument of perigee and therefore the divergence in $\omega$ is not visible when MEE are used to compute the FTLE. In addition, \fref{fig:40890_eccIncl_FTLE_COE_aeiOw} shows that the value of the FTLE considering all elements, i.e. $(a,e,i,\Omega,\omega)$, is completely determined by the divergence in $\Omega$ and $\omega$, since the divergence in $e$ and $i$ is very small compared to the divergence in $\Omega$ and $\omega$. Furthermore, the FTLE plot considering only the mean anomaly $M$ justifies that $M$ should not be considered when computing the FTLE, because the divergence in $M$ would dominate the value of the FTLE and make the behaviour of the more relevant orbital elements invisible.

Regarding the dependence of the FTLE on time, we can see that both the structures in the phase space and the values of the FTLE depend on time. On the other hand, after 200 years the FTLE plots for COE and MEE coordinates look more similar (compare Figs. \ref{fig:40890_EccIncl_FTLE_COE_200y} and \ref{fig:40890_EccIncl_FTLE_MEE_200y}) than after 100 years, which suggests convergence of the FTLE. In addition, the FTLE plots considering only $e$, $i$, $\Omega$, $\omega$ look more similar after 200 years, even though the FTLE values are different, see \fref{fig:40890_EccIncl_FTLE_COE_aeiOw_200y}.

\begin{figure*}[tbp]
     \centering
     \subfigure[][FTLE using COE after 100 years]{\includegraphics[width=0.35\textwidth]{input40890_case1_200y_1stZonal1stSunMoon_RAAN260-380_argPer0-355_FTLEtime_100y_NEW} \label{fig:40890_RAANArgP_FTLE_COE_100y}}
     \hspace{4pt}%
     \subfigure[][FTLE using COE after 200 years]{\includegraphics[width=0.35\textwidth]{input40890_case1_200y_1stZonal1stSunMoon_RAAN260-380_argPer0-355_FTLEtime_200y_NEW} \label{fig:40890_RAANArgP_FTLE_COE_200y}}
     
     \subfigure[][FTLE using MEE after 100 years]{\includegraphics[width=0.35\textwidth]{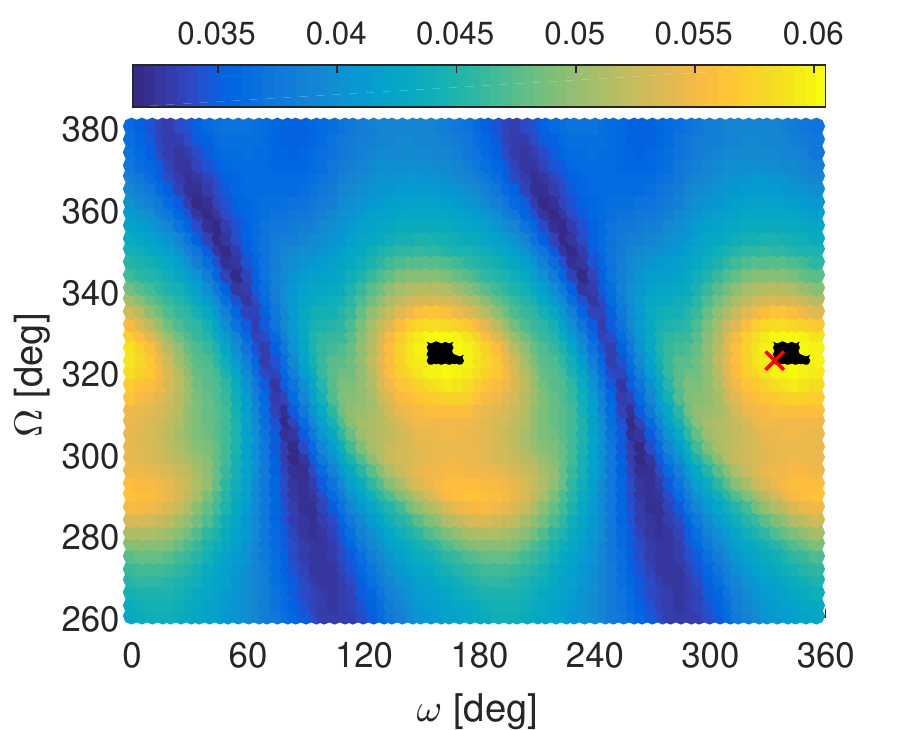} \label{fig:40890_RAANArgP_FTLE_MEE_100y}}
     \hspace{4pt}%
     \subfigure[][FTLE using MEE after 200 years]{\includegraphics[width=0.35\textwidth]{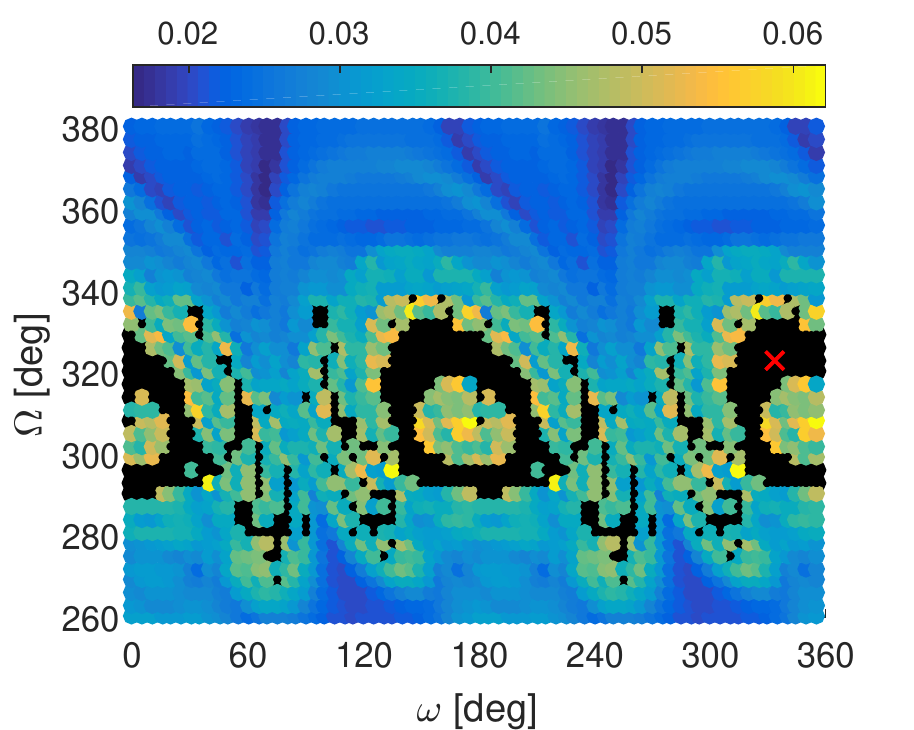} \label{fig:40890_RAANArgP_FTLE_MEE_200y}}
     
     \subfigure[][FTLE for only $e$ and $i$ after 100 years]{\includegraphics[width=0.35\textwidth]{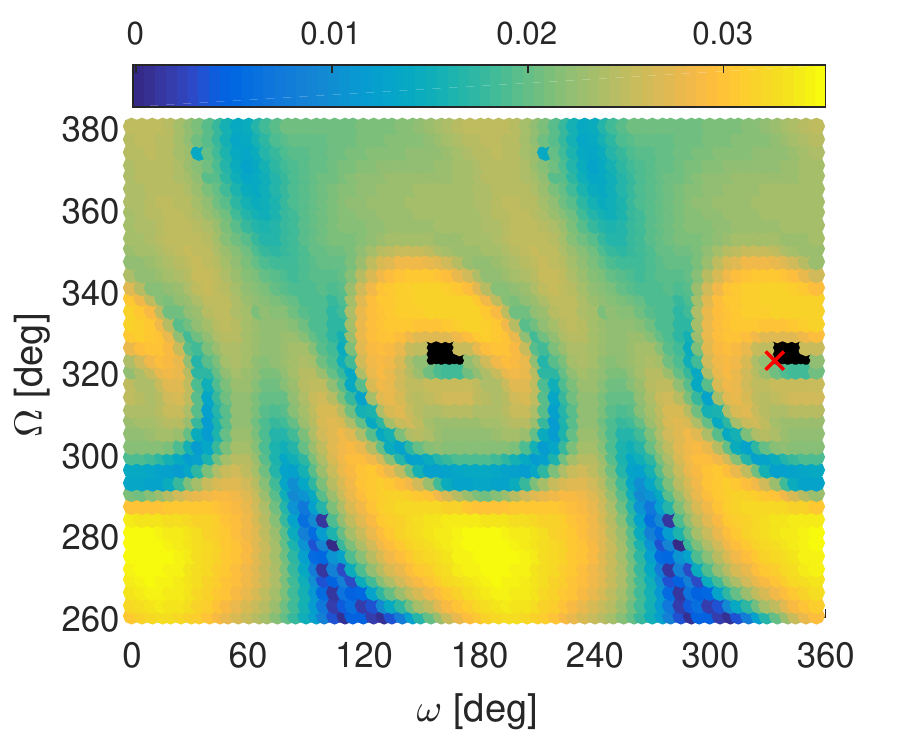} \label{fig:40890_RAANArgP_FTLE_eccIncl_100y}}
     \hspace{4pt}%
     \subfigure[][FTLE for only $e$ and $i$ after 200 years]{\includegraphics[width=0.35\textwidth]{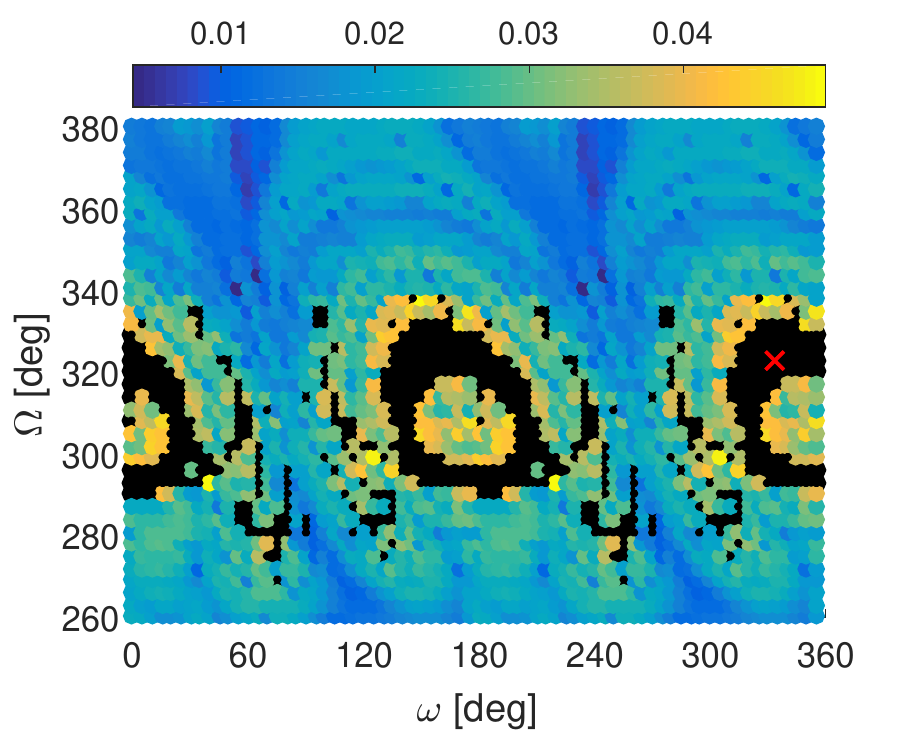} \label{fig:40890_RAANArgP_FTLE_eccIncl_200y}}
     
     \subfigure[][Maximum eccentricity after 100 years]{\includegraphics[width=0.35\textwidth]{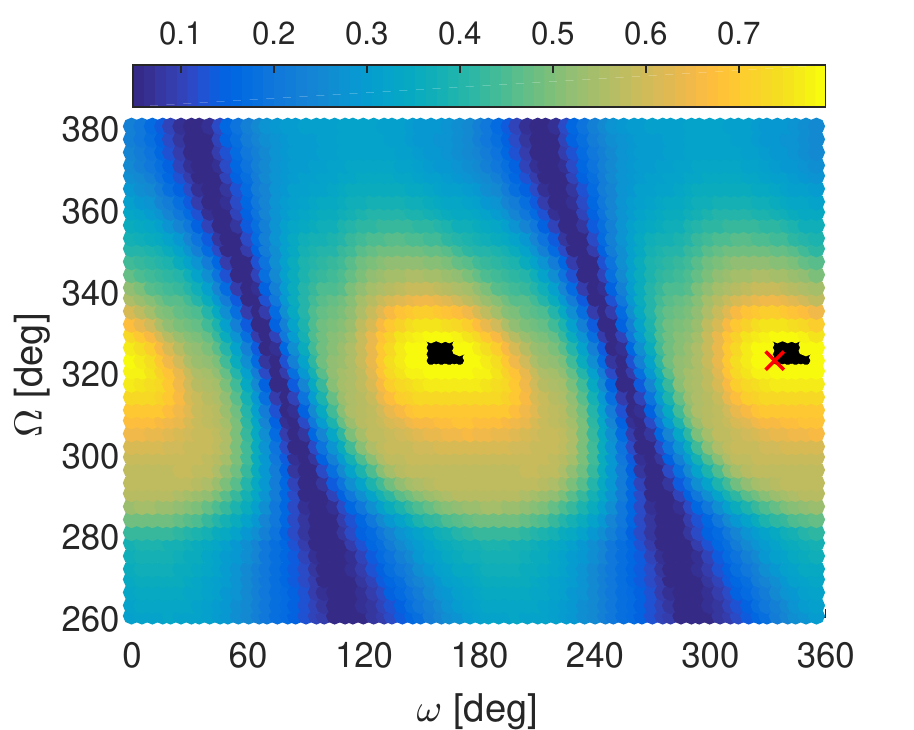} \label{fig:40890_RAANArgP_maxEcc_100y}}
     \hspace{4pt}%
     \subfigure[][Maximum eccentricity after 200 years]{\includegraphics[width=0.35\textwidth]{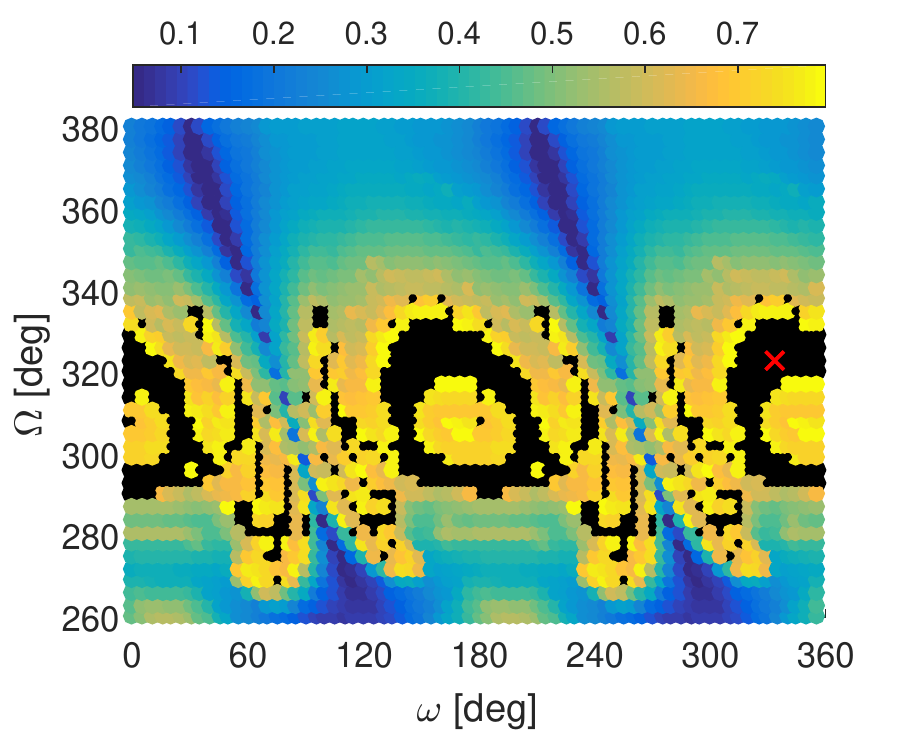} \label{fig:40890_RAANArgP_maxEcc_200y}}
     
     \caption{FTLE computed using COE and MEE and considering only $e$ and $i$, and maximum eccentricity for different initial $\Omega$ and $\omega$ for case 40890 after 100 (left) and 200 years (right)}
     \label{fig:40890_RAANargP_FTLE}
\end{figure*}

The FTLE computed for different initial $\Omega$ and $\omega$ for case 40890 are shown in \fref{fig:40890_RAANargP_FTLE}.
Again, we see increased FTLE values in the region where the eccentricity growth is small when COE are used (see the yellow and green dots in \fref{fig:40890_RAANArgP_FTLE_COE_100y} where $\omega \in [40,120]\degr$ and $\omega \in [220,300]\degr$). These high FTLE values are again caused by divergence in the argument of perigee and are therefore not present when MEE are used (see \fref{fig:40890_RAANArgP_FTLE_MEE_100y}). On the other hand, the FTLE plots computed using COE and MEE after 200 years look similar (compare Figs. \ref{fig:40890_RAANArgP_FTLE_COE_200y} and \ref{fig:40890_RAANArgP_FTLE_MEE_200y}). The values of the FTLE are again dominated by divergence in $\Omega$ and $\omega$, since the FTLE values considering only $e$ and $i$, see Figs. \ref{fig:40890_RAANArgP_FTLE_eccIncl_100y} and \ref{fig:40890_RAANArgP_FTLE_eccIncl_200y}, are smaller than the FTLE considering all elements.

\begin{figure*}[tbp]
     \centering
     \subfigure[][$i$, $\Omega$ and $\omega$ in radians and $a$ unitless]{\includegraphics[width=0.35\textwidth]{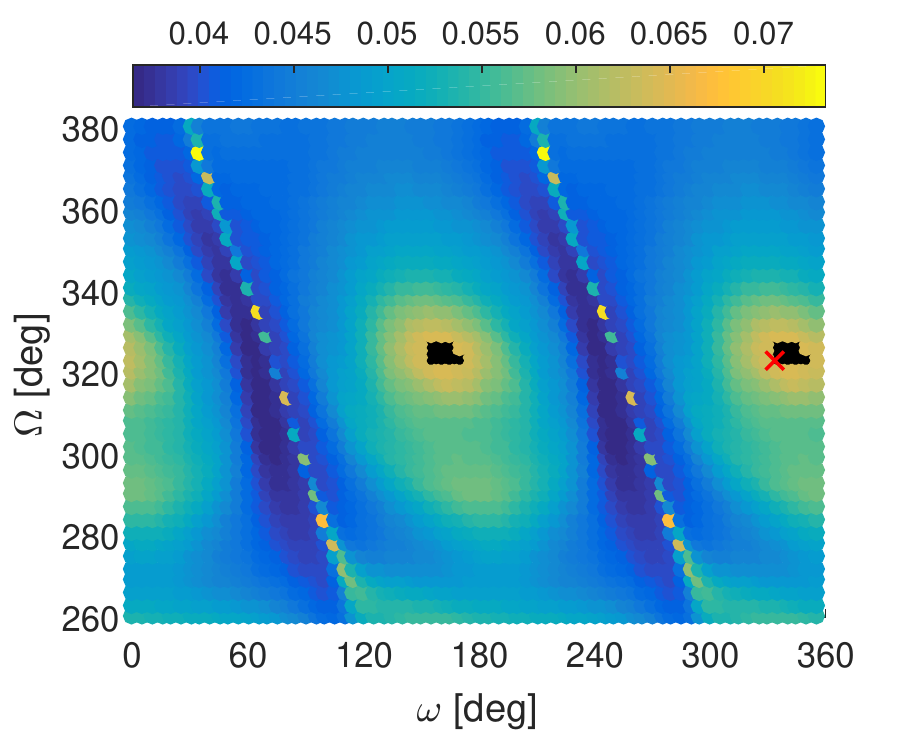} \label{fig:40890_FTLEtime_100y}}
     \hspace{8pt}%
     \subfigure[][$i$, $\Omega$ and $\omega$ in radians and $a$ unitless]{\includegraphics[width=0.35\textwidth]{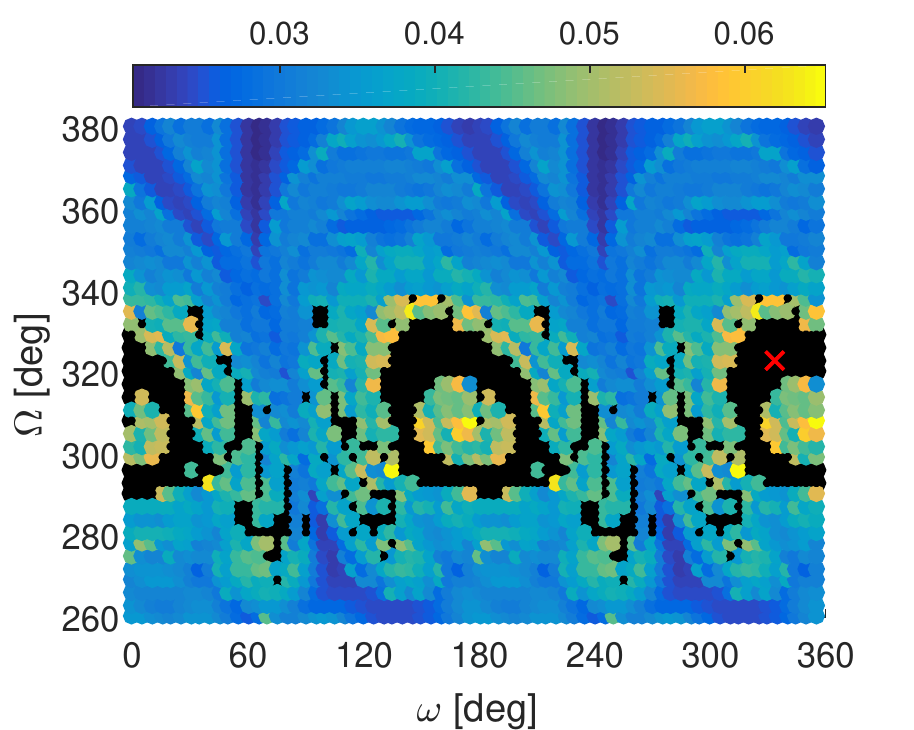} \label{fig:40890_FTLEtime_200y}}
     
     \subfigure[][$i$, $\Omega$ and $\omega$ in degrees and $a$ unitless]{\includegraphics[width=0.35\textwidth]{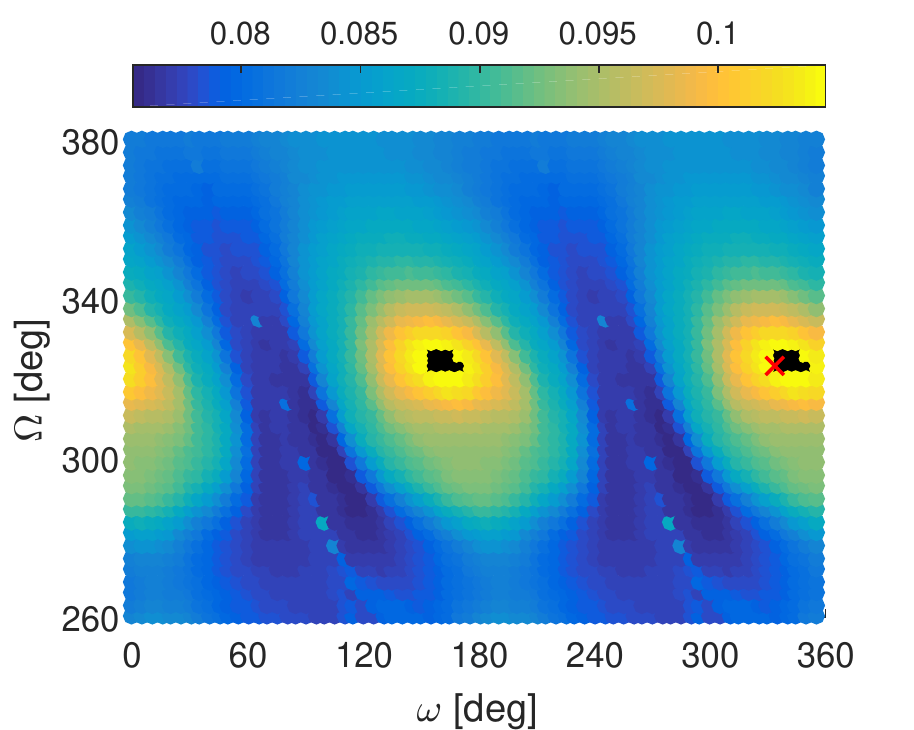} \label{fig:40890_DEGREES_FTLEtime_100y}}
     \hspace{8pt}%
     \subfigure[][$i$, $\Omega$ and $\omega$ in degrees and $a$ unitless]{\includegraphics[width=0.35\textwidth]{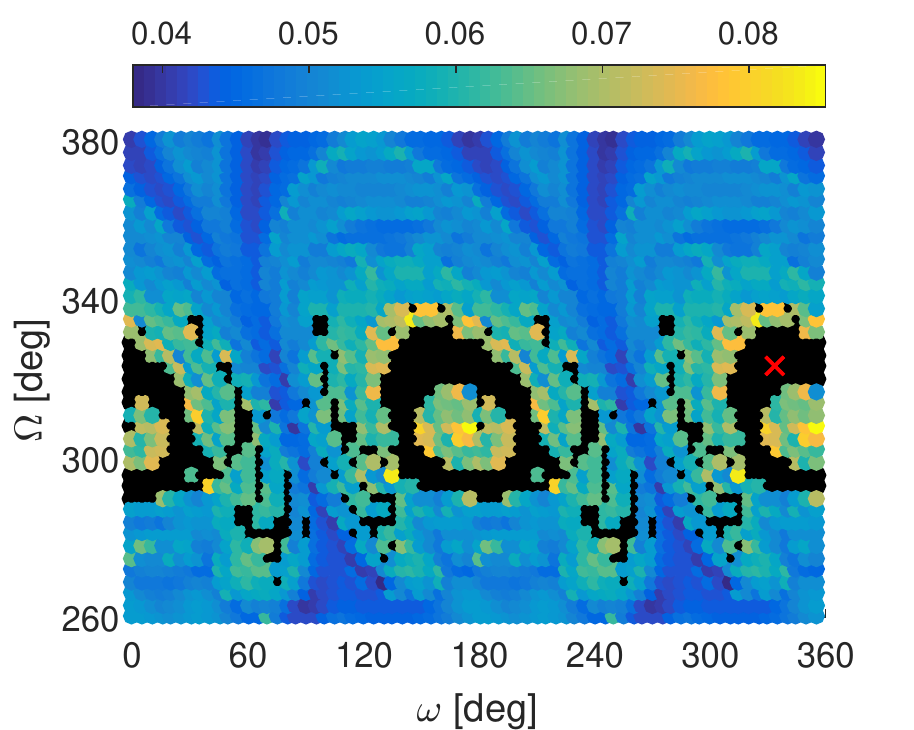} \label{fig:40890_DEGREES_FTLEtime_200y}}
     
     \subfigure[][$i$, $\Omega$ and $\omega$ in radians and $a$ in km]{\includegraphics[width=0.35\textwidth]{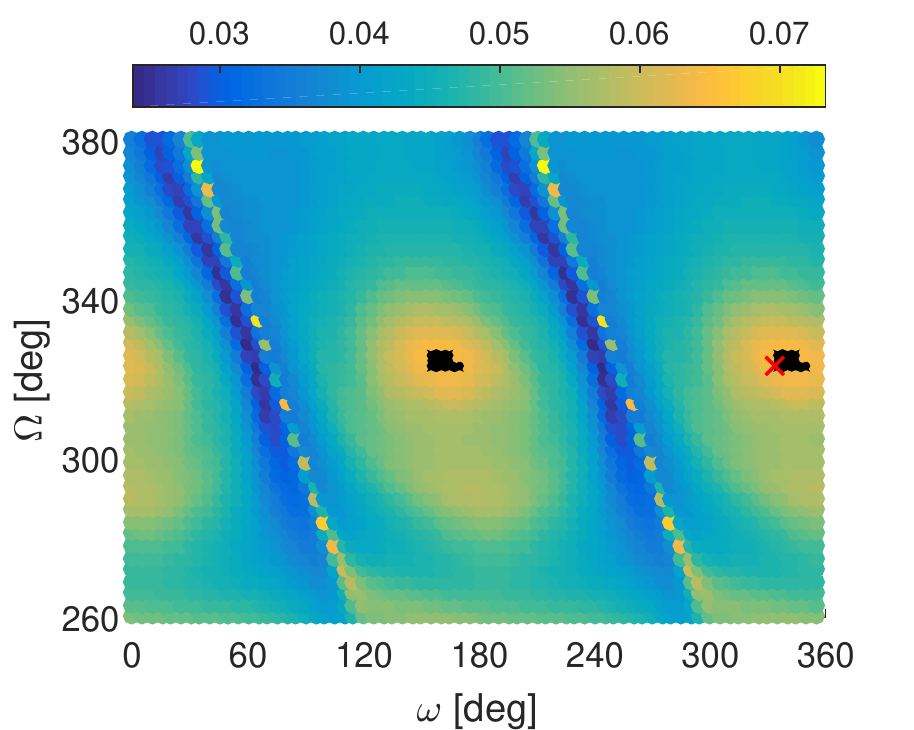} \label{fig:40890_KM_FTLEtime_100y}}
     \hspace{8pt}%
     \subfigure[][$i$, $\Omega$ and $\omega$ in radians and $a$ in km]{\includegraphics[width=0.35\textwidth]{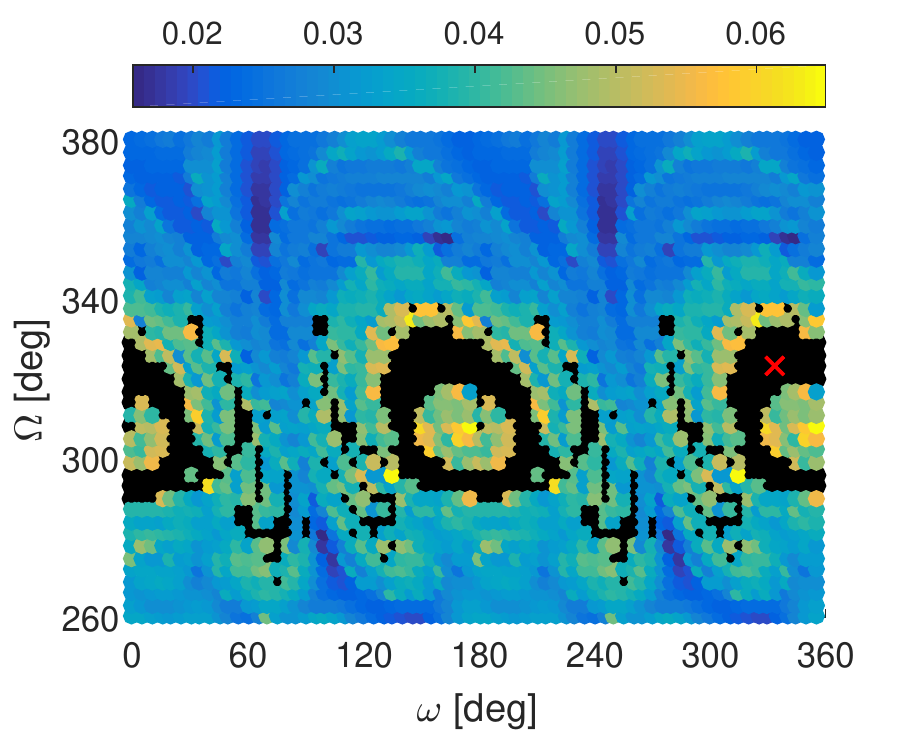} \label{fig:40890_KM_FTLEtime_200y}}
     
     \subfigure[][$i$, $\Omega$ and $\omega$ in degrees and $a$ in km]{\includegraphics[width=0.35\textwidth]{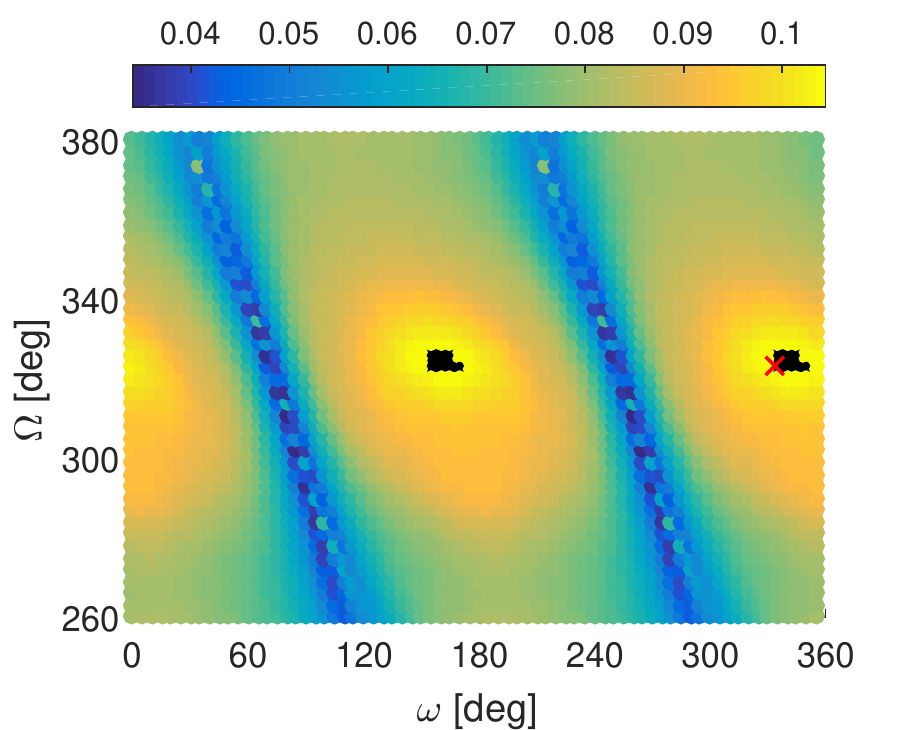} \label{fig:40890_DEGREES_KM_FTLEtime_100y}}
     \hspace{8pt}%
     \subfigure[][$i$, $\Omega$ and $\omega$ in degrees and $a$ in km]{\includegraphics[width=0.35\textwidth]{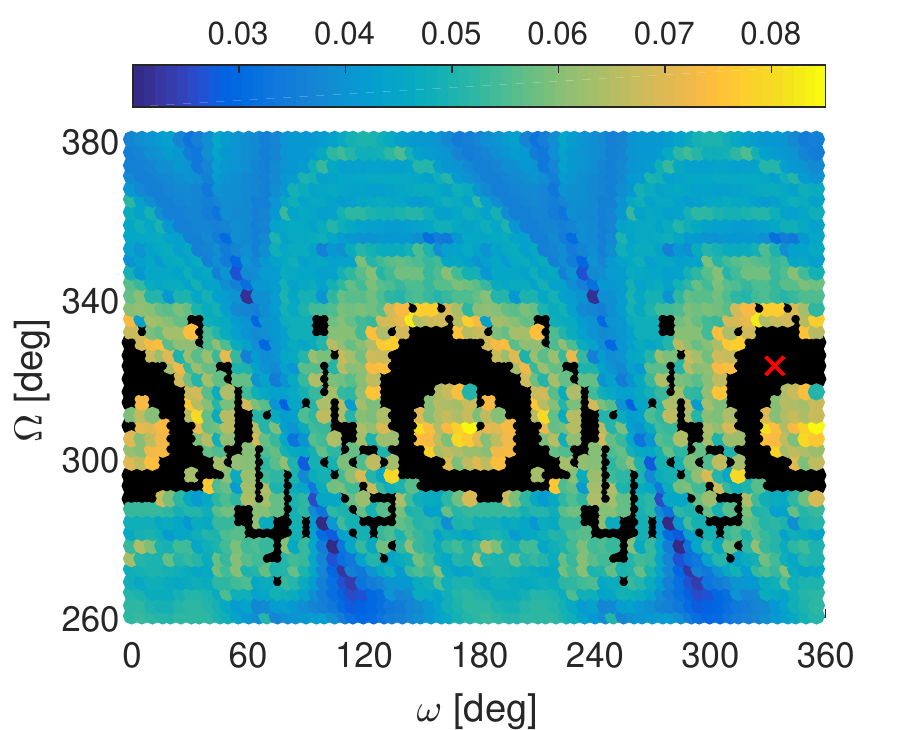} \label{fig:40890_DEGREES_KM_FTLEtime_200y}}
     
     \caption{FTLE for different initial $\Omega$ and $\omega$ for case 40890 computed using different units for $a$, $i$, $\Omega$ and $\omega$ after 100 (left) and 200 years (right). The unit used for $i$, $\Omega$ and $\omega$ is either radian or degree and for $a$ the unit is km or it is unitless, that is $a/a_0$.}
     \label{fig:40890_FTLE_diffCoord}
\end{figure*}

The previous results show that the FTLE depends on the choice of coordinates and on time. We can also compute the FTLE using different units for the orbital elements, e.g. degrees instead of radians. \fref{fig:40890_FTLE_diffCoord} shows the FTLE computed using different units for the semi-major axis $a$ and angles $i$, $\Omega$ and $\omega$ (here Figs. \ref{fig:40890_FTLEtime_100y} and \ref{fig:40890_FTLEtime_200y} are equal to Figs. \ref{fig:40890_RAANArgP_FTLE_COE_100y} and \ref{fig:40890_RAANArgP_FTLE_COE_200y}. The results show that also the choice of units has an impact of the values of the FTLEs and consequently changes the look of the FTLE plot. For comparison, also the FLI has been computed using different units, see \fref{fig:40890_FLI_diffUnits} in the Appendix.

\begin{figure*}[tbp]
     \centering
     \subfigure[][40890]{\includegraphics[width=0.31\textwidth,trim={0 0 0.6cm 0},clip]{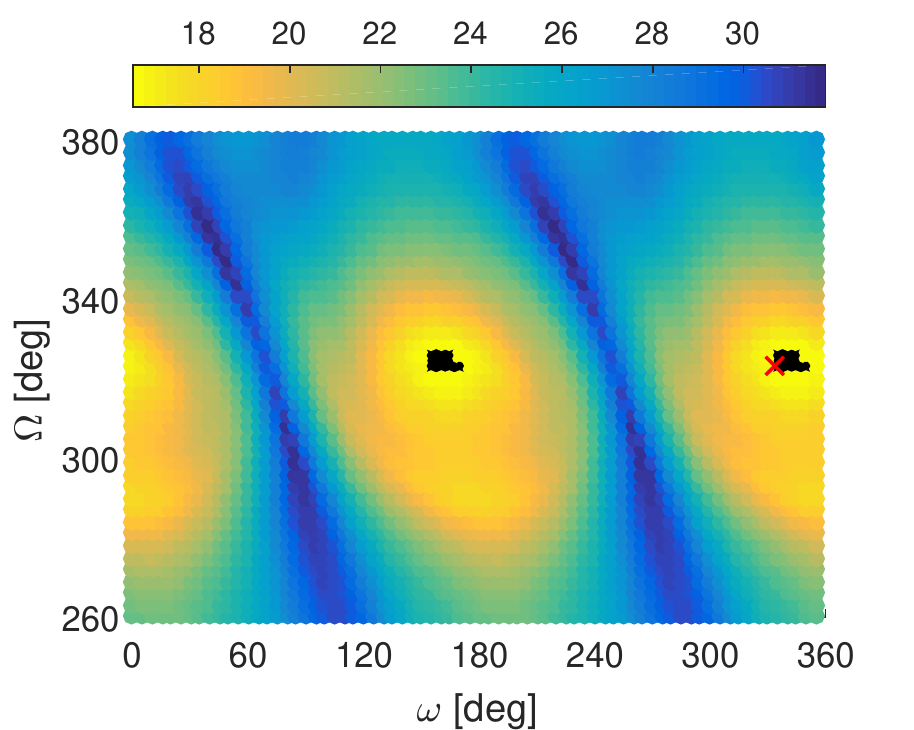} \label{fig:40890_lyapunovtime_RAANargP_MEE}}
     \hspace{4pt}%
     \subfigure[][37846]{\includegraphics[width=0.31\textwidth,trim={0 0 0.6cm 0},clip]{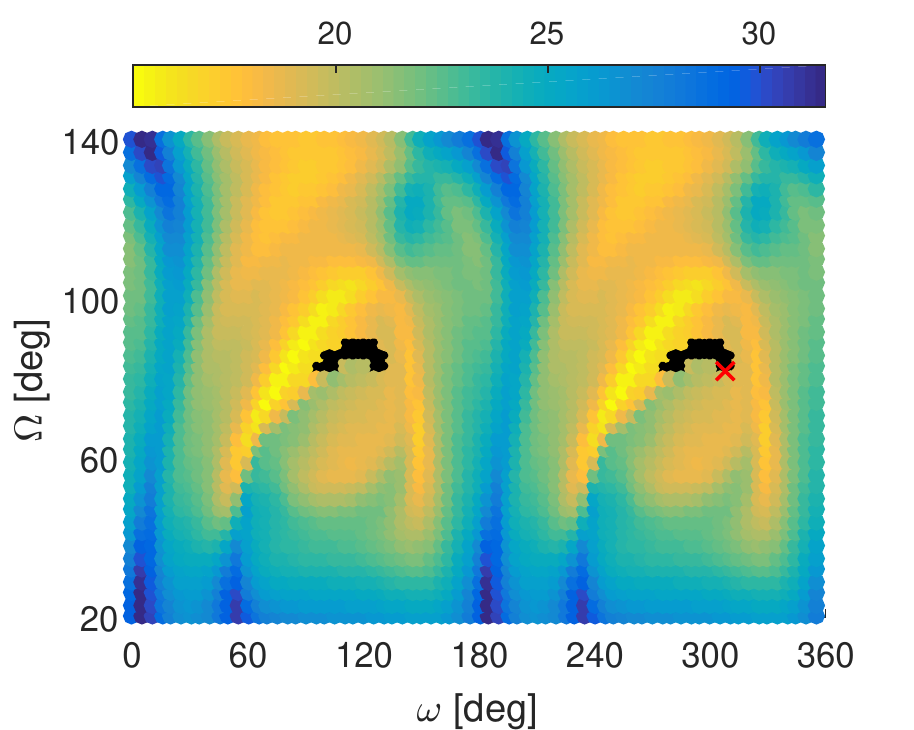} \label{fig:37846_lyapunovtime_RAANargP_MEE}}
     \hspace{4pt}%
     \subfigure[][41175]{\includegraphics[width=0.31\textwidth,trim={0 0 0.6cm 0},clip]{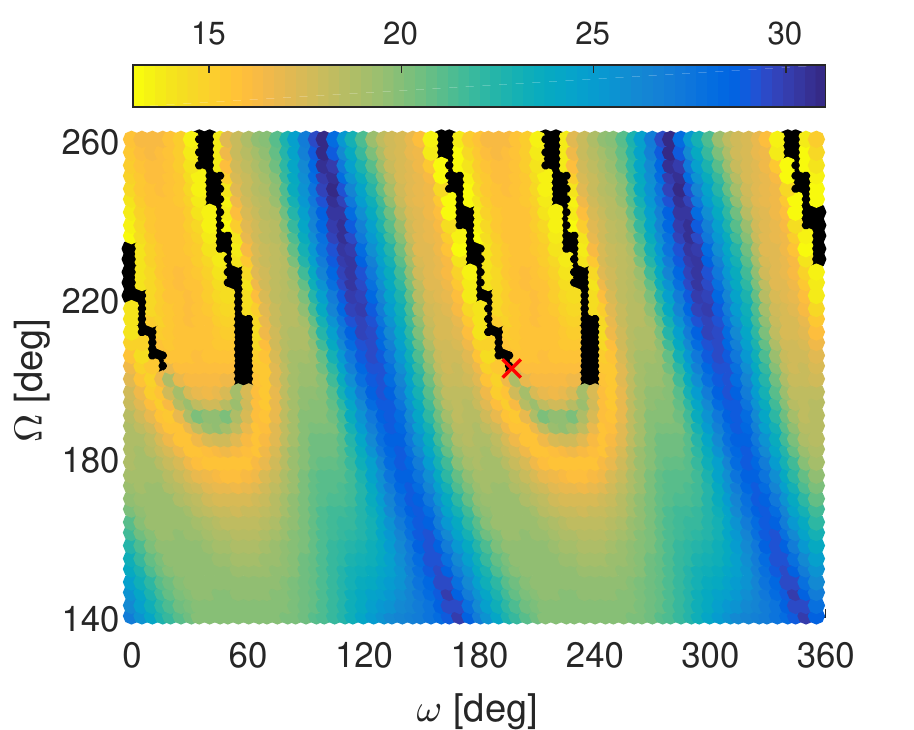} \label{fig:41175_lyapunovtime_RAANargP_MEE}}
     
     \caption{Lyapunov time computed using MEE for different initial $\Omega$ and $\omega$ for cases 40890, 37846 and 41175 after 100 years}
     \label{fig:LyapunovTimeReentryOrbits}
\end{figure*}

From Figs. \ref{fig:40890_RAANArgP_FTLE_COE_100y} and \ref{fig:40890_RAANArgP_FTLE_MEE_100y} it is clear that the re-entry disposal orbit is located in a chaotic region with high FTLE values after 100 years. \fref{fig:40890_lyapunovtime_RAANargP_MEE} shows the Lyapunov time for different initial $\Omega$ and $\omega$ for object 40890 after 100 years computed using MEE. The re-entry orbit is located in a region with very short Lyapunov times and the Lyapunov time of the re-entry orbit is only 16.7 years whereas we are propagating for 100 years. This suggests that we are propagating the orbit beyond its limit of predictability.

On the other hand, Figs. \ref{fig:40890_EccIncl_maxEcc_100y} and \ref{fig:40890_RAANArgP_maxEcc_100y} show that the behaviour of the eccentricity until 100 years is very smooth. All neighbouring orbits have similar values of maximum eccentricity, which suggests that the evolution of the eccentricity is not very sensitive to changes in the initial conditions until 100 years. In addition, all neighbouring orbits have re-entered after 200 years, see Figs. \ref{fig:40890_EccIncl_maxEcc_200y} and \ref{fig:40890_RAANArgP_maxEcc_200y}.

For re-entry cases 37846 and 41175 we find similar results. The Lyapunov times computed using MEE for different initial $\Omega$ and $\omega$ for these two test cases are shown in Figs. \ref{fig:37846_lyapunovtime_RAANargP_MEE} and \ref{fig:41175_lyapunovtime_RAANargP_MEE}. The Lyapunov time of the disposal orbits is much smaller than 100 years, namely 19.4 and 17.9 years\footnote{The Lyapunov time of the re-entry orbits of objects 37846, 40890 and 41175 are 16.5, 15.6 and 15.5 years, respectively, when computed using COE.} for the orbits of 37846 and 41175, respectively. 

\fref{fig:40890_eccIncl_FTLE_COE_aeiOw} has shown that the value of the FTLE is mainly determined by divergence in $\Omega$ and $\omega$. In addition, structures in the phase space visible after 100 years due to the behaviour of $\omega$ are not visible any more after 200 years. Also, the strong divergence in $\omega$ is not visible when MEE instead of COE coordinates are used to compute the FTLE. Finally, the re-entry orbits are located in chaotic regions and have a Lyapunov time smaller than 20 years, whereas 100-year propagations are required.

\subsubsection{Graveyard}
\begin{figure*}[tbp]
     \centering
     \subfigure[][FTLE using COE after 100 years]{\includegraphics[width=0.45\textwidth,trim={2.7cm 3.8cm 15.8cm 0},clip]{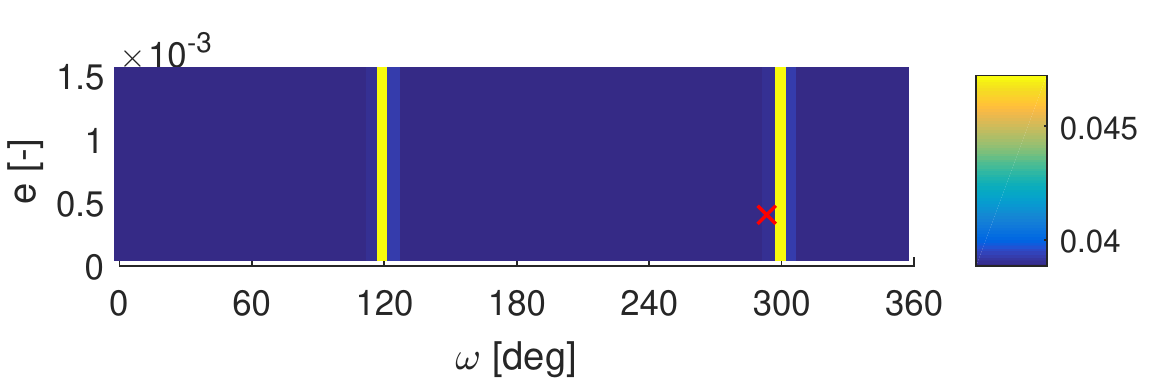} \label{fig:38858_eccArgP_FTLE_COE_100y}}
     \hspace{1pt}%
     \subfigure[][FTLE using COE after 200 years]{\includegraphics[width=0.45\textwidth,trim={15.8cm 3.8cm 2.7cm 0},clip]{input38858_case2_200y_1stZonal1stSunMoon_DAexpansionAEIOo_1stO_every10years_compact_sc1e-3_ecc0_0001-0_0015_argPer0-355_FTLEtime_maxEcc_100-200yv2.pdf} \label{fig:38858_eccArgP_FTLE_COE_200y}}
     
     \subfigure[][FTLE using MEE after 100 years]{\includegraphics[width=0.45\textwidth,trim={2.7cm 3.8cm 15.8cm 0},clip]{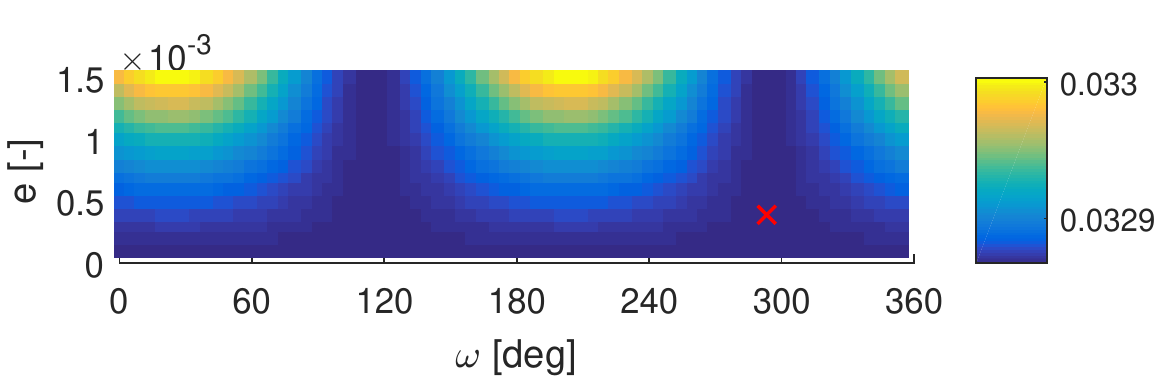} \label{fig:38858_eccArgP_FTLE_MEE_100y}}
     \hspace{1pt}%
     \subfigure[][FTLE using MEE after 200 years]{\includegraphics[width=0.45\textwidth,trim={15.8cm 3.8cm 2.7cm 0},clip]{input38858_case2_200y_1stZonal1stSunMoon_DAexpansionAEIOo_1stO_every10years_compact_sc1e-3_ecc0_0001-0_0015_argPer0-355_MEE_FTLEtime_maxEcc_100-200yv2.pdf} \label{fig:38858_eccArgP_FTLE_MEE_200y}}
     
     \subfigure[][FTLE for $e$ and $i$ only after 100 years]{\includegraphics[width=0.45\textwidth,trim={2.7cm 3.8cm 15.8cm 0},clip]{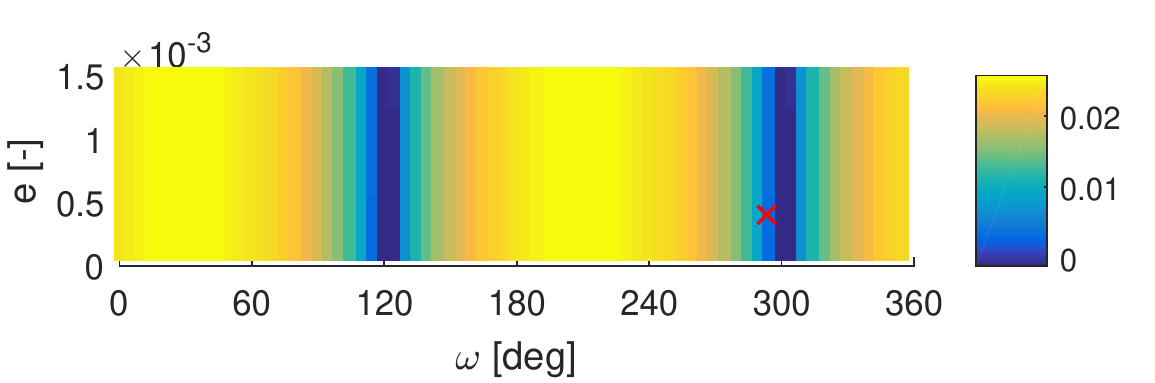} \label{fig:38858_eccArgP_FTLE_eccIncl_100y}}
     \hspace{1pt}%
     \subfigure[][FTLE for $e$ and $i$ only after 200 years]{\includegraphics[width=0.45\textwidth,trim={15.8cm 3.8cm 2.7cm 0},clip]{input38858_case2_200y_1stZonal1stSunMoon_DAexpansionAEIOo_1stO_every10years_compact_sc1e-3_ecc0_0001-0_0015_argPer0-355_FTLEtime_maxEcc_100-200y_FTLEonlyEccIncv2.pdf} \label{fig:38858_eccArgP_FTLE_eccIncl_200y}}
     
     \subfigure[][Maximum eccentricity after 100 years]{\includegraphics[width=0.45\textwidth,trim={2.7cm 0 15.8cm 3.9cm},clip]{input38858_case2_200y_1stZonal1stSunMoon_DAexpansionAEIOo_1stO_every10years_compact_sc1e-3_ecc0_0001-0_0015_argPer0-355_FTLEtime_maxEcc_100-200yv2.pdf} \label{fig:38858_eccArgP_maxEcc_100y}}
     \hspace{1pt}%
     \subfigure[][Maximum eccentricity after 200 years]{\includegraphics[width=0.45\textwidth,trim={15.8cm 0 2.7cm 3.9cm},clip]{input38858_case2_200y_1stZonal1stSunMoon_DAexpansionAEIOo_1stO_every10years_compact_sc1e-3_ecc0_0001-0_0015_argPer0-355_FTLEtime_maxEcc_100-200yv2.pdf} \label{fig:38858_eccArgP_maxEcc_200y}}
     
     \caption{FTLE computed using COE and MEE and considering only $e$ and $i$, and maximum eccentricity for different initial $e$ and $\omega$ for case 38858 after 100 (left) and 200 years (right)}
     \label{fig:38858_eccArgP_COE}
\end{figure*}

\fref{fig:38858_eccArgP_COE} shows the FTLE computed using COE and MEE and considering only $e$ and $i$, and the maximum eccentricity after 100 and 200 years for different initial $e$ and $\omega$ for the graveyard orbit scenario. The actual graveyard orbit is indicated by the red cross. Note that all the plots have different colour scales, so yellow, green and blue colours indicate different FLTE values in different plots.

As expected the graveyard orbit is located in a region of low eccentricity growth. However, according to the FTLE plot computed using COE after 100 years, the orbit is positioned close to a chaotic region at $\omega=300\degr$, see \fref{fig:38858_eccArgP_FTLE_COE_100y}. On the other hand, after 200 years this region of increased FTLE at $\omega=300\degr$ has disappeared, see \fref{fig:38858_eccArgP_FTLE_COE_200y}, and instead has become a region of low FTLE that corresponds to the region of low eccentricity growth, see \fref{fig:38858_eccArgP_maxEcc_200y}.

When we compute the FTLE using MEE, the region of increased FTLE at $\omega$ is $120\degr$ and $300\degr$ is not present after 100 years, see \fref{fig:38858_eccArgP_FTLE_MEE_100y}, and the values of the FTLE and the maximum eccentricity seem to be correlated, compare Figs. \ref{fig:38858_eccArgP_FTLE_MEE_200y} and \ref{fig:38858_eccArgP_maxEcc_200y}. 

The increased FTLE values computed using COE after 100 years are caused by the behaviour of the argument of perigee $\omega$. The system contains separatrices located close to $\omega$ equal to 120$\degr$ and 300$\degr$ that separate regions of different behaviour of the argument of perigee. Neighbouring orbits starting close to the separatrix can diverge strongly, hence the increased FTLE values. The mechanics behind this behaviour of the argument of perigee will be investigated in future work.
As in the 40890 test case, the strong divergence in the argument of perigee $\omega$ corresponds to minimum growth and divergence in the eccentricity.

\begin{figure*}[tbp]
     \centering
     \subfigure[][FTLE using COE after 100 years]{\includegraphics[width=0.35\textwidth]{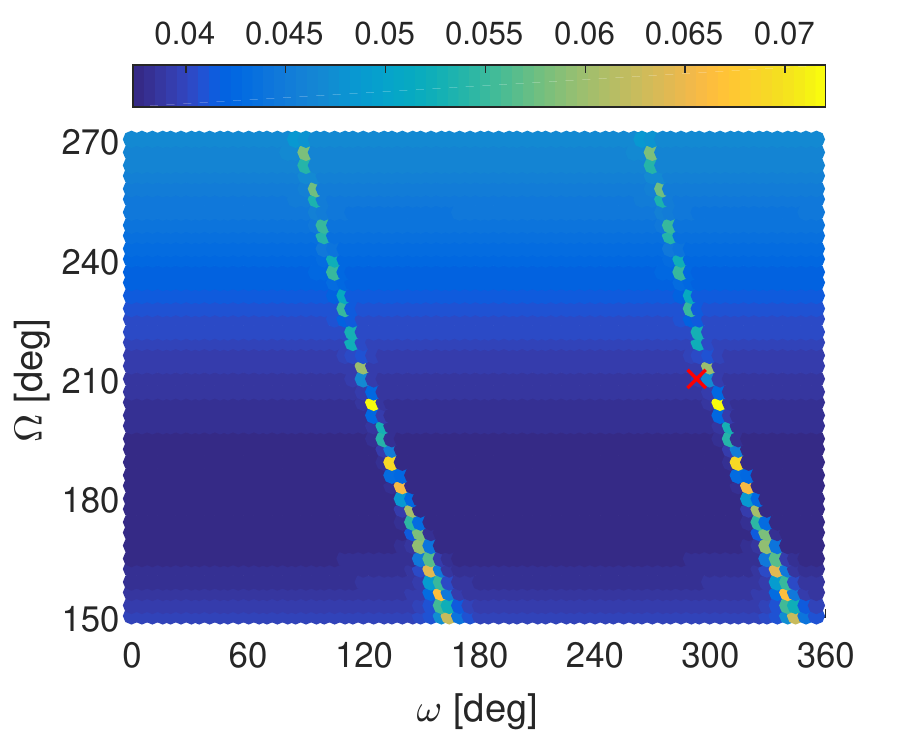} \label{fig:38858_RAANArgP_FTLE_COE_100y}}
     \hspace{8pt}%
     \subfigure[][FTLE using COE after 200 years]{\includegraphics[width=0.35\textwidth]{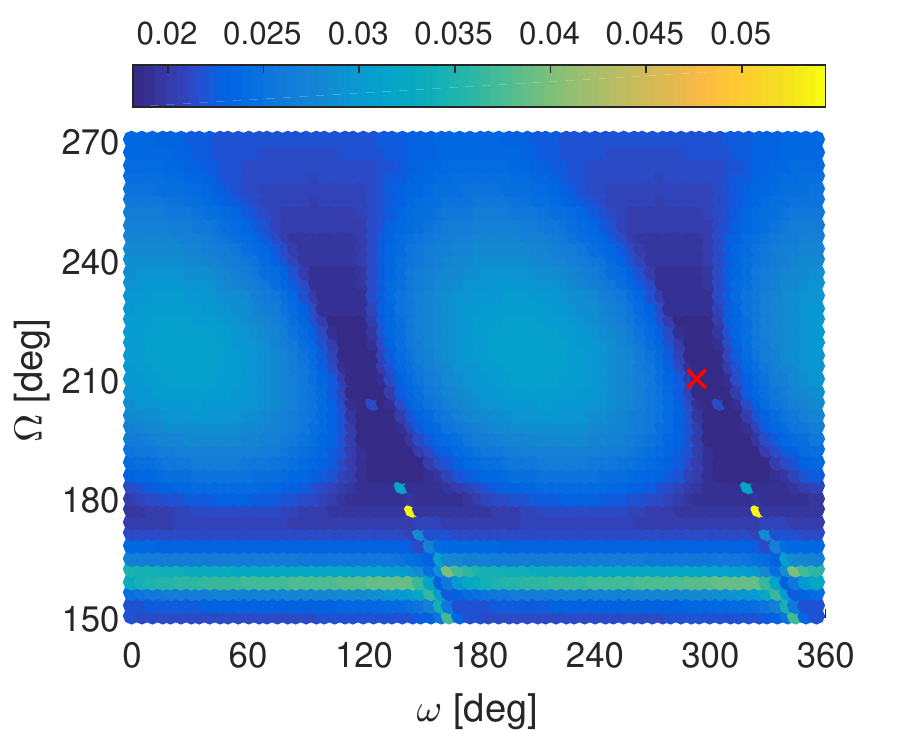} \label{fig:38858_RAANArgP_FTLE_COE_200y}}
     
     \subfigure[][FTLE using MEE after 100 years]{\includegraphics[width=0.35\textwidth]{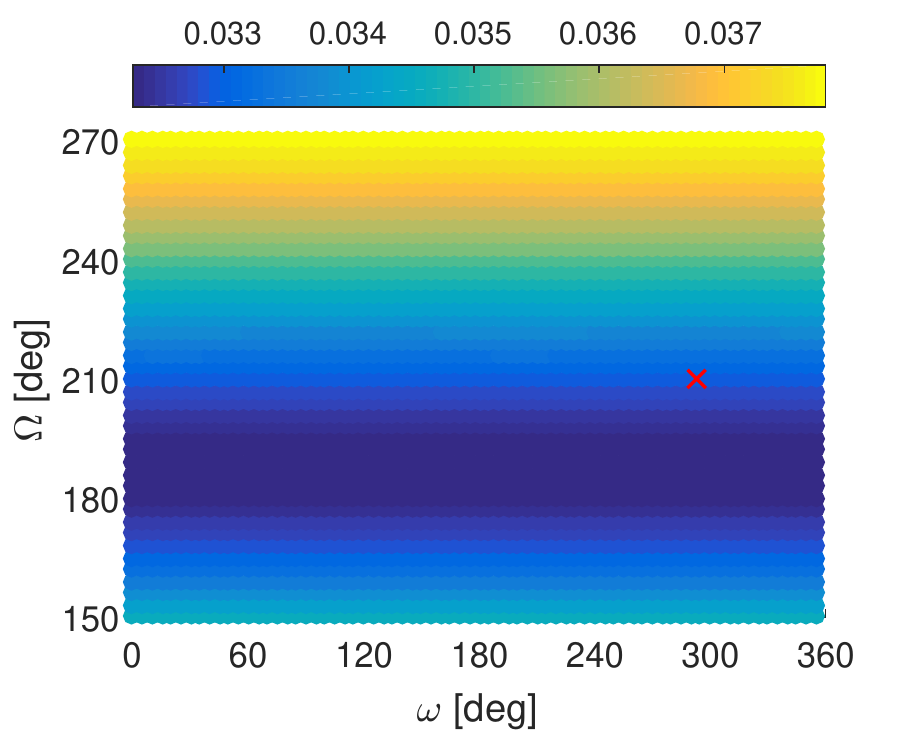} \label{fig:38858_RAANArgP_FTLE_MEE_100y}}
     \hspace{8pt}%
     \subfigure[][FTLE using MEE after 200 years]{\includegraphics[width=0.35\textwidth]{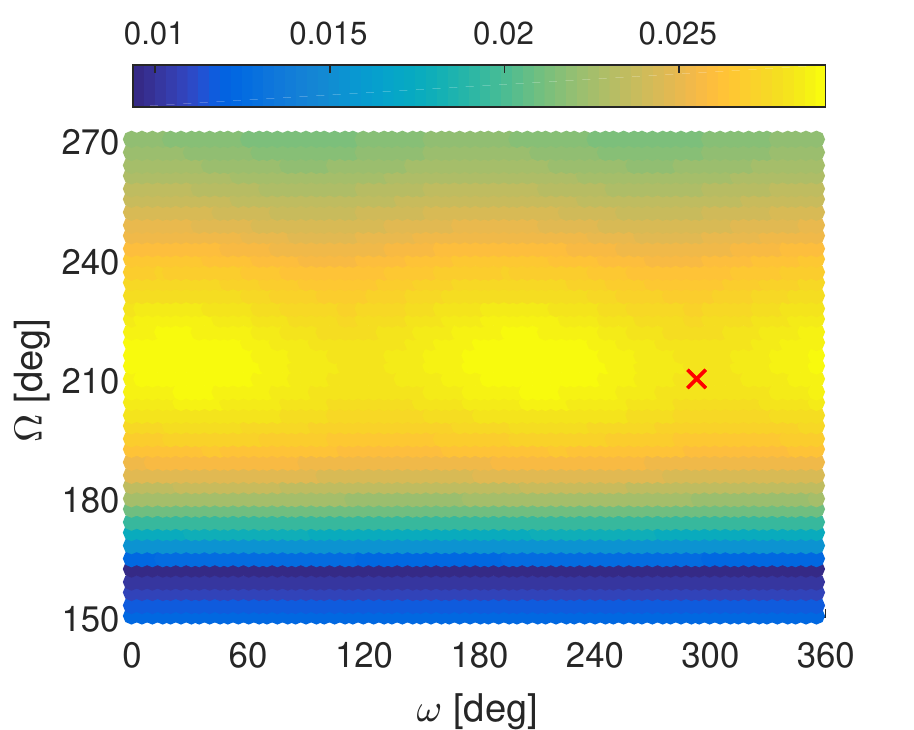} \label{fig:38858_RAANArgP_FTLE_MEE_200y}}
     
      \subfigure[][FTLE for only $e$ and $i$ after 100 years]{\includegraphics[width=0.35\textwidth]{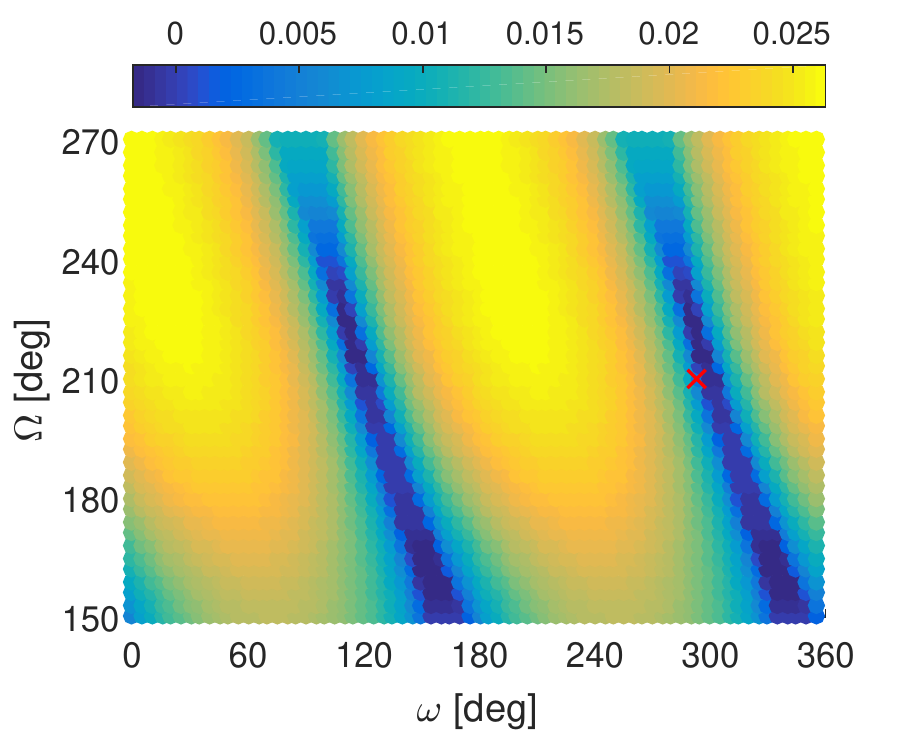} \label{fig:38858_RAANArgP_FTLE_eccIncl_100y}}
     \hspace{8pt}%
     \subfigure[][FTLE for only $e$ and $i$ after 200 years]{\includegraphics[width=0.35\textwidth]{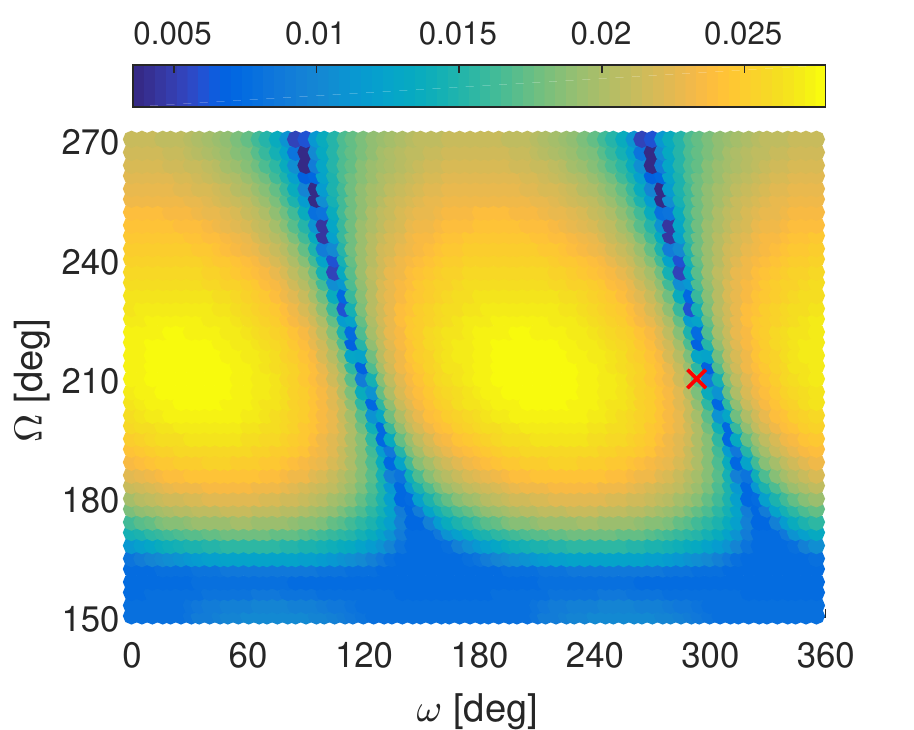} \label{fig:38858_RAANArgP_FTLE_eccIncl_200y}}
     
     \subfigure[][Maximum eccentricity after 100 years]{\includegraphics[width=0.35\textwidth,trim={0 0 0 0.08cm},clip]{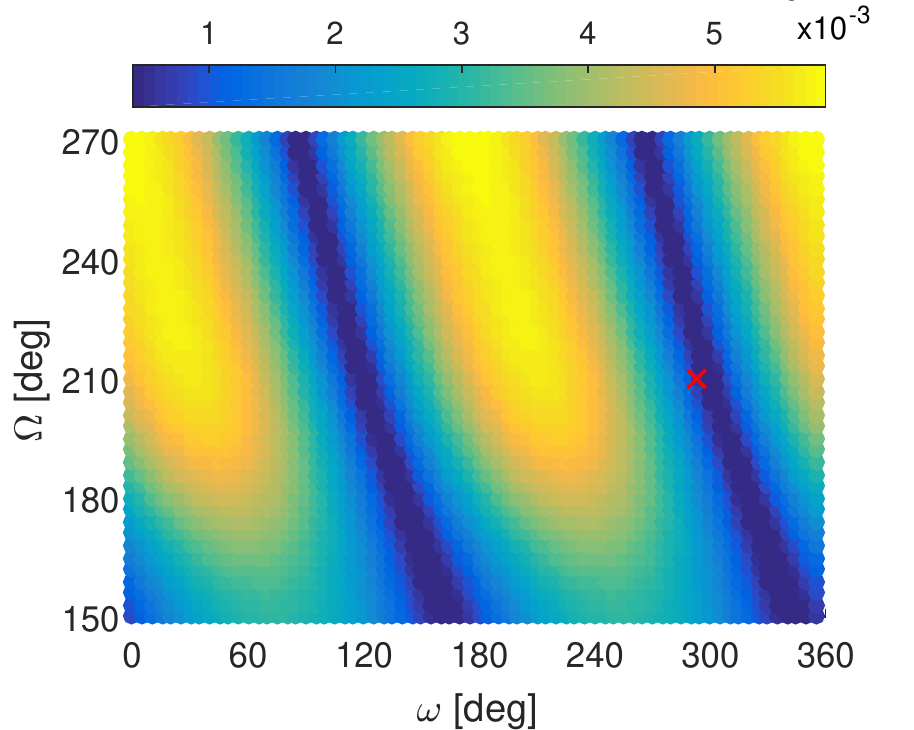} \label{fig:38858_RAANArgP_maxEcc_100y}}
     \hspace{8pt}%
     \subfigure[][Maximum eccentricity after 200 years]{\includegraphics[width=0.35\textwidth,trim={0 0 0 0.08cm},clip]{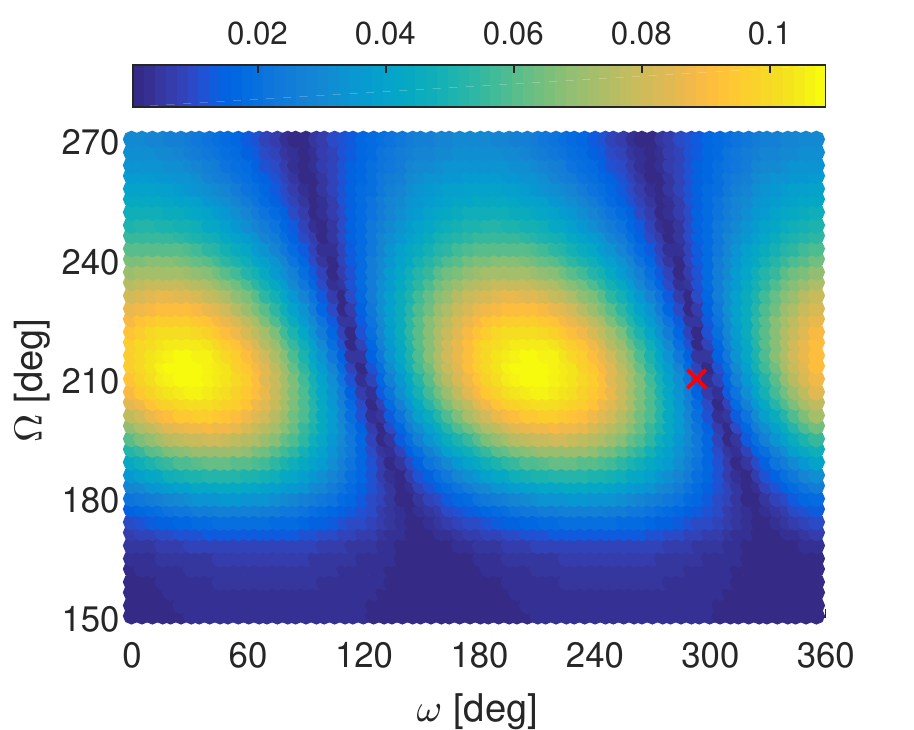} \label{fig:38858_RAANArgP_maxEcc_200y}}
     
     \caption{FTLE computed using COE and MEE and considering only $e$ and $i$, and maximum eccentricity for different initial $\Omega$ and $\omega$ for case 38858 after 100 (left) and 200 years (right)}
     \label{fig:38858_RAANargP_COE}
\end{figure*}

\fref{fig:38858_RAANargP_COE} shows the FTLE and maximum eccentricity for different initial $\Omega$ and $\omega$ for the graveyard orbit. 
The graveyard orbit's initial $\Omega$ of 210$\degr$ results in large eccentricity growth for most initial values of the argument of perigee except in two narrow blue valleys at $\omega=120\degr, 300\degr$, see Figs. \ref{fig:38858_RAANArgP_maxEcc_100y} and \ref{fig:38858_RAANArgP_maxEcc_200y}. This means that a small perturbation in the initial argument of perigee will result in a significantly larger eccentricity on the long term. This sensitivity of the eccentricity to changes in the initial $\omega$ is undesirable, because the error in initial $\omega$ due to manoeuvre uncertainties may be as large as 4.5$\degr$, see Table~\ref{tab:manoeuvreStateUncertainty}. The sensitivity of the eccentricity to the initial $\omega$ can be reduced by changing the initial $i$ or $\Omega$, e.g. changing $\Omega$ to $150\degr$ (see \fref{fig:38858_RAANArgP_maxEcc_200y}). This is, however, impractical in terms of required $\Delta V$ or waiting time needed for $\Omega$ to change sufficiently due to natural precession\footnote{It takes approximately 6 years for $\Omega$ to change from 210$\degr$ to 150$\degr$ by natural precession.}.

Furthermore, we again find high FTLE values computed using COE in the region where the eccentricity growth is low as a result of divergence in the argument of perigee, see Figs. \ref{fig:38858_RAANArgP_FTLE_COE_100y} and \ref{fig:38858_RAANArgP_FTLE_COE_200y}. Besides, opposite to previous results using MEE, the relation between the FTLE computed using MEE after 100 years and maximum eccentricity seems absent, compare Figs. \ref{fig:38858_RAANArgP_FTLE_MEE_100y} and \ref{fig:38858_RAANArgP_maxEcc_100y}. Only after 200 years, we see high FLTE values in the range of initial $\Omega$ that corresponds to large eccentricity growth, compare Figs. \ref{fig:38858_RAANArgP_FTLE_MEE_200y} and \ref{fig:38858_RAANArgP_maxEcc_200y}. On the other hand, if we compute the FTLE considering only $e$ and $i$ then the correlation between FTLE and eccentricity growth seems clear, see Figs. \ref{fig:38858_RAANArgP_FTLE_eccIncl_100y} and \ref{fig:38858_RAANArgP_FTLE_eccIncl_200y}. Note that after 100 years, we find orbits whose FTLE considering only $e$ and $i$ is negative, see \fref{fig:38858_RAANArgP_FTLE_eccIncl_100y}, which means that the eccentricity and inclination of neighbouring orbits do not diverge exponentially fast.

\begin{figure*}[tbp]
     \centering
     \subfigure[][Lyapunov time using COE]{\includegraphics[width=0.35\textwidth]{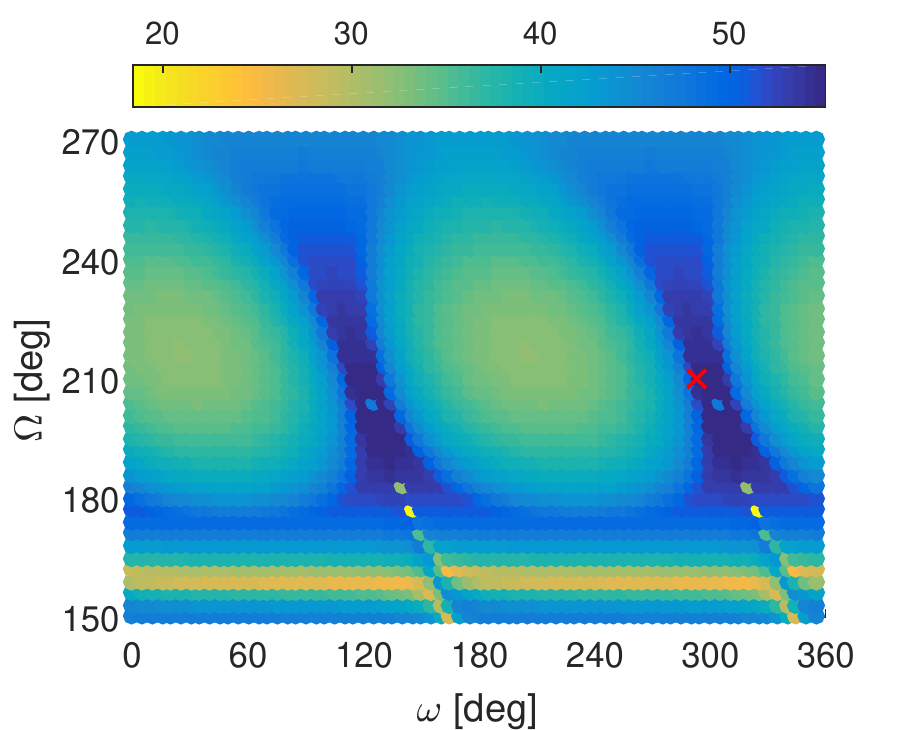} \label{fig:38858_RAANArgP_LyapunovTime_COE_200y}}
     \hspace{8pt}%
     \subfigure[][Lyapunov time using MEE]{\includegraphics[width=0.35\textwidth]{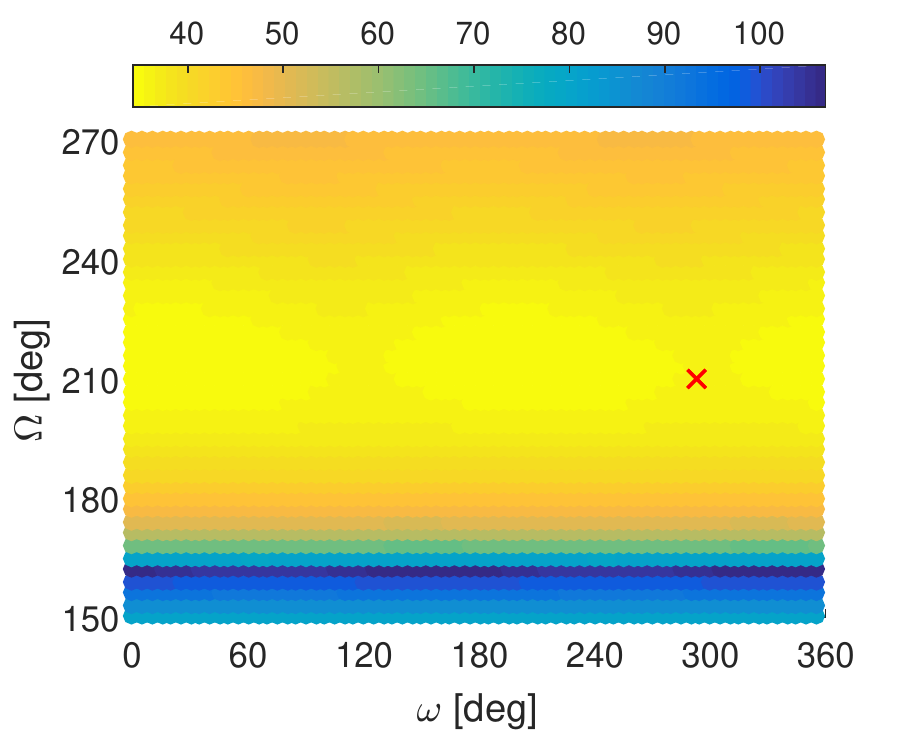} \label{fig:38858_RAANArgP_LyapunovTime_MEE_200y}}
     
     \caption{Lyapunov time computed using COE (left) and MEE (right) for different initial $\Omega$ and $\omega$ for case 38858 after 200 years}
     \label{fig:38858_RAANArgP_LyapunovTime}
\end{figure*}

Finally, Figs. \ref{fig:38858_eccArgP_COE} and \ref{fig:38858_RAANargP_COE} show that all FTLE in the investigated domain of the phase space are larger than zero. This indicates that all orbits are chaotic, which is undesirable for a graveyard orbit.
\fref{fig:38858_RAANArgP_LyapunovTime} shows the Lyapunov time computed using COE and MEE for different initial $\Omega$ and $\omega$ for the graveyard orbit scenario after 200 years. According to the plot computed using COE, the graveyard orbit is located in a region with the longest Lyapunov times in the domain, see \fref{fig:38858_RAANArgP_LyapunovTime_COE_200y}, whereas the plot using MEE indicates that the graveyard orbit is in a region with the shortest Lyapunov times, see \fref{fig:38858_RAANArgP_LyapunovTime_MEE_200y}. The Lyapunov time of the graveyard orbit computed using COE and MEE after 200 years is 54.6 and 35.5 years, respectively, which means that the Lyapunov time computed at finite time depends significantly of the choice of coordinates. Moreover, both Lyapunov times suggests that we are propagating the orbit beyond the limits of predictability.

To summarise, the FTLE plots for different initial $e$ and $\omega$ and different initial $\Omega$ and $\omega$ have shown that the results of the FTLE analysis depend on the choice of coordinates. In addition, dynamical structures (e.g. regions of high FTLE) that are appear after 100 years are not visible after 200 years (the volume of the regions shrunk such that they are not visible any more). Furthermore, there is not always a clear correlation between the FTLE and the growth in eccentricity. Finally, the Lyapunov time of the graveyard orbit is much smaller than the required propagation time of 200 years.

\subsection{Sensitivity analysis}
To determine if the disposal orbits satisfy the disposal requirements when subject to manoeuvre uncertainties, we investigate the effect of manoeuvre errors on the evolution of the disposal orbits using sensitivity analysis. This also allows us to see if the orbits display chaotic behaviour and to explicitly compute the divergence between orbits due to manoeuvre errors. For this, 225 different manoeuvre errors are sampled (by combining nine different $\Delta V$ errors $\in [-1,1]\%$ and five different errors $\in [-1,1]\degr$ in both $\alpha$ and $\delta$) and the resulting initial conditions are propagated for 110 years for re-entry orbits and for 200 years for the graveyard orbit.

\subsubsection{Re-entry}
\begin{figure*}
\centering
\includegraphics[width=0.8\textwidth]{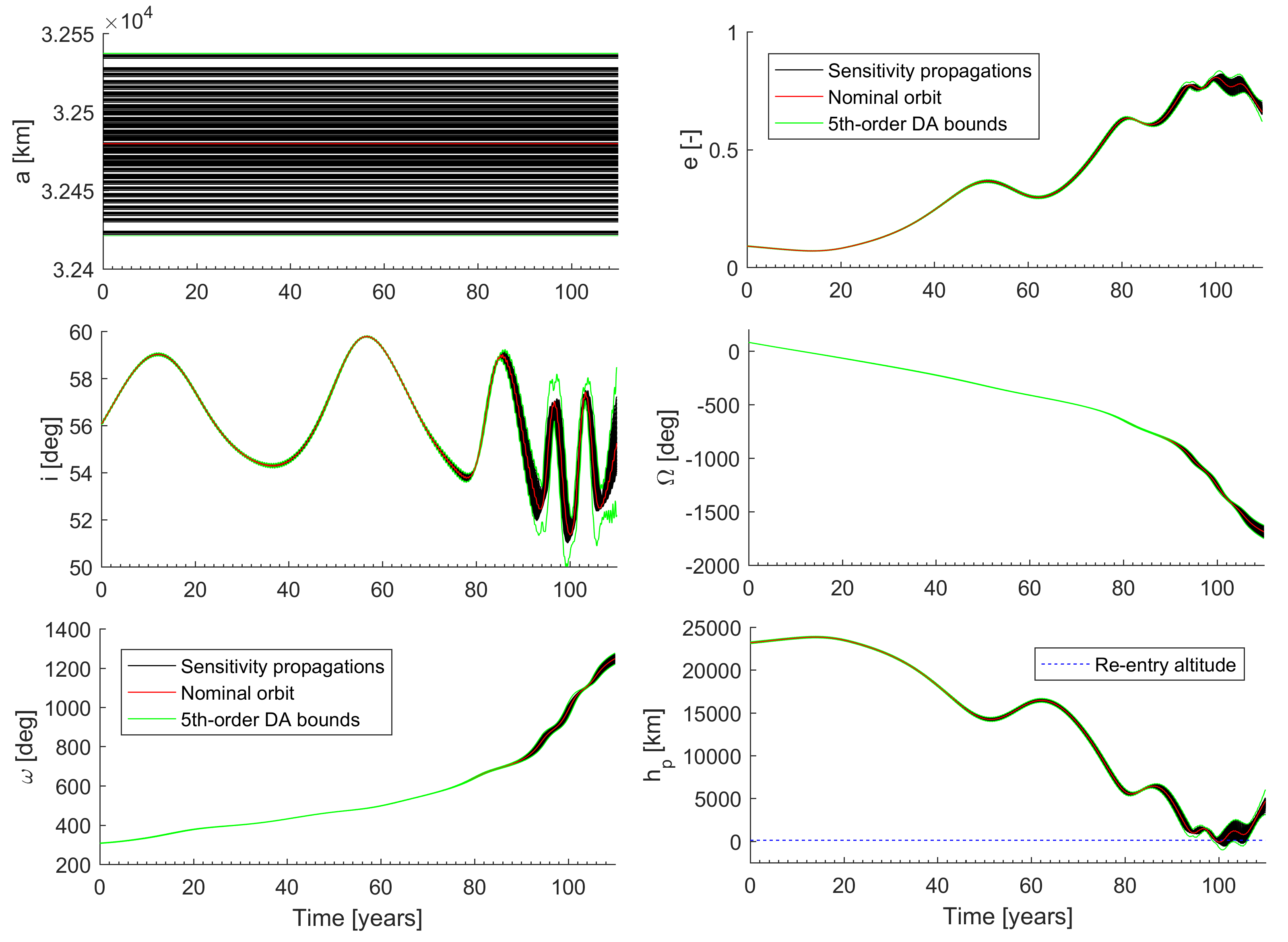}
\caption{200-year evolution of 225 different orbits (black lines) in the uncertainty domain of object 37846. The red curve is the nominal orbit and the green curves are the bounds of the uncertainty domain computed using a 5th-order DA expansion. The blue dashed line is the re-entry altitude.}
\label{fig:37846_case1_sensitivity}
\end{figure*}

\fref{fig:37846_case1_sensitivity} shows the orbital evolution of 225 orbits (black curves) that start in the initial uncertainty domain due to manoeuvre uncertainties for reentry case 37846 over 110 years. The green curves indicate the bounds of the uncertainty set computed using a 5th-order DA expansion and the red curve is the nominal orbit. The dashed blue line shows the re-entry altitude of 120 km. All orbits in the uncertainty set are within the bounds computed using the 5th-order DA expansion. The bounds overestimate the domain of the uncertainty set, but are a good estimate of size of the domain. 

\begin{figure}
\centering
\includegraphics[width=\columnwidth]{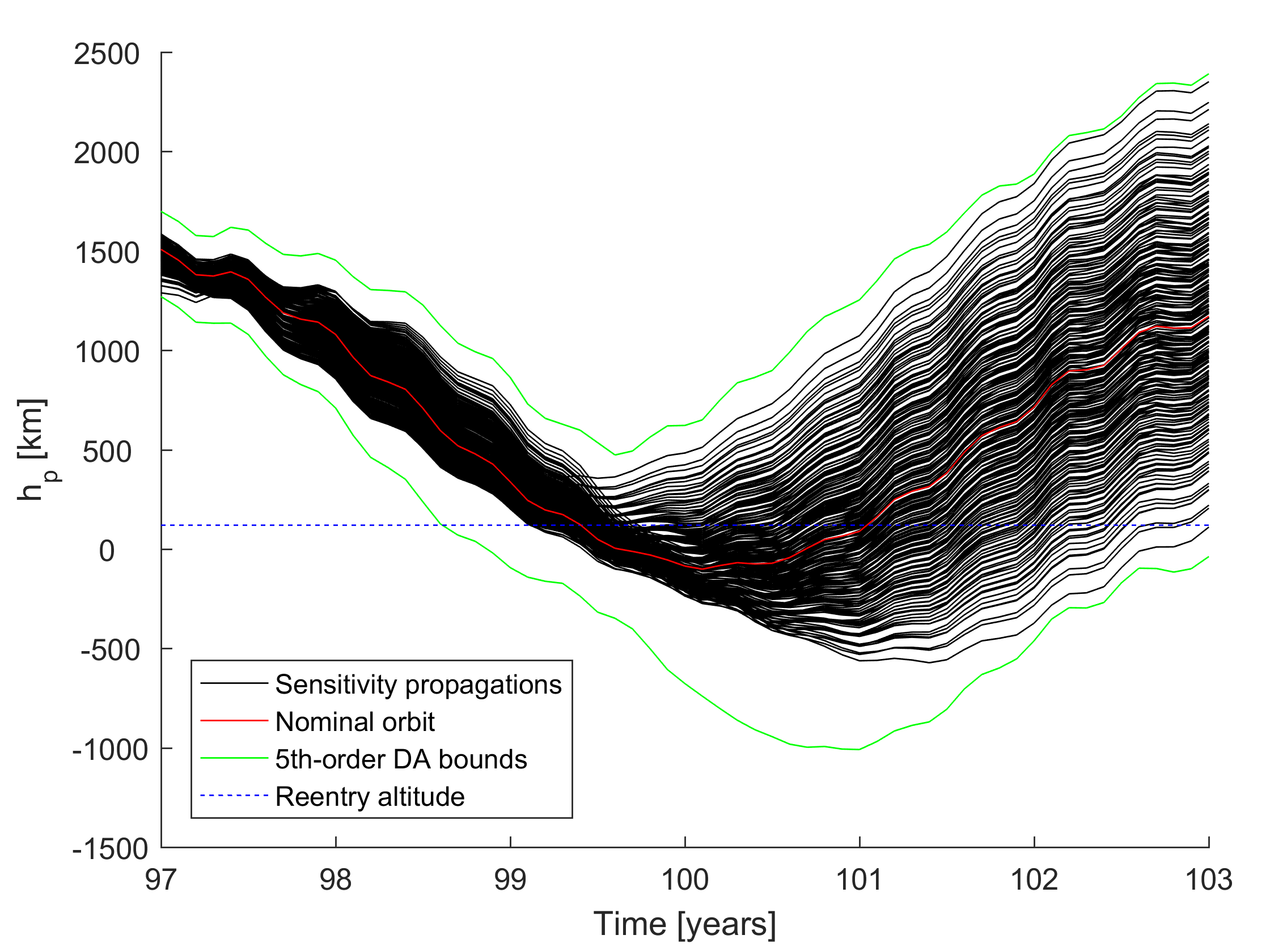}
\caption{Evolution of perigee altitude of 225 different orbits (black lines) due to manoeuvre uncertainties for object 37846 around the nominal re-entry epoch}
\label{fig:37846_case1_sensitivity_perigee}
\end{figure}

A close-up of the perigee altitude around 100 years shows that not all orbits in the uncertainty domain reach the re-entry altitude of 120 km, see \fref{fig:37846_case1_sensitivity_perigee}. This re-entry orbit is therefore not reliable in case of manoeuvring errors. The 5th-order DA bounds correctly indicate that part of the uncertainty set does not reach an 120 km altitude.

\begin{figure}
\centering
\includegraphics[width=\columnwidth]{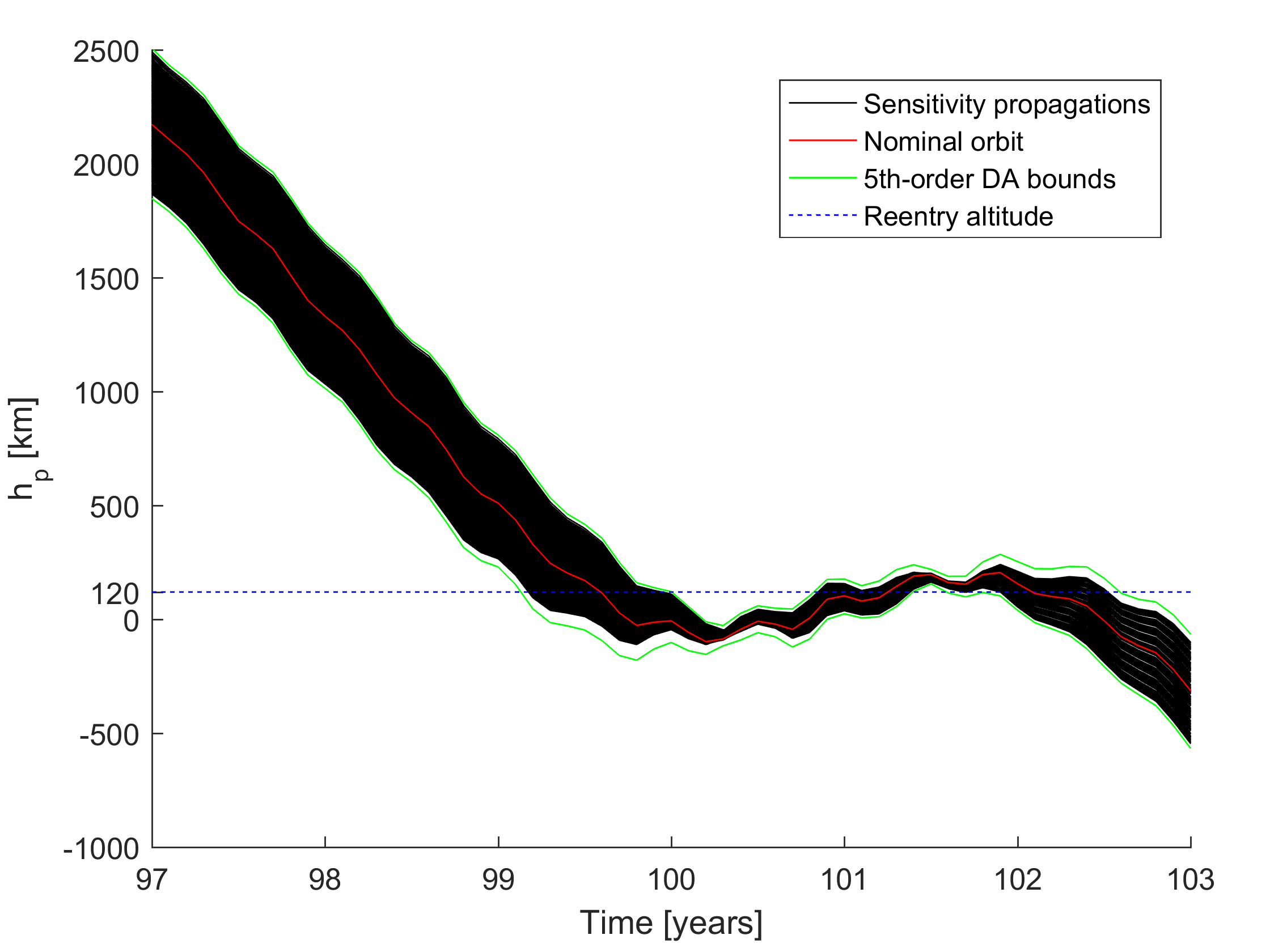}
\caption{Evolution of perigee altitude of 225 different orbits (black lines) due to manoeuvre uncertainties for object 40890 around the nominal re-entry epoch}
\label{fig:40890_case1_sensitivity_perigee}
\end{figure}

\begin{figure}
\centering
\includegraphics[width=\columnwidth]{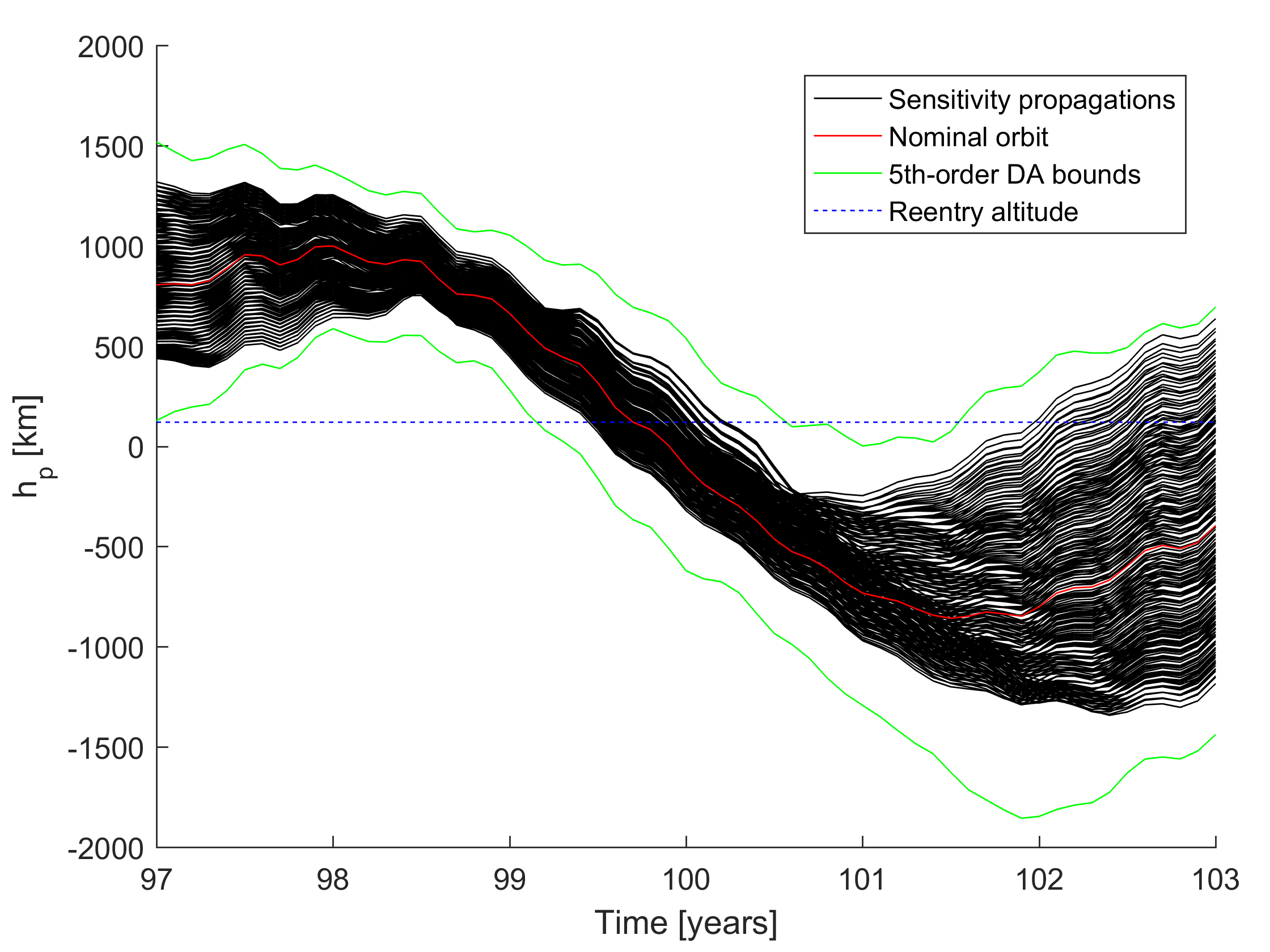}
\caption{Evolution of perigee altitude of 225 different orbits due to manoeuvre uncertainties for object 41175 around the nominal re-entry epoch}
\label{fig:41175_case1_sensitivity_perigee}
\end{figure}

Figs. \ref{fig:40890_case1_sensitivity_perigee} and \ref{fig:41175_case1_sensitivity_perigee} show close-ups of the perigee height around 100 years for the sensitivity analyses for objects 40890 and 41175. In these cases all orbits in the uncertainty set re-enter within 101 years. Therefore, these orbits are reliable disposal options considering realistic manoeuvre uncertainties. Furthermore, Figs. \ref{fig:37846_case1_sensitivity} to \ref{fig:41175_case1_sensitivity_perigee} show that neighbouring orbits diverge from the nominal trajectory and that the distance between orbits grow faster over time. However, up to re-entry the behaviour seems regular, since all orbits in the uncertainty domain evolve similarly to the nominal orbit.

\subsubsection{Graveyard}
\begin{figure*}
\centering
\includegraphics[width=0.8\textwidth]{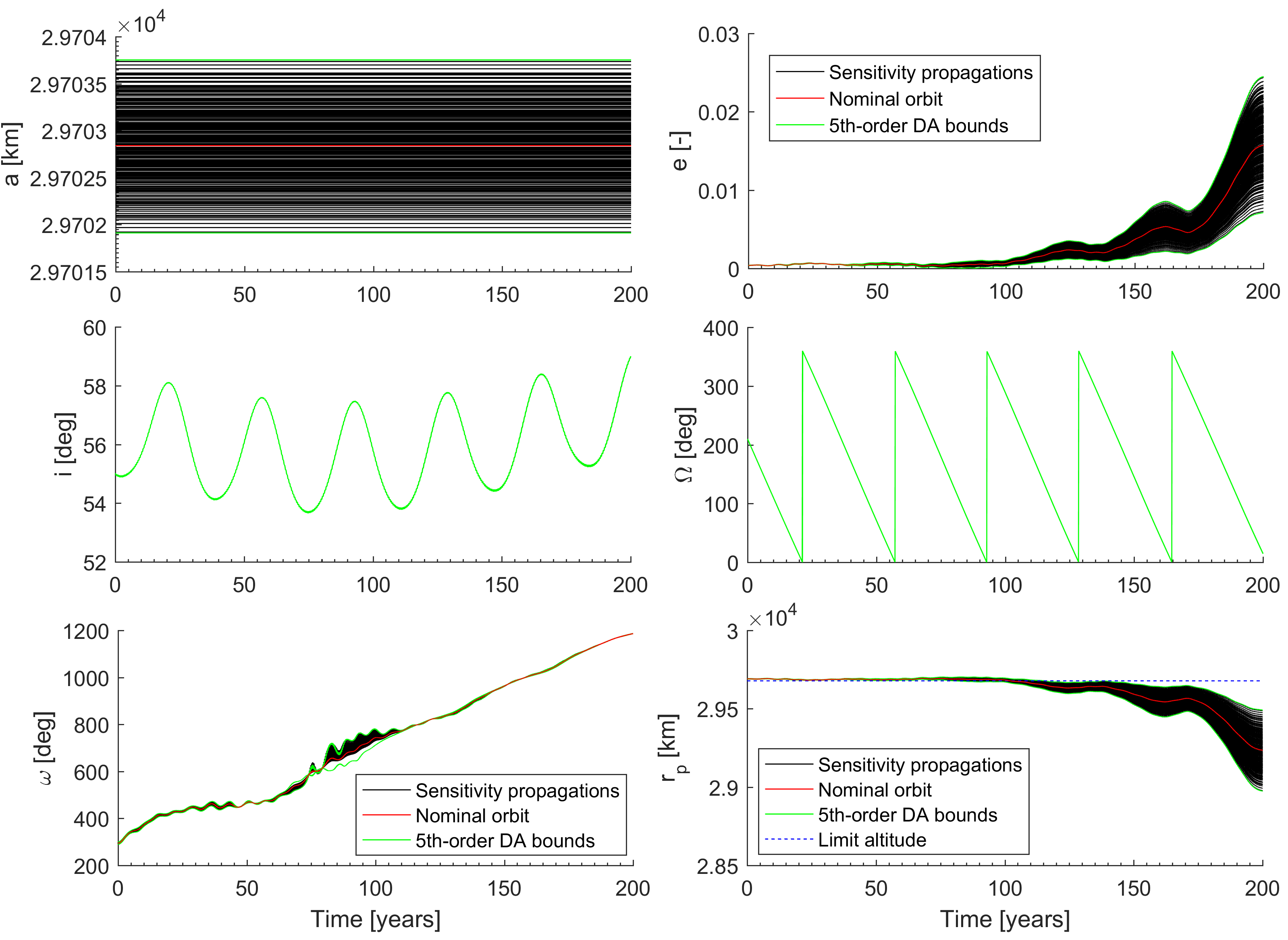}
\caption{200-year evolution of 225 different orbits (black lines) due to manoeuvre uncertainties for object 38858. The red curve is the nominal orbit and the green curves are the bounds of the uncertainty domain computed using a 5th-order DA expansion. The blue dashed line is the limit altitude for safe disposal.}
\label{fig:38858_case2_sensitivity}
\end{figure*}

\begin{figure}
\centering
\includegraphics[width=\columnwidth]{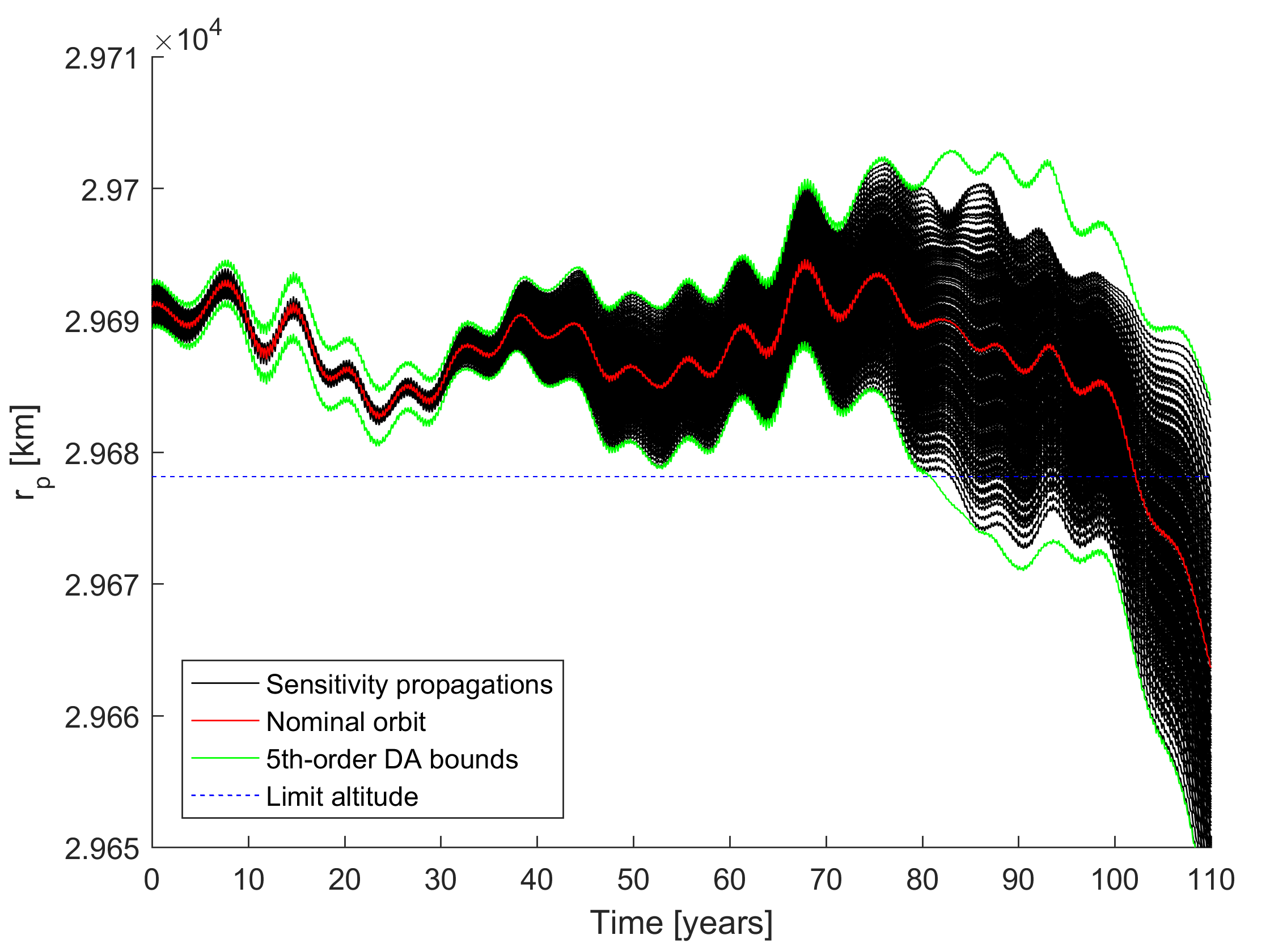}
\caption{110-year evolution of perigee altitude of 225 different orbits (in black) due to manoeuvre uncertainties for object 38858}
\label{fig:38858_case2_sensitivity_perigee}
\end{figure}

The evolution of the uncertainty domain for the graveyard disposal orbit of object 38858 is shown in \fref{fig:38858_case2_sensitivity}. The dashed blue line indicates the limit of the protected GNSS region, that is 100 km above the Galileo operational altitude. A close-up of the perigee radius for the first 110 years is shown in \fref{fig:38858_case2_sensitivity_perigee}. As expected, the nominal orbit remains above the limit altitude for 100 years, since that was the requirement during design. However, some orbits in the uncertainty set cross the limit altitude before 100 years. This means that the graveyard orbit is not reliable and could be improved such that all orbits in the uncertainty domain remain above the safe-distance altitude for at least 100 years.

It can be noted that all orbits in the initial domain are within the bounds computed using the DA expansion. Only when the eccentricity becomes very small around 70 years the bounds underestimate the range of $\omega$, see \fref{fig:38858_case2_sensitivity}. This can be contributed to strong non-linearities close to the singularity in the dynamical model at zero eccentricity. 
For the other orbital elements, the limits of the uncertainty set were estimated accurately over time using the bounds computed via a single propagation in DA. 

\begin{figure*}[htp]
     \centering
     \subfigure[Estimated truncation error]{\includegraphics[width=0.38\textwidth]{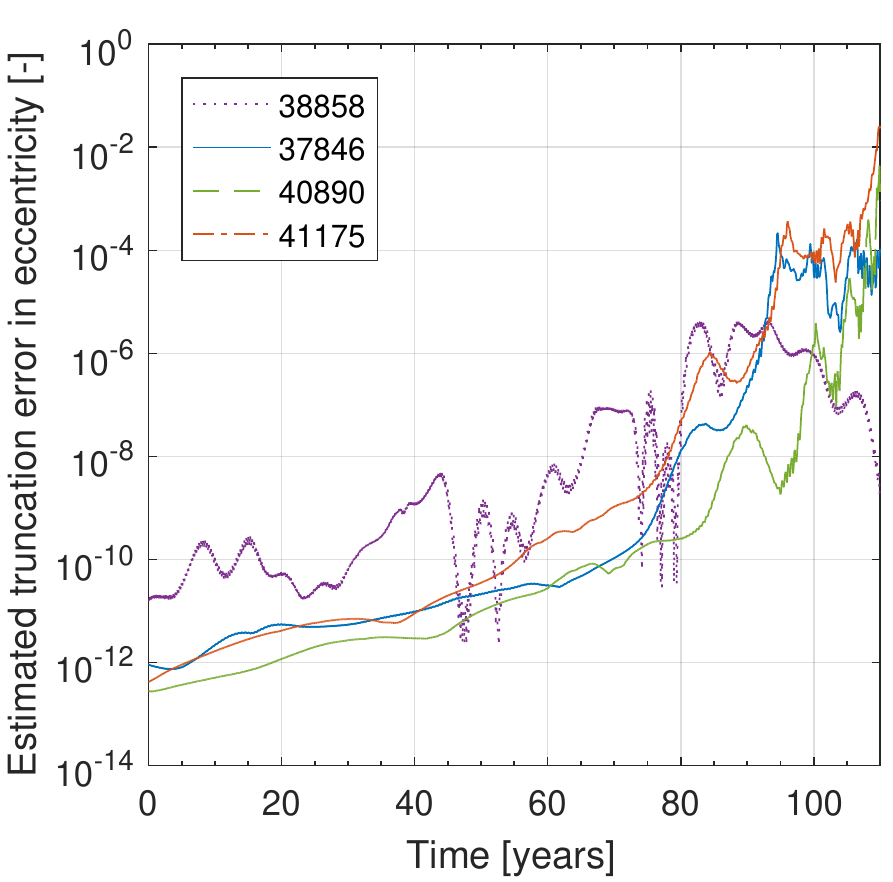} \label{fig:truncationErrorEcc}}
     \hspace{1pt}%
     \subfigure[Radius of domain where estimated truncation error $\le 10^{-5}$]{\includegraphics[width=0.38\textwidth]{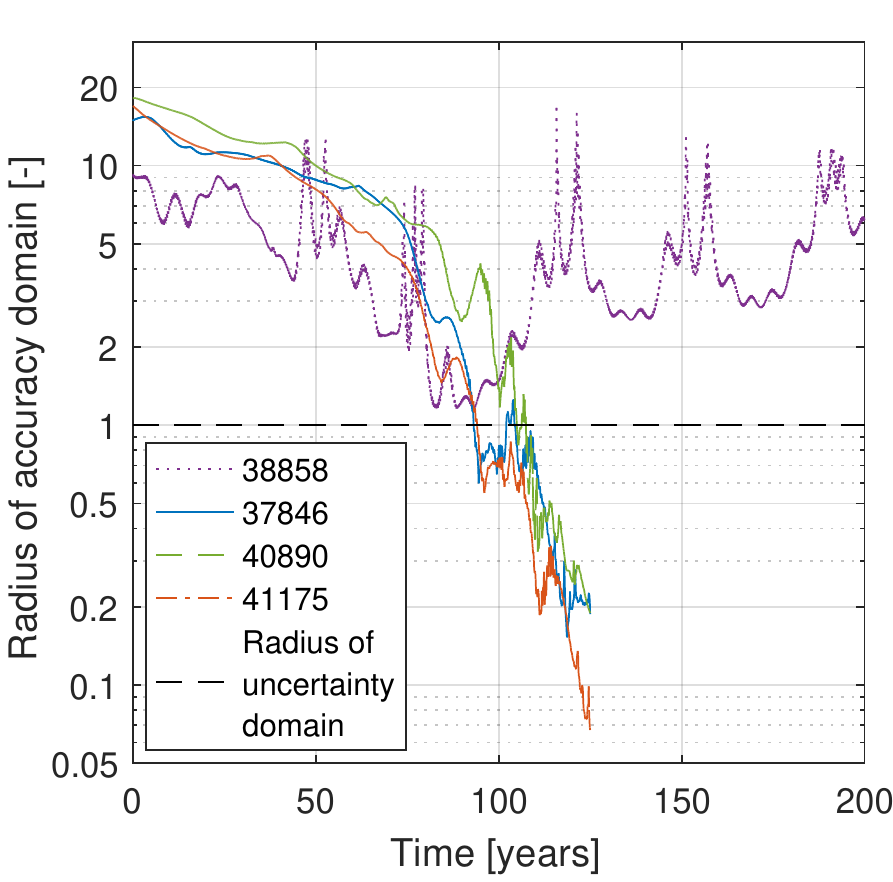} \label{fig:radConvEcc}}
     
     \caption{Estimated truncation error (left) and radius of domain where estimated truncation error $\le 10^{-5}$ (right) for 5th-order expansion of eccentricity with respect to initial manoeuvre uncertainties}
     \label{fig:TruncErrRadConvEcc}
\end{figure*}

To check the accuracy of the DA expansions, the estimated truncation errors of the 5th-order Taylor expansions of the eccentricity for the four disposal orbits are shown in \fref{fig:truncationErrorEcc}. The truncation error is at most $10^{-4}$ in the entire uncertainty set for all disposal orbits for the first 100 years. In addition, for all re-entry orbits the truncation error is less than $10^{-5}$ for at least the first 93 years, whereas for the graveyard orbit the error is never larger than $10^{-5}$. Considering that the semi-major axes of the orbits are constant and equal to approximately 30,000 km, an error in the eccentricity of $10^{-4}$ results a 3 km error in perigee altitude. Furthermore, the size of the domain where the estimated truncation error of the eccentricity expansion is smaller than $10^{-5}$ is shown in \fref{fig:radConvEcc}. For the graveyard orbit, the size of this domain is always larger than the uncertainty domain. These results show that Taylor expansions computed using DA can be used to accurately compute the evolution of orbits in the entire uncertainty domain due to manoeuvre errors. In addition, the small truncation errors indicate that the orbits do not behave chaotically on the reference time scale. Strong exponential divergence of neighbouring orbits would result in large high-order coefficients and would thus strongly increase the truncation error of the Taylor expansion. Still, although a small truncation error indicates that the orbit behaves regularly, it is no proof that all orbits in the uncertainty domain are regular.

To summarise, the sensitivity analysis results of the four disposal orbits have shown that the four disposal orbits do not exhibit chaotic behaviour on the reference time scale. The orbits starting in the uncertainty domain diverge over time but this seems to happen smoothly until re-entry for re-entry orbits and for 200 years for the graveyard orbit. In addition, the results showed that the evolution of the uncertainty set due to orbit insertion errors can be estimated accurately using a DA expansion by computing the bounds of the set. Furthermore, two of the three re-entry disposal options were found to be reliable in the sense that the satellite will re-enter even under manoeuvre uncertainties. The other re-entry orbit and the graveyard orbit do not satisfy the disposal constraints in case of manoeuvre errors.

\subsection{Sensitivity to dynamical model}
In the previous sections, the FTLE and Lyapunov time have been computed using the simple gravitational model, whereas for the sensitivity analysis we used the full gravitational model, see Section~\ref{sec:dynamics}. To verify that the simple model is sufficiently accurate for investigating the predictability of Galileo disposal orbits and to analyse the sensitivity of model uncertainties, two sensitivity analyses were repeated using different dynamical models.

\begin{figure*}[htp]
     \centering
     \subfigure[Perigee radius.]{\includegraphics[width=0.47\textwidth]{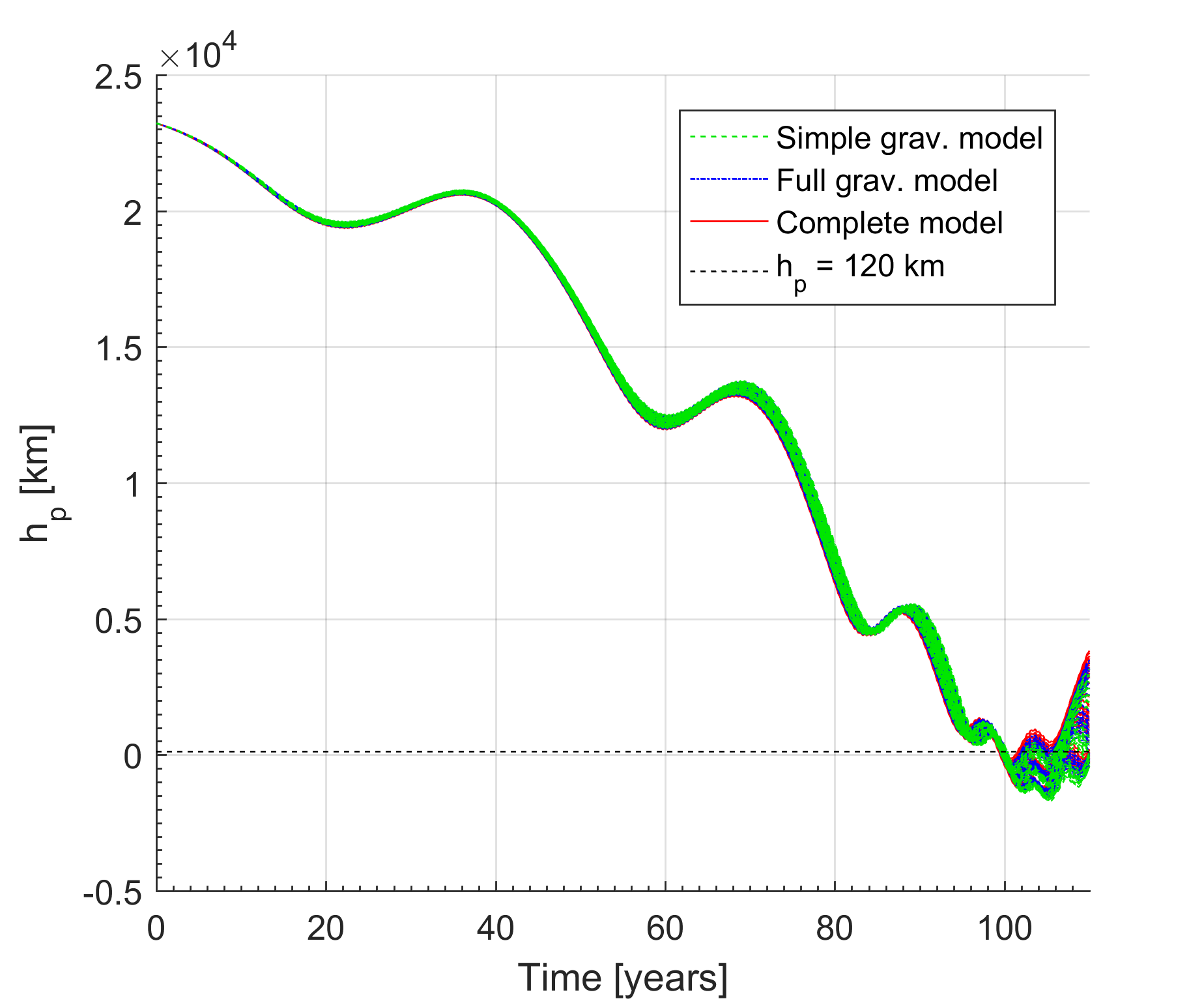} \label{fig:41175sensitivityDiffDynamicsPerigee}}
     \hspace{1pt}%
     \subfigure[Perigee radius around 100 years.]{\includegraphics[width=0.47\textwidth]{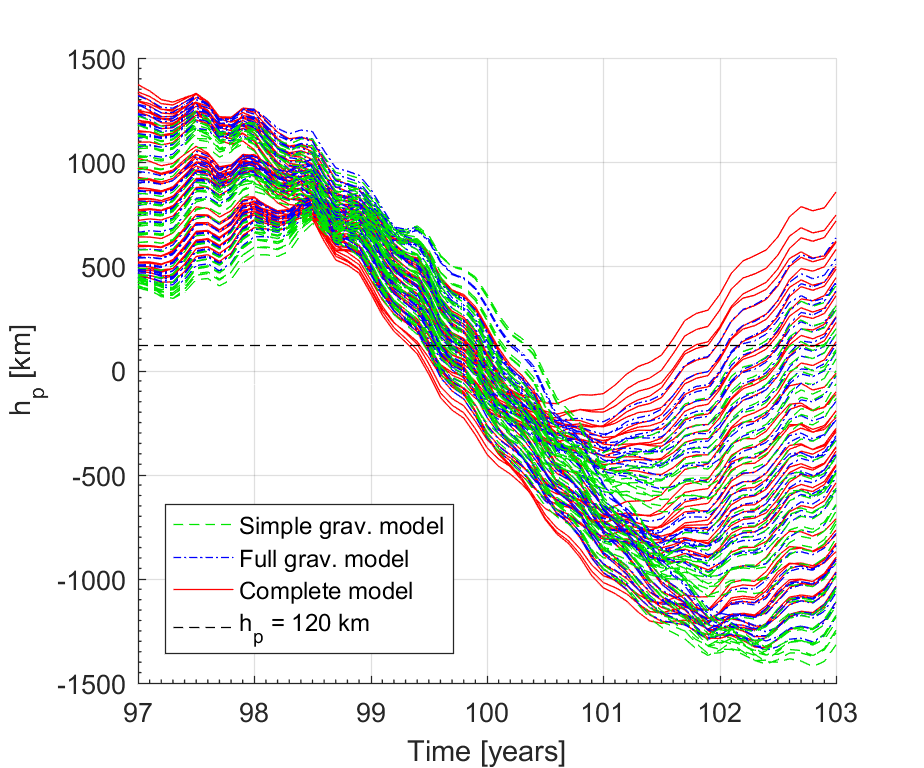} \label{fig:41175sensitivityDiffDynamicsPerigeeZoom}}
     
     \caption{Evolution of the perigee radius for object 41175 according to different dynamical models: simple gravitational model, full gravitational model and complete model}
     \label{fig:41175sensitivityDiffDynamics}
\end{figure*}

First, the sensitivity analysis for the re-entry orbit 41175 was carried out using the simple, full and complete model. \fref{fig:41175sensitivityDiffDynamicsPerigee} shows the evolution of the perigee altitude according to the three different dynamical models over 110 years for 45 orbits in the initial uncertainty domain. At 100 years, the maximum difference between the sets is 157 km in perigee altitude. This difference is significant regarding the strict re-entry altitude of 120 km and in comparison with the effect of manoeuvre uncertainties. On the other hand, the difference is small compared to the total change in perigee altitude, which is more than 23,000 km. Also, the divergence in perigee altitude due to uncertainties is similar for all models. The simple model can therefore be used to both qualitatively and quantitatively study the orbit (which is in agreement with \cite{daquin2016}) if highly accurate results are not required.

\begin{figure*}
\centering
\includegraphics[width=0.8\textwidth]{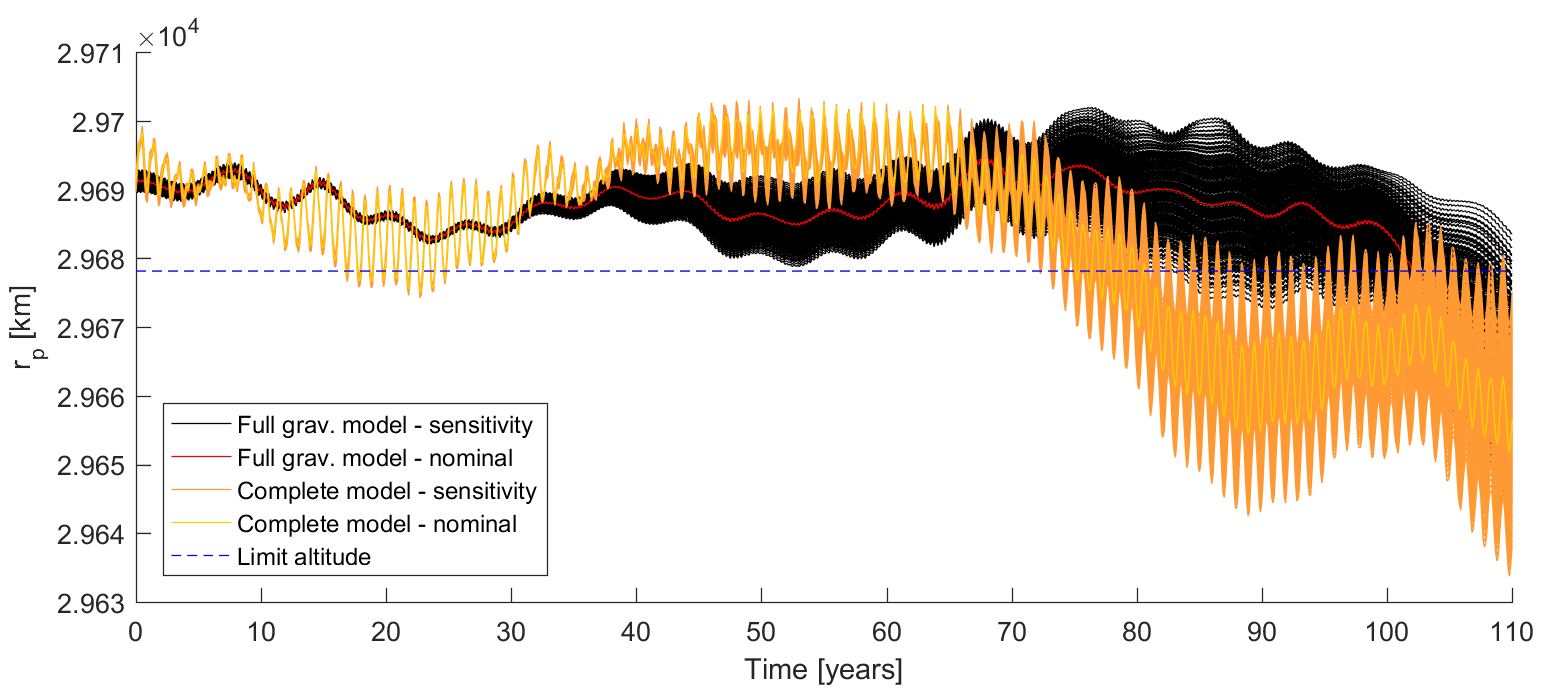}
\caption{110-year evolution of perigee radius of 225 different orbits in the uncertainty domain of object 38858 computed using the full gravitational model (black lines) and complete model (orange lines). The blue dashed line is the limit altitude.}
\label{fig:38858case2sensitivityDiffDynamicsPerigeeZoom}
\end{figure*}

However, for graveyard orbits the effect of SRP is more significant and should be not be neglected. \fref{fig:38858case2sensitivityDiffDynamicsPerigeeZoom} shows the evolution of the perigee radius for 225 orbits in the initial uncertainty domain of the graveyard orbit according to the full gravitational model and the complete model that also includes SRP. When SRP is included the nominal graveyard orbit crosses the safe-distance altitude after just 16.8 years due to oscillations in the eccentricity caused by SRP. This shows that for accurate predictions the effects of SRP and higher-order gravitational perturbations due to the Earth and Moon cannot be neglected and should at least be considered as an additional uncertainty in the orbit's evolution.

\begin{figure*}
\centering
\includegraphics[width=0.8\textwidth]{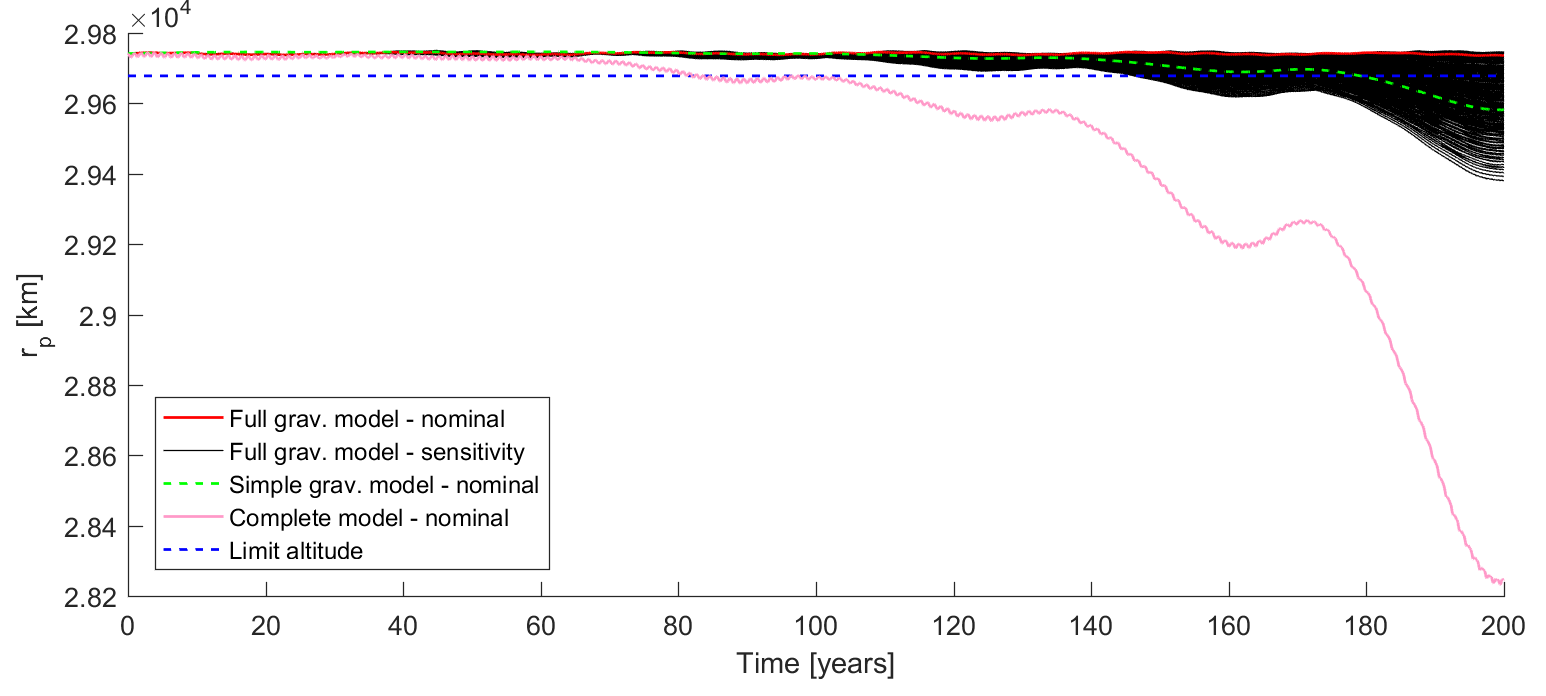}
\caption{200-year evolution of perigee radius computed using simple (green) and full gravitational model (red) and complete model (pink) starting at initial conditions: $a=29749.52$ km, $e=0.0002656$, $i=54.9752\degr$, $\Omega=210.6650\degr$, $\omega=116.6427\degr$, $M=-69.6767\degr$ and JD = 2457250.5345. Black lines show perigee of 225 orbits due to manoeuvre uncertainties in full gravitational model and blue lines indicates the limit altitude.}
\label{fig:38858case11sensitivityDiffDynamicsPerigeeZoom}
\end{figure*}

A more dramatic example of the influence of the dynamical model is shown in \fref{fig:38858case11sensitivityDiffDynamicsPerigeeZoom}. The graveyard orbit in this example is very stable in the full model; it stays 58 km above the safe distance for 200 years and the eccentricity remains smaller than 0.00046. In addition, in case of manoeuvre errors the safe-distance altitude is not crossed for 146 years. However, when SRP is included in the model, the safe-distance requirement is violated by the nominal orbit after only 81 years and the eccentricity grows to 0.05 in 200 years. 
This large difference in orbital evolution is related to behaviour of the argument of perigee. The SRP perturbs the orbit and due to the sensitivity of $\omega$, $\omega$ diverges strongly from the evolution it followed in the full gravitational model. As a consequence of the different evolution of $\omega$, the growth in eccentricity is much larger. This also happens when the simple model instead of the full model is used, but the effect on the eccentricity growth is smaller. This means that in the full gravitational model the orbit is stable in the sense of eccentricity change, but due to the sensitivity of the argument of perigee the orbital evolution is very sensitive to perturbations, such as SRP. Indeed, in Figs. \ref{fig:38858_eccArgP_maxEcc_200y} and \ref{fig:38858_RAANArgP_maxEcc_200y}, we have seen that the region of low eccentricity growth can be very small and a small perturbation can therefore move the orbit into a region of large eccentricity growth. For that reason, a large region of low eccentricity growth is required to ensure that a graveyard remains stable when subject to uncertainties in the dynamical model.

\section{Discussion}
\label{sec:discussion}

\subsection{FTLE analysis versus sensitivity analysis}
The FTLE analyses showed that all investigated disposal orbits are located in chaotic regions in the initial phase space (i.e. in regions where the FTLE is larger than zero) and that the Lyapunov time of the orbits is much shorter than the required prediction time. This suggests that the orbits behave chaotically and that predicting the orbits for 100 years or more is beyond the limit of predictability. However, the sensitivity analyses showed that the uncertainty domains due to manoeuvre errors grow over time but the evolutions of orbits in the uncertainty sets show no signs of chaotic behaviour. So, based on the FTLE and sensitivity analyses we can draw different conclusions about the predictability of the orbits on the time scale of interest.

Four features of the FTLE can explain the difference in conclusions:
\begin{enumerate}
\item The FTLE is computed in finite time, whereas LEs should be studied in infinite time;
\item The FTLE is based on linearised dynamics, whereas finite deviations behave non-linearly;
\item The FTLE looks at the direction of maximum growth, whereas in finite time a finite deviation in another direction may grow more in absolute terms;
\item The FTLE considers all orbital elements, whereas only the behaviour of the eccentricity is of interest.
\end{enumerate}

From various examples in literature, it is known that chaos indicators provide proper information about the chaotic behaviour of orbits in infinite time, but can show different results in finite time, see e.g. the discussion in \cite[p. 40-42]{lega2016}.
Indeed, we saw that FTLE plots tend to look very different after 100 and 200 years. In addition, FTLE plots computed using different coordinates often look different after 100 years, but tend to look similar on the long term (200 years). Moreover, it should be noted that the value of an FTLE after finite time does not provide conclusive information about chaotic behaviour, because we may find a positive Lyapunov exponent even if the growth is linear in time due to the ansatz \eqref{eq:exponentialGrowthOfDeviation}. Therefore, to be sure that the growth is exponential we should propagate for infinite time. However, in general, the FTLE is expected to converge to a constant value after a sufficiently long finite time.

\begin{figure*}
\centering
\includegraphics[width=0.8\textwidth]{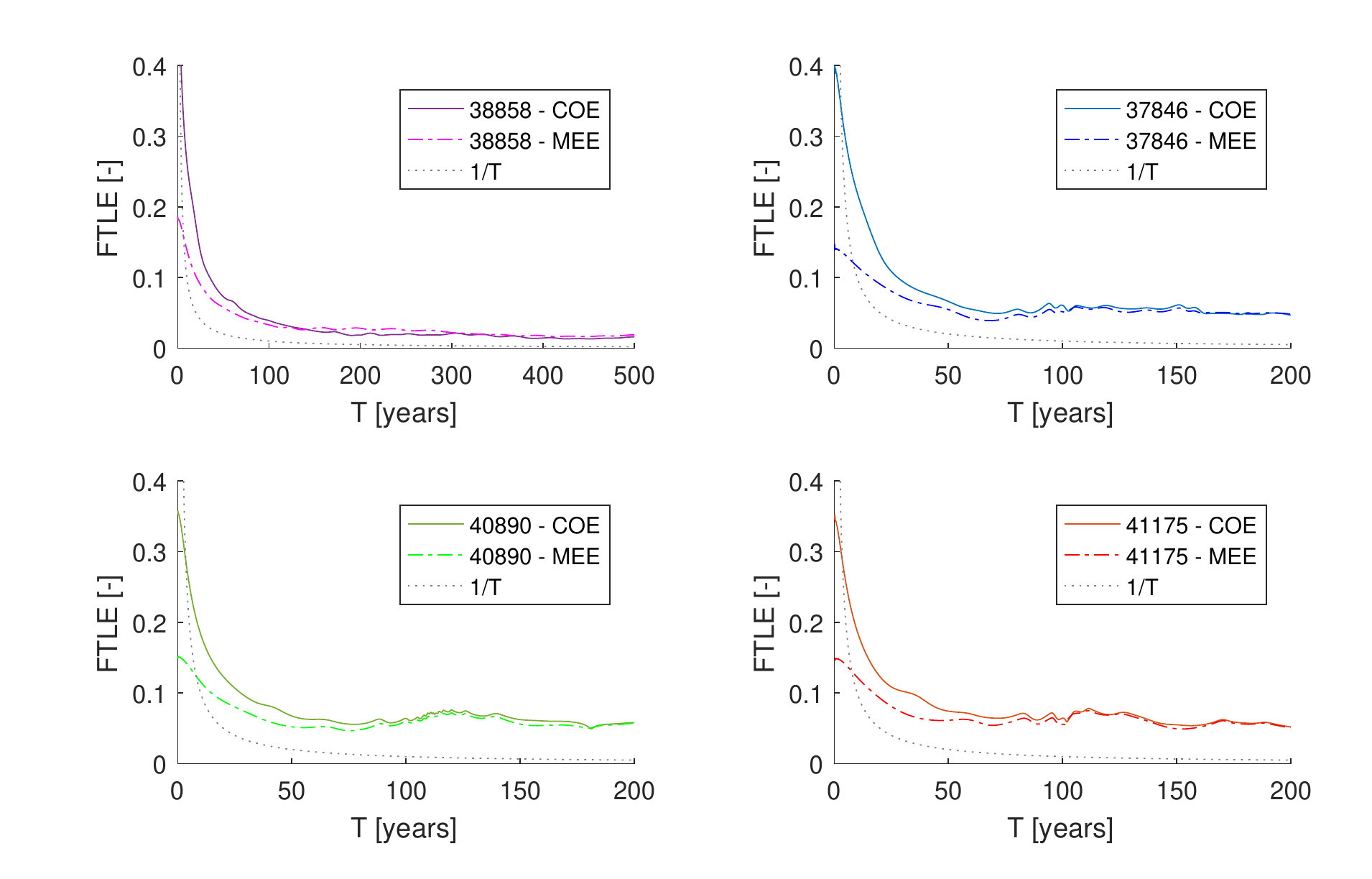}
\caption{Evolution of FLTE over time for four disposal orbits computed using COE and MEE coordinates}
\label{fig:evolutionFTLEoverTimeDiffCoord}
\end{figure*}

To check if the FTLE of the disposal orbits have converged, we computed their evolution over time. \fref{fig:evolutionFTLEoverTimeDiffCoord} shows the evolution of the FTLE for the four disposal orbit computed using COE and MEE. From these plots it is clear that the FTLEs computed using COE and MEE converge to the same value on the long term. In addition, the FTLEs for the re-entry disposal orbits (37846, 40890 and 41175) seem to have already converged after 100 years, whereas the FTLE of the graveyard orbit keeps decreasing for the first 500 years. The figures also contain a plot of the inverse of the time $T^{-1}$ that indicates the boundary between exponential divergence and slower divergence or contraction. The FTLEs of all orbits is larger $1/T$ except for the first few years, so all orbits are supposedly chaotic. Furthermore, the FTLE of the graveyard orbit is smaller than the FTLEs of the re-entry orbits, which suggests that the graveyard orbit behaves less chaotic than the re-entry orbits. Because the FTLE for the re-entry disposal orbits has already converged after 100 years, the finite time property of the FTLE does not fully explain the difference between the sensitivity and FTLE analysis.

The FTLE is based on linearised dynamics that are only valid for infinitesimal deviations. This approximation is sufficient to determine whether an orbit behaves chaotically or not. However, in reality we deal with finite uncertainties and therefore non-linear terms cannot be neglected. This means that the growth of a finite deviation can be larger or smaller than computed by the FTLE. 

In addition to this, the fact that the FTLE considers the direction of maximum growth could explain the difference in conclusions. 
The FTLE and Lyapunov time are based on the maximum stretching between neighbouring orbits, which is computed via the maximum eigenvalue of the Cauchy-Green deformation tensor. The eigenvector that corresponds to the maximum eigenvalue is the direction of maximum expansion. This means that an initial deviation will grow most (relative to its initial size) when it points in the direction of maximum growth. Deviations that are not aligned with the direction of maximum expansion will grow slower or may even shrink over time. 

Deviations from the nominal initial state due to manoeuvre uncertainties may only have a small component in the direction of the maximum growth. On the long term, this component in the direction of the maximum growth will become the largest component of the deviation, because it grows fastest. However, in finite time this component does not necessarily become the largest component of the deviation vector. Instead, another component of the deviation vector that grows relatively slower but is initially larger can increase more in absolute size. Moreover, the FTLE computed in this work considers growth in all orbital elements (except the fast angle), whereas we are mainly interested in the deviation in eccentricity. Therefore, we can find a high FTLE due to strong divergence in one of the state variables while the deviation in eccentricity only grows little. This was indeed the case for the FTLEs of both re-entry and graveyard orbits whose value was dominated by the deviation in $\Omega$ and $\omega$ (see e.g. Figs. \ref{fig:40890_eccIncl_FTLE_COE_aeiOw} and \ref{fig:38858_RAANargP_COE}).

To compare the maximum growth with the growth of finite deviations, let us consider the FTLE computed using COE for the re-entry orbit of object 40890 after 100 years. The value of the FTLE is 0.064, which corresponds to a growth factor of 603.3 (computed using the simple model). The Jacobian of the flow map corresponding to this FTLE is:
\begin{align}
\frac{d\phi^{t}}{dx_0} &=
\begin{bmatrix}
	\partial_a a & \partial_e a & \partial_i a & \partial_{\Omega} a & \partial_{\omega} a \\
	\partial_a e & \partial_e e & \partial_i e & \partial_{\Omega} e & \partial_{\omega} e \\
	\partial_a i & \partial_e i & \partial_i i & \partial_{\Omega} i & \partial_{\omega} i \\
	\partial_a \Omega & \partial_e \Omega & \partial_i \Omega & \partial_{\Omega} \Omega & \partial_{\omega} \Omega \\
	\partial_a \omega & \partial_e \omega & \partial_i \omega & \partial_{\Omega} \omega & \partial_{\omega} \omega 
\end{bmatrix}
\label{eq:Jacobian40890_100y} \\
&=
\begin{bmatrix}
	1 &	0 &	0 &	0 &	0 \\
	2.50 &	2.01 &	-0.52 &	0.07 &	0.02 \\
	0.84 &	9.66 &	0.21 &	-0.05 &	0.07 \\
	-104.80 &	-534.80 &	16.21 &	-0.84 &	-4.02 \\
	88.00 &	214.17 &	-37.99 &	3.65 &	1.62 
\end{bmatrix} \notag
\end{align}
Here, the angles $i$, $\Omega$ and $\omega$ are in radian and the semi-major axis is scaled by it's nominal value $a_{\mathrm{nom}} = 31086$ km. The Jacobian shows that the largest growth in deviation occurs in $\Omega$ and $\omega$ (see values in the fourth and fifth row). Consequently, the value of the FTLE after 100 years is dominated divergence in $\Omega$ and $\omega$. 

The corresponding direction of maximum expansion, that is, the eigenvector $\boldsymbol{u}$ corresponding to the maximum eigenvalue $\lambda_{\mathrm{max}}$ of $\Delta$, is: 
\begin{align*}
	\boldsymbol{u}_{\lambda_{\mathrm{max}}} = [ &0.206190, 0.977595, -0.041636, \\
	&0.002684, 0.007248]
\end{align*}
The largest possible deviation due to manoeuvre errors for object 40890 (see Table~\ref{tab:manoeuvreStateUncertainty}) that is aligned with the direction of maximum expansion is:
\begin{align*}
	\Delta \boldsymbol{x}_{\boldsymbol{u}} = [ &0.0001193 a_{\mathrm{nom}},	0.0005658,	-2.410\times10^{-5}\mathrm{~rad}, \\
	&1.553\times10^{-6}\mathrm{~rad},	4.195\times10^{-6}\mathrm{~rad}]
\end{align*}
According to the Jacobian \eqref{eq:Jacobian40890_100y} this deviation vector grows by a factor 603.3 and the magnitude of the deviation in $e$ grows to 0.001762 after 100 years. 
If, however, we apply this deviation to the nominal initial state and propagate the orbit for 100 years using the simple model, then the deviation grows by a factor 608 and causes a divergence from the nominal orbit of 3.7 km in $a$ and 0.000853 in $e$ after 100 years, which is equivalent to a deviation of 25.8 km in perigee altitude. 

On the other hand, the error in the disposal manoeuvre that causes the largest change in eccentricity after 100 years is an error of $-1\%$ in $\Delta V$ and $-1\degr$ in $\alpha$ and $\delta$, which corresponds to a deviation $\Delta \boldsymbol{x}_{e}$ in the initial orbital state of the disposal orbit of:
\begin{align*}
	\Delta\boldsymbol{x}_{e} = [ &-0.0005389 a_{\mathrm{nom}}, -0.0005139, \\
	& -7.737\times10^{-5}\mathrm{~rad}, 4.591\times10^{-5}\mathrm{~rad}, \\
	&-0.01219\mathrm{~rad}]
\end{align*}
This deviation grows by a factor of only 34, but causes a change in $a$ and $e$ of 16.8 km and 0.00426, respectively, resulting in a deviation in perigee altitude of 129 km after 100 years according to propagation in the simple model. 
Although the deviation is clearly not aligned with the maximum expansion direction and has only a small component in the direction of maximum expansion, it causes a larger change in eccentricity and perigee altitude. In addition, it can be noted that according to the Jacobian \eqref{eq:Jacobian40890_100y} the divergence in $e$ due to this initial perturbation $\Delta \boldsymbol{x}_{e}$ is only 0.00287, which shows that non-linear terms cannot be neglected.

This example demonstrates that large deviations due to manoeuvre uncertainties are generally not aligned with the direction of maximum growth. In addition, because the FTLE does not consider non-linear terms, the growth of the deviation in eccentricity according to FTLE is not accurate. Moreover, the largest divergence takes place in $\Omega$ and $\omega$ that are of little interest for re-entry and graveyard disposal orbits.

On the other hand, in the sensitivity analyses we did not look at the maximum possible growth of a deviation, but considered the growth of the whole uncertainty domain and focused on the uncertainty in eccentricity. The advantage is that here non-linear effects are taken into account and the magnitude of the actual divergence due to initial uncertainties is computed directly. So, instead of having an estimate of the divergence and chaotic behaviour, we directly calculate the effect of initial uncertainties on the disposal orbit.

\subsection{Practicality of chaos indicators}
We have just shown that the FTLE is not a suitable tool to estimate the divergence in orbital elements that are relevant for investigating the predictability of disposal orbits. In addition, some other drawbacks of the chaos indicators FTLE and Lyapunov time are:
\begin{enumerate}
\item The values of the FTLE and Lyapunov time depend on the choice of coordinates and units;
\item Chaos indicators only give information about a single orbit and not about nearby orbits, so a domain of orbits must be investigated;
\item FTLE plots do not include any information about required $\Delta V$ for disposal manoeuvre;
\item The FTLE does not provide information about the sensitivity to uncertainties in dynamical model;
\item It is not clear how to interpret the values of FTLE and Lyapunov time as measure of predictability of orbit.
\end{enumerate}
Because the values of the FTLE and Lyapunov time depend on the choice of coordinates and units, it cannot be used as an absolute measure for predictability. Instead, it should always be compared with the FTLE and Lyapunov time of other orbits in the phase space. This means that multiple orbits must be investigated and therefore there are no advantage with respect to a sensitivity analysis regarding computation time. 

The strongest divergence observed in the work was found in the evolution of the argument of perigee. This chaos was correctly detected by FTLE analysis, however, only by FTLE that explicitly considers $\omega$ (compare e.g. the FTLE computed using COE and MEE in Figs. \ref{fig:38858_eccArgP_FTLE_COE_100y} and \ref{fig:38858_eccArgP_FTLE_MEE_100y}, respectively). Moreover, the FTLE plots do not clearly show the chaos in $\omega$ after long periods of time, but only for short times (see \fref{fig:38858_eccArgP_FTLE_COE_200y}). Therefore, the chaotic behaviour of $\omega$ is only detected in finite time and if $\omega$ is a state variable. Knowledge of the behaviour of $\omega$ is important, because $\omega$ affects the eccentricity evolution due to lunisolar resonances and thus chaotic behaviour of $\omega$ is undesirable for safe graveyard disposal, see e.g. \fref{fig:38858case11sensitivityDiffDynamicsPerigeeZoom}. Fortunately, the effect of the behaviour of $\omega$ on the eccentricity is also visible by simply looking at the eccentricity evolution of different orbits, since a long-term small eccentricity is only possible in very narrow region of initial $\omega$, see \fref{fig:38858_eccArgP_maxEcc_200y}.

FTLE plots may be used for finding suitable graveyard orbits, because orbits that do not sensitively depend on the initial conditions are favourable candidates for safe disposal. However, FTLE plots do not provide any information about required $\Delta V$ to achieve a certain initial condition. Therefore, pruning the phase space by optimising the disposal manoeuvre while satisfying specified disposal requirements seems a more efficient approach for obtaining suitable disposal orbits.

Some researchers have argued that there exists a relation between the Lyapunov time and the time until strong event happens, e.g. an escape or a close encounter with another body \cite{morbidelli1995}. We have not investigated if such a relationship exists between the Lyapunov time and time to re-entry for MEO orbits. However, even if such a relation is found, then the correlation is likely to be too weak to draw conclusions for individual orbits.

We have mentioned several drawbacks of the use of FTLE for analysing the predictability of orbits. 
Nevertheless, it must be stressed that chaos indicators are very useful tools, because they enable the detection of dynamical structures in the phase space, which is not possible by simply propagating orbits.

\subsection{DA-based sensitivity analysis}
As an alternative for point-wise propagation we have used a high-order Taylor expansion of the flow. The truncation error of a 5th-order DA expansion was estimated to result in a 3 km error in perigee altitude after 100 years. This error is small compared to the 157 km difference in perigee altitude as result of the use of different dynamical models. Furthermore, the estimated bounds of the Taylor polynomials contained the entire set of orbits due to initial uncertainties. However, the DA bounds were also found to significantly overestimate the uncertainty domain. On the other hand, the fact that we can accurately estimate the evolution of orbits in the uncertainty domain using a 5th-order Taylor expansion gives confidence that the orbit does not behave chaotically. If the flow would be strongly chaotic then the truncation error of Taylor expansion would grow rapidly because strong exponential divergence cannot be approximated accurately using polynomial expansions. This was, however, not the case for all tested re-entry orbits until re-entry and for the graveyard orbit for 200 years. 

\subsection{Reliability of disposal orbits}
The results have shown that reliable disposal via re-entry is possible. In two of the three considered re-entry cases, the disposal resulted in re-entry for all considered initial conditions due to manoeuvre uncertainties. On the other hand, for test case 37846, re-entry under manoeuvre uncertainties was not guaranteed. Therefore, this orbit should be modified to ensure re-entry. In addition, it was found that uncertainties in the dynamical model have little effect on the orbital evolution of the re-entry orbits. Therefore, there is no reason for concern about the predictability of these orbits. 

For the investigated graveyard orbit, we found that there is only a very narrow range of initial argument of perigee that results in safe disposal for 100 years. Manoeuvre errors can cause deviations in the initial argument of perigee of several degrees that result in increased eccentricity growth and violation of the safe disposal requirements. In addition, it was shown that neglected perturbations in the dynamical model may cause the orbit to evolve differently and cross the safe-distance altitude much earlier than required. Therefore, reliable disposal in graveyard orbits is not always feasible with a limited $\Delta V$ budget. To guarantee safe disposal, a higher $\Delta V$ budget is needed to move the satellite further away from the operational altitude or to change the initial inclination or longitude of the node to reduce the sensitivity to initial $\omega$.

\section{Conclusions}
We have studied the predictability of Galileo disposal orbits by computing their FTLE and Lyapunov time and analysing the sensitivity of the orbital evolution due to manoeuvre uncertainties using both a numerical and DA approach. 
The results showed that the studied Galileo disposal orbits are chaotic according to FTLE. However, according to sensitivity analysis the orbits are predictable on the reference time scale and no chaotic behaviour is exhibited. In addition, two of the three studied re-entry disposal orbits re-enter for all considered errors in the disposal manoeuvre. The studied graveyard orbit is not a reliable disposal option, because it may not keep the required safe distance from the Galileo constellation due to uncertainties in the disposal manoeuvre and dynamical model.

We argue that sensitivity analysis is the proper way to study the predictability of disposal orbits. Chaos indicators such as the FTLE and Lyapunov time do not provide good indications of the limits of predictability and tend to indicate chaotic behaviour long before it manifests. Another drawback of the FTLE and Lyapunov time is their dependence on choice of coordinates and units in finite time which implies that the value of the FTLE and Lyapunov time cannot be taken as absolute measure of chaos and should be compared against other values in the phase space. In addition, the FTLE and Lyapunov time indicate chaos in all orbital elements while we are mainly interested in the behaviour of the eccentricity. Furthermore, the FTLE and Lyapunov time do not consider the uncertainties in the initial orbital manoeuvre, but assume any perturbation or uncertainty to be equally likely, which is not the case in practice. Moreover, using sensitivity analysis we can quantify the growth in uncertainties, which is what we are interested in, whereas FTLE only provide a linear approximation of their relative growth. 

These results show that chaos indicators should be used for qualitative analysis, such as detecting the dynamical structures in the phase space, but do not provide sufficient information for determining whether or not an orbit behaves chaotically on a specific finite time scale.  Orbits have to be propagated for times longer than the reference time scales considered in this work for the FTLE to converge and provide useful information. When searching for stable graveyard orbits or feasible re-entry disposal orbits it may be more efficient to prune the phase space using an optimization algorithm than to analyse the full phase space using FTLE plots. 

Finally, because the disposal orbits were shown to be predictable, there seems to be no reason for concern about the predictability of orbits during disposal orbit design. Re-entry disposal of Galileo satellites can therefore be considered a reliable and feasible option for end-of-life disposal.

\begin{acknowledgements}
The authors are grateful to Dr Juan Felix San Juan and Dr Martin Lara of Universidad de La Rioja for their support and for providing the HEOSAT propagation model. 
David Gondelach was funded by the Surrey Space Centre (SSC) to undertake his PhD at the University of Surrey. Finally, the use of the Differential Algebra Computational Engine (DACE) developed by Dinamica Srl is acknowledged.
\end{acknowledgements}

\noindent \small \textbf{Conflicts of interest} The authors declare that there are no conflicts of interest regarding the publication of this paper.

\section*{Appendix: FLI comparison}
\label{app:FLIdiffUnits}
\begin{figure*}[htp]
     \centering
     \subfigure[][$i$, $\Omega$ and $\omega$ in radians and $a$ unitless]{\includegraphics[width=0.35\textwidth]{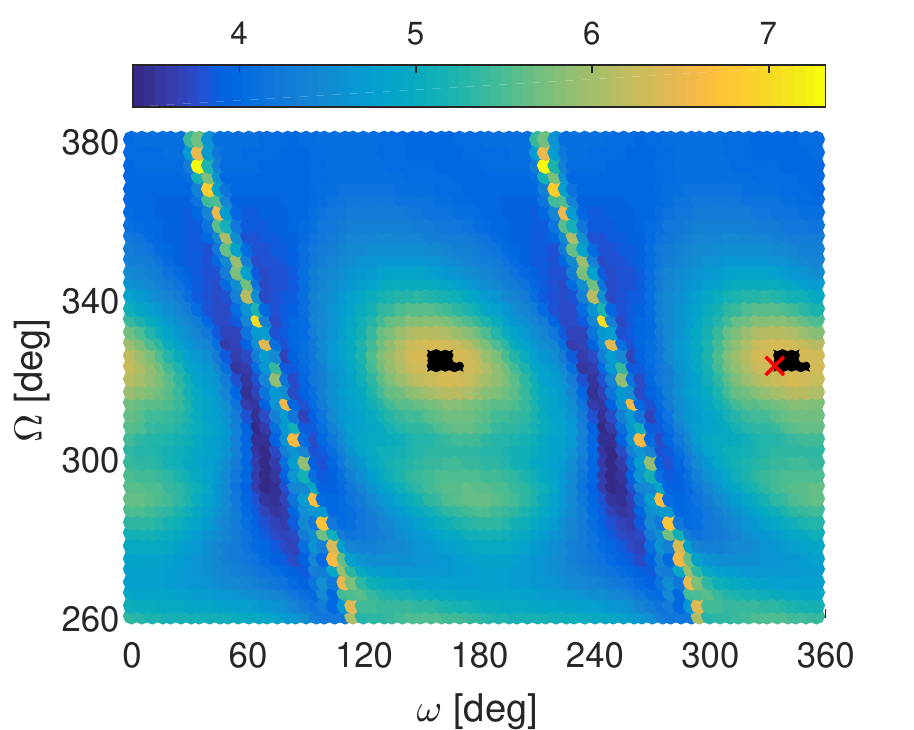} \label{fig:40890_FLI_100y}}
     \hspace{8pt}%
     \subfigure[][$i$, $\Omega$ and $\omega$ in radians and $a$ unitless]{\includegraphics[width=0.35\textwidth]{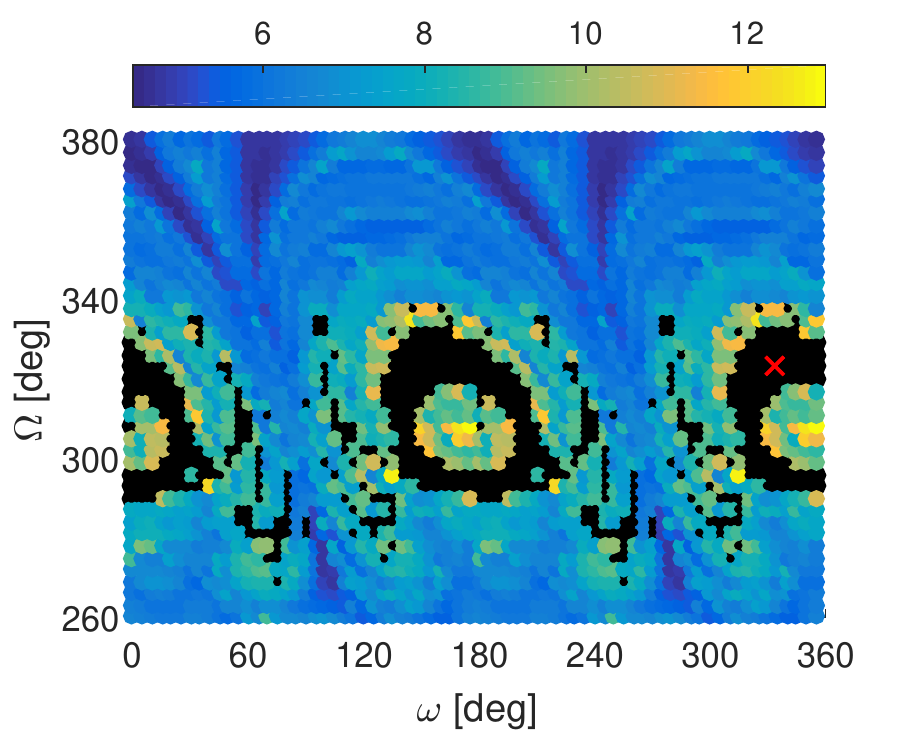} \label{fig:40890_FLI_200y}}
     
     \subfigure[][$i$, $\Omega$ and $\omega$ in degrees and $a$ unitless]{\includegraphics[width=0.35\textwidth]{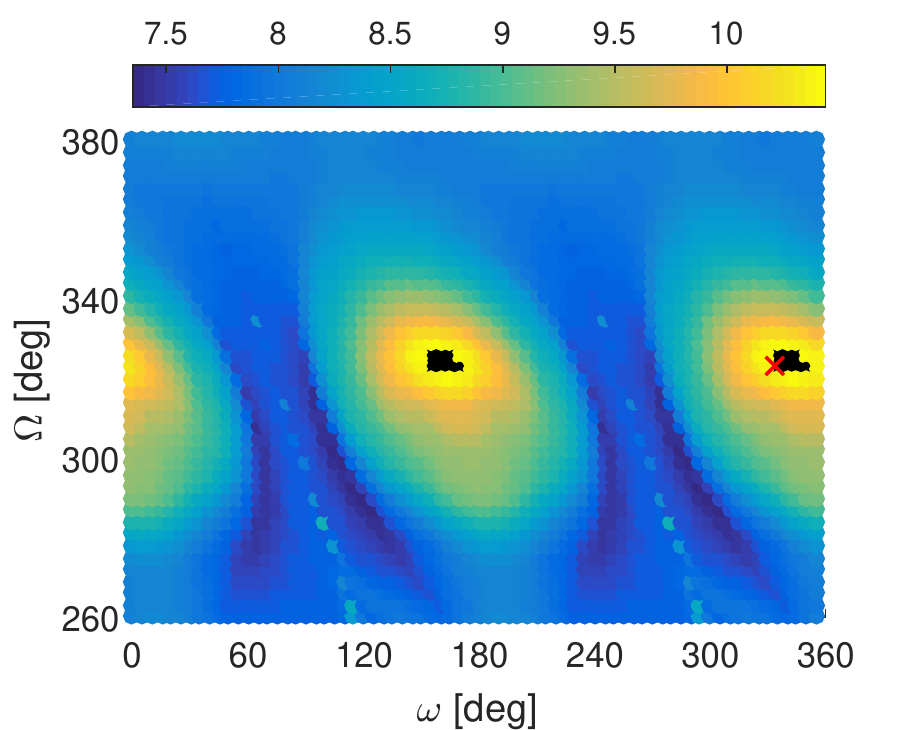} \label{fig:40890_DEGREES_FLI_100y}}
     \hspace{8pt}%
     \subfigure[][$i$, $\Omega$ and $\omega$ in degrees and $a$ unitless]{\includegraphics[width=0.35\textwidth]{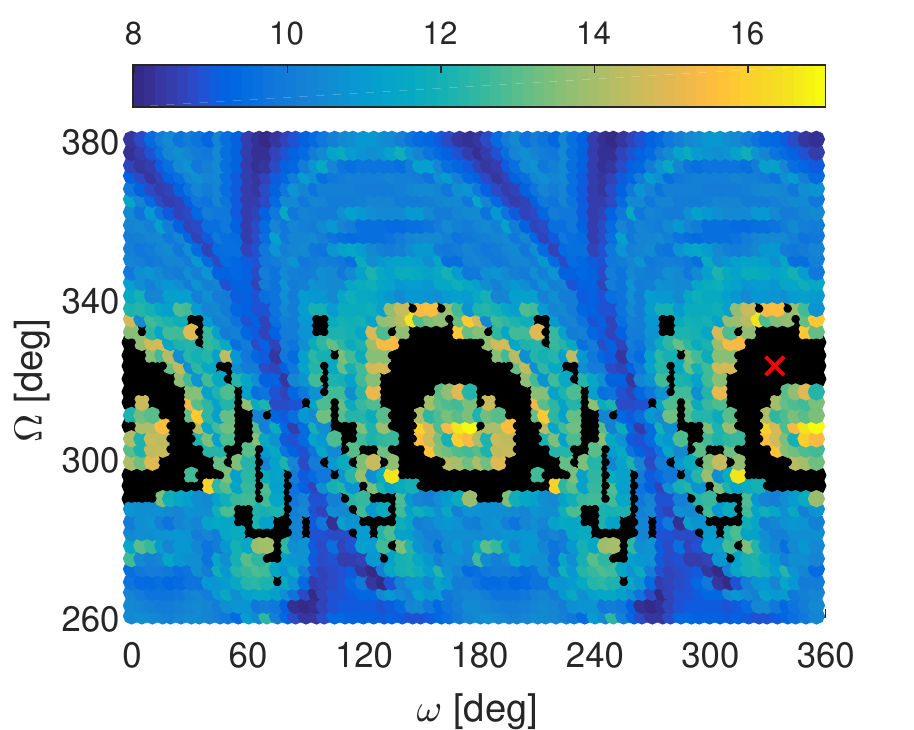} \label{fig:40890_DEGREES_FLI_200y}}
     
     \subfigure[][$i$, $\Omega$ and $\omega$ in radians and $a$ in km]{\includegraphics[width=0.35\textwidth]{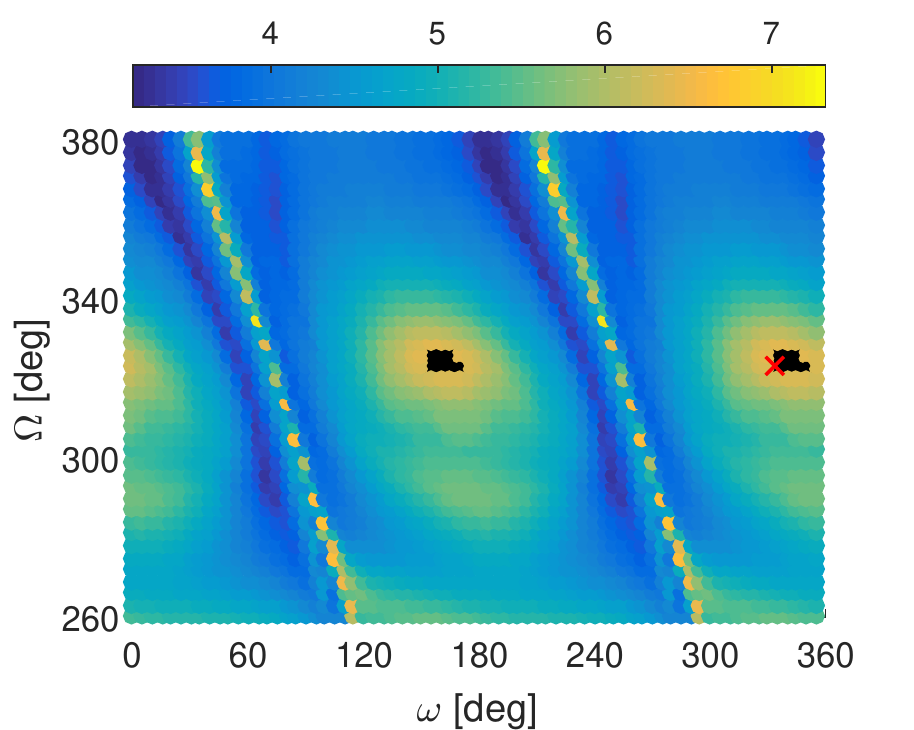} \label{fig:40890_KM_FLI_100y}}
     \hspace{8pt}%
     \subfigure[][$i$, $\Omega$ and $\omega$ in radians and $a$ in km]{\includegraphics[width=0.35\textwidth]{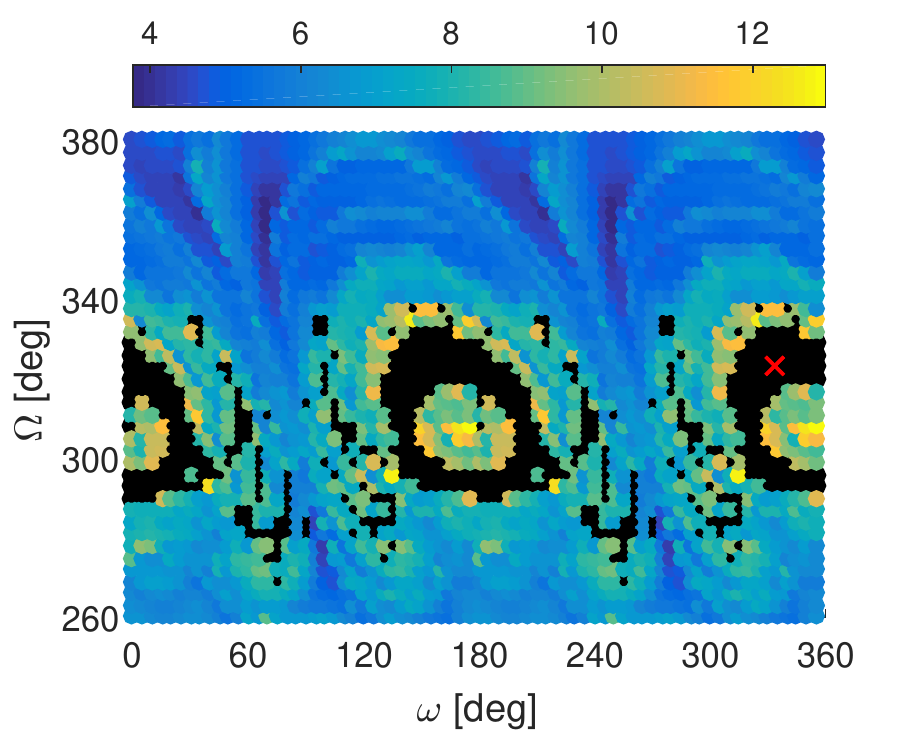} \label{fig:40890_KM_FLI_200y}}
     
     \subfigure[][$i$, $\Omega$ and $\omega$ in degrees and $a$ in km]{\includegraphics[width=0.35\textwidth]{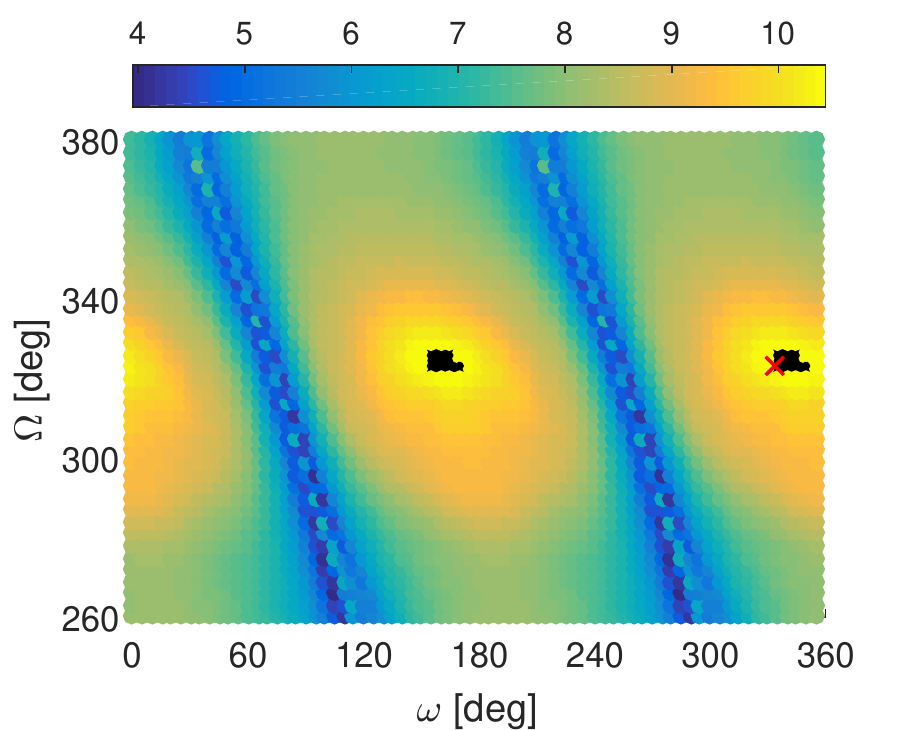} \label{fig:40890_DEGREES_KM_FLI_100y}}
     \hspace{8pt}%
     \subfigure[][$i$, $\Omega$ and $\omega$ in degrees and $a$ in km]{\includegraphics[width=0.35\textwidth]{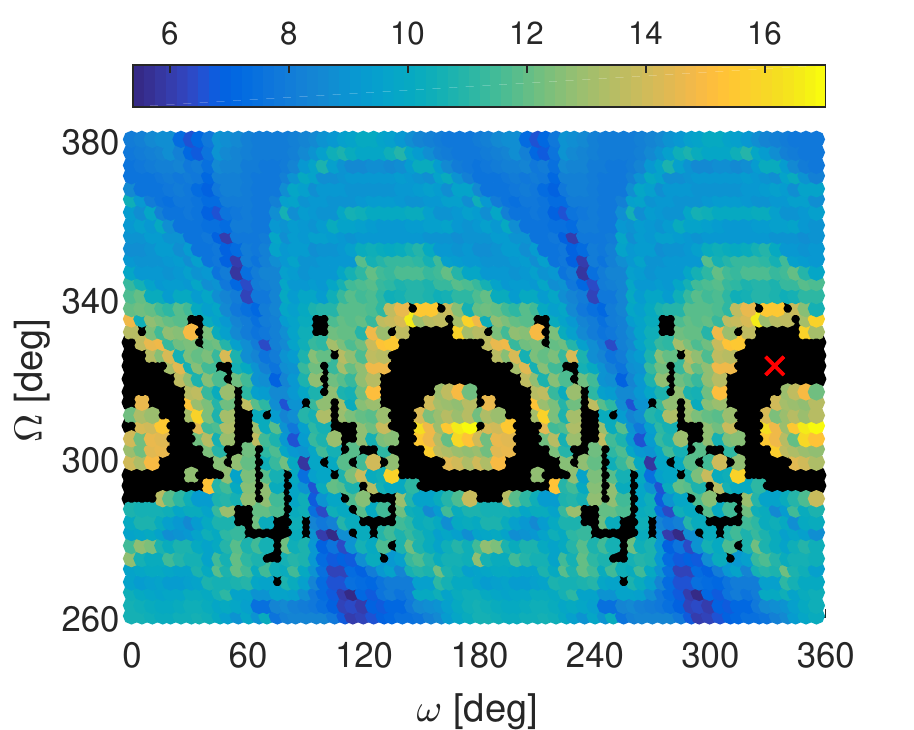} \label{fig:40890_DEGREES_KM_FLI_200y}}
     
     \caption{FLI (computed using five different initial deviation vectors to evaluate Eq.~\eqref{eq:FLI} and taking the maximum); for different initial $\Omega$ and $\omega$ for case 40890 computed using different units for $a$, $i$, $\Omega$ and $\omega$ after 100 (left) and 200 years (right). The unit used for $i$, $\Omega$ and $\omega$ is either radian or degree and for $a$ the unit is km or it is unitless, that is $a/a_0$.}
     \label{fig:40890_FLI_diffUnits}
\end{figure*}

For comparison, \fref{fig:40890_FLI_diffUnits} shows the FLI computed using different units for the semi-major axis $a$ and angles $i$, $\Omega$ and $\omega$. As for the FTLE, the choice of units affects both the absolute values of the FLI and their relative values.

\bibliographystyle{acm}
\bibliography{library2}

\end{document}